\documentclass[journal, 10pt]{IEEEtran}
\usepackage{amsmath}
\usepackage{amsfonts}
\usepackage{amssymb}
\usepackage{algorithm}
\usepackage{url}
\usepackage{algpseudocode}
\usepackage{graphicx}
\usepackage{epstopdf}
\usepackage{subcaption}
\usepackage[numbers,sort&compress]{natbib}
\usepackage{cases}
\usepackage{color}
\usepackage{tikz}
\usetikzlibrary{shapes,arrows}
\usepackage{verbatim}
\usetikzlibrary{fit,backgrounds,positioning}
\usepackage{graphbox}
\usepackage[font=footnotesize]{caption}
\usepackage{etoolbox}\AtBeginEnvironment{algorithmic}{\footnotesize}
\algdef{SE}[DOWHILE]{Do}{DoWhile}{\algorithmicdo}[1]{\algorithmicwhile\ #1}
\DeclareMathOperator*{\argmin}{\arg\!\min}
\DeclareMathOperator*{\argmax}{\arg\!\max}

\ifCLASSINFOpdf
\else
\fi

\algdef{SE}[SUBALG]{Indent}{EndIndent}{}{\algorithmicend\ }%
\algtext*{Indent}
\algtext*{EndIndent}

\begin{document}

\title{Greedy Approximate Projection for Magnetic Resonance Fingerprinting with Partial Volumes}




\author{Roberto~Duarte,
        Audrey~Repetti,
        Pedro~A. G\'omez,
        Mike Davies
        and Yves Wiaux
\thanks{R. Duarte, A. Repetti and Y. Wiaux are with the Institute of Sensors Signals and Systems, Heriot-Watt University, Edinburgh, UK (e-mail: rd8@hw.ac.uk; a.repetti@hw.ac.uk; y.wiaux@hw.ac.uk).}
\thanks{P. A. G\'omez is with the Munich School of Bioengineering, Technische Universit\"at M\"unchen, Munich, Germany (e-mail: pedro.gomez@tum.de).}
\thanks{M. Davies is with the Institute for Digital Communications (IDCom), The University of Edinburgh, EH9 3JL, UK (e-mail: mike.davies@ed.ac.uk)}
\thanks{Manuscript received xxxx; revised xxxx.}}

\maketitle

\begin{abstract}
In quantitative Magnetic Resonance Imaging, traditional methods suffer from the so-called Partial Volume Effect (PVE) due to spatial resolution limitations. As a consequence of PVE, the parameters of the voxels containing more than one tissue are not correctly estimated. Magnetic Resonance Fingerprinting (MRF) is not an exception. The existing methods addressing PVE are neither scalable nor accurate. We propose to formulate the recovery of multiple tissues per voxel as a non-convex constrained least-squares minimisation problem. To solve this problem, we develop a memory efficient, greedy approximate projected gradient descent algorithm, dubbed GAP-MRF. Our method adaptively finds the regions of interest on the manifold of fingerprints defined by the MRF sequence. We generalise our method to compensate for phase errors appearing in the model, using an alternating minimisation approach. We show, through simulations on synthetic data with PVE, that our algorithm outperforms state-of-the-art methods. Our approach is validated on the EUROSPIN phantom and on \textit{in vivo} datasets.

\end{abstract}


\section{Introduction}
Magnetic Resonance Imaging (MRI) is a powerful tool for diagnosis in medicine. Its main advantage over other medical imaging modalities is that MRI acquisitions are non-ionising and non-invasive. Nevertheless, the main drawback of MRI is that it produces qualitative images whose intensity values are a nonlinear response to underpinning physical parameters. Quantitative MRI (qMRI) is a particular modality that aims to produce spatial quantitative maps of parameters related to the tissues under investigation, such as $T_1$ and $T_2$ relaxation times \cite{jara2013theory}. Unfortunately, due to prohibitively long acquisition times, qMRI is not the standard for diagnosis. 
To overcome this difficulty, Magnetic Resonance Fingerprinting (MRF) was introduced to accelerate qMRI acquisitions \cite{MRF}, inspired by Compressive Sensing (CS) theory \cite{CS1}. MRF uses a combination of random excitation pulse sequences and $k$-space (i.e. Fourier space) undersampling to simultaneously acquire all relevant quantitative information. These random excitation sequences are used to produce unique temporal patterns called \textit{fingerprints}, which are compared to the ones predicted by the model to extract the parameters of interest. More recently, a full CS strategy was formulated in \cite{CSMRF} for MRF. In this work, the authors developed an iterative projection algorithm (also known as projected gradient descent, or forward-backward algorithm \cite{FISTA,Tseng2000}), dubbed BLoch response recovery via Iterative Projection (BLIP), reconstructing MRF signal with less acquisitions than the traditional MRF method \cite{MRF}.

\begin{figure}[!t]
\centering\scalebox{1}{
\begin{tikzpicture}

\node[inner sep=0pt] (pv1) at (0,0)
  {\includegraphics[width=.35\linewidth]{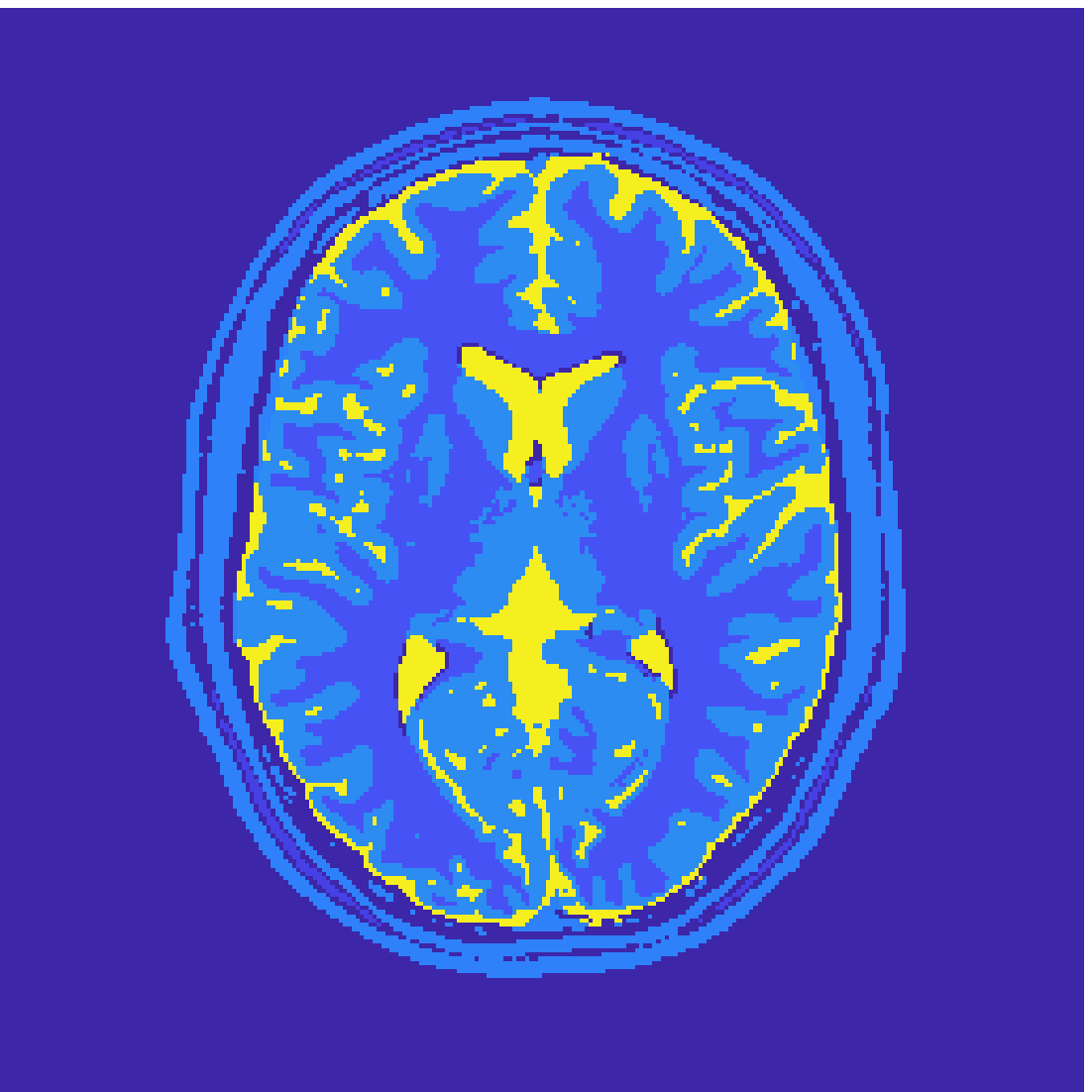}};
\node[inner sep=0pt] (pv2) at (4,0)
  {\includegraphics[width=.35\linewidth]{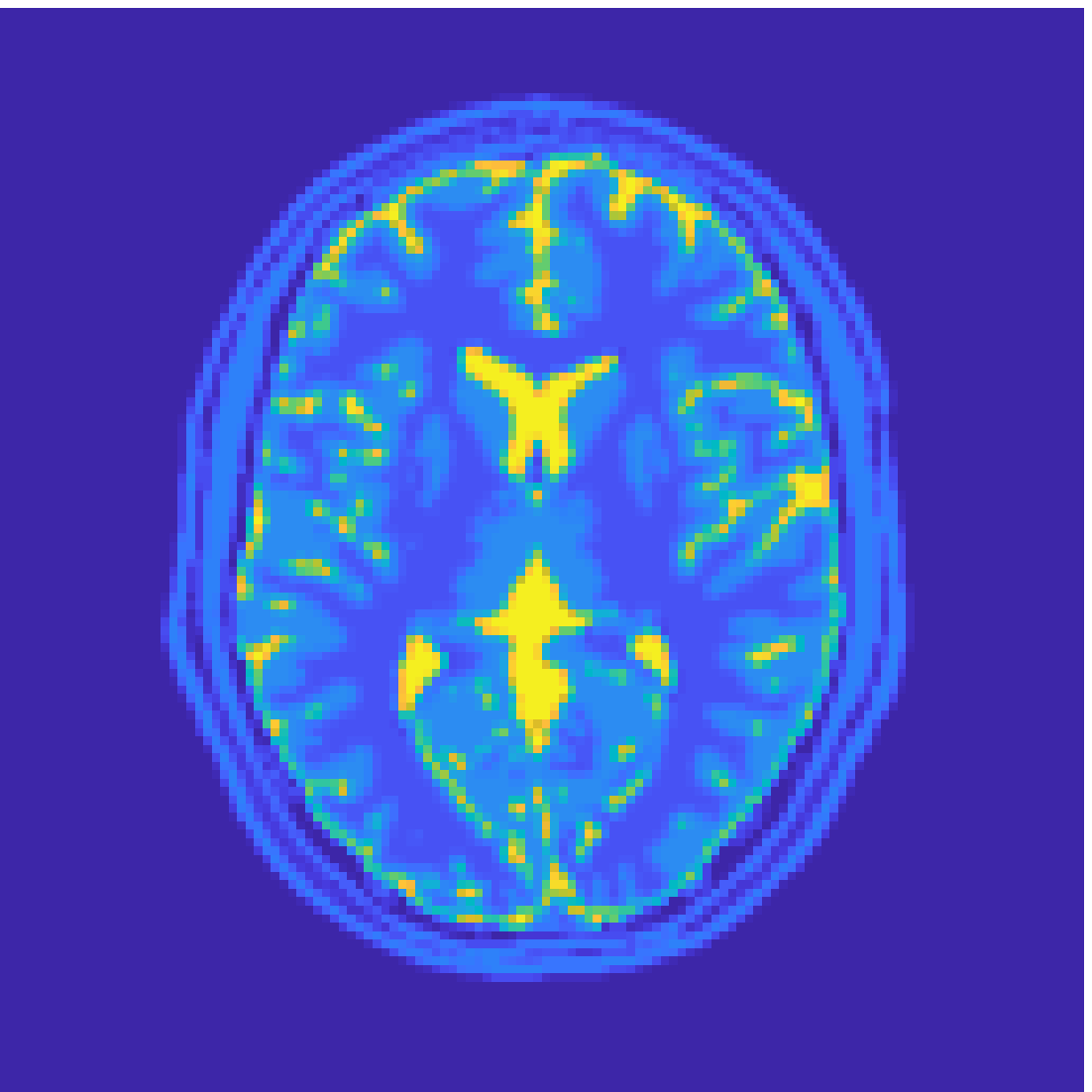}};
\node[inner sep=0pt] (pv1s) at (2,-1.5)
 {\includegraphics[width=.2\linewidth]{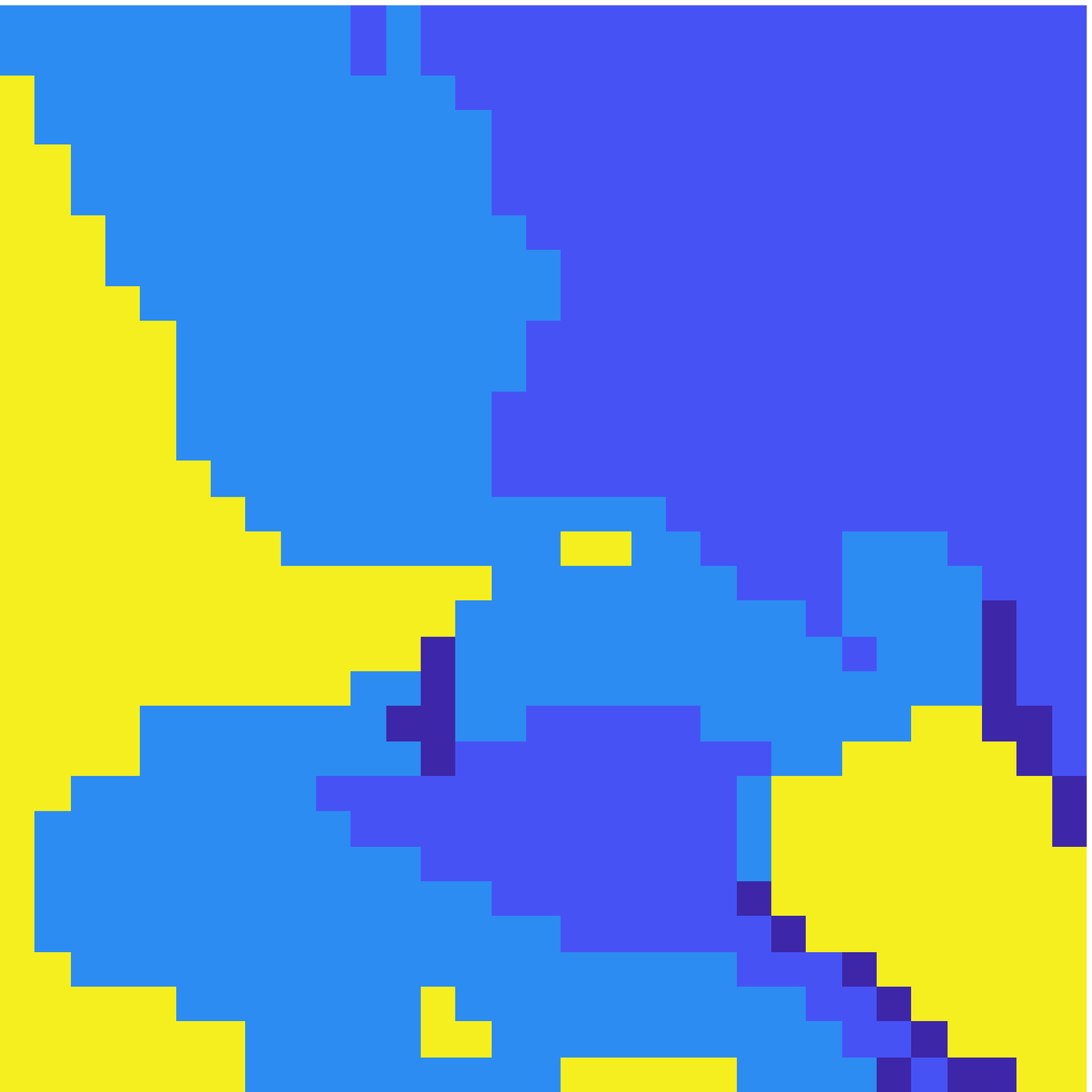}};
 \draw (pv1.center) rectangle +(.3,-.3);
\draw[-,thick] (pv1.center) -- (pv1s.north west);
\draw[-,thick] (pv1.center)+(.3,0) -- (pv1s.north east);
\draw[-,thick] (pv1.center)+(0,-.3) -- (pv1s.south west);

\node[inner sep=0pt] (pv2s) at (6,-1.5)
 {\includegraphics[width=.2\linewidth]{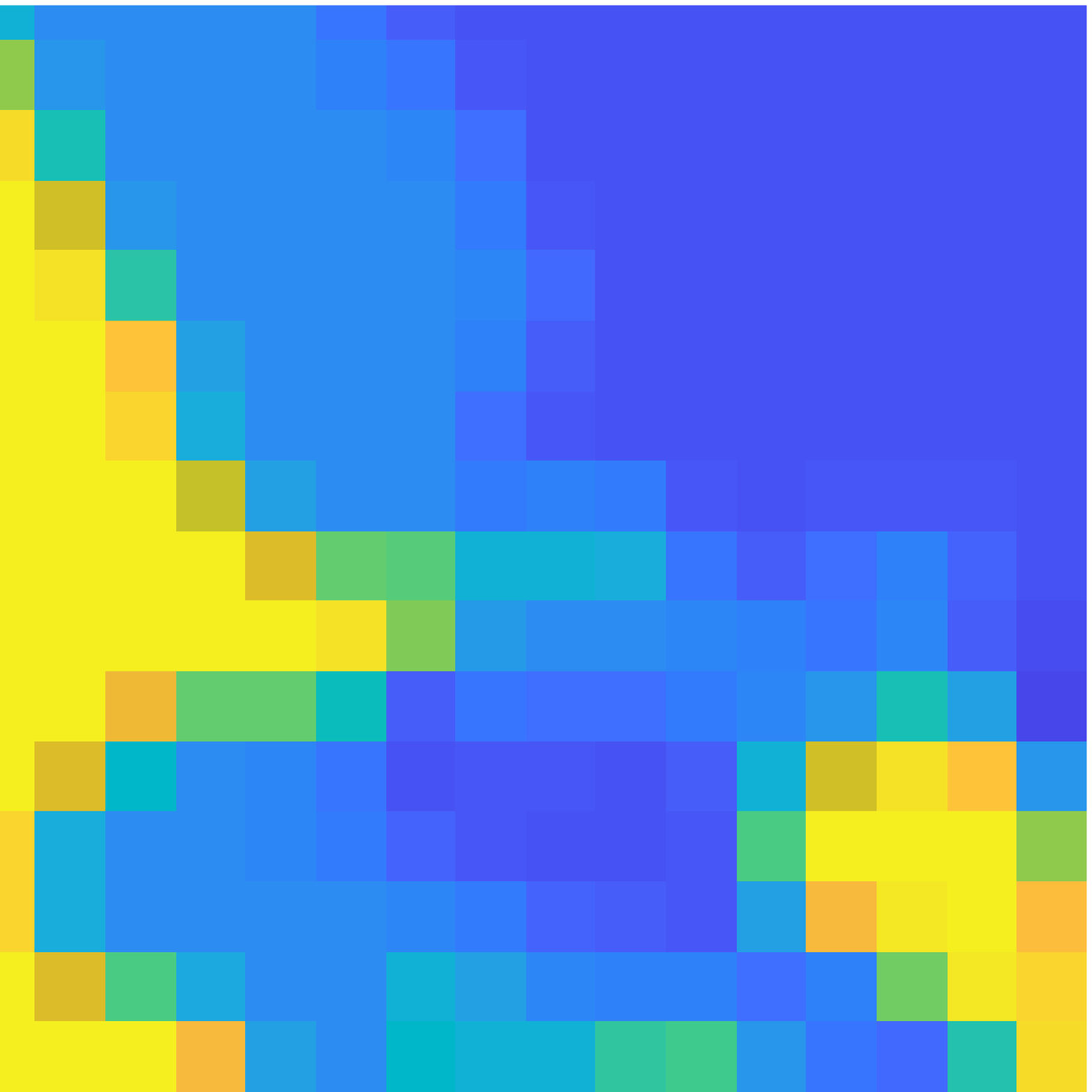}};
 \draw (pv2.center) rectangle +(.3,-.3);
\draw[-,thick] (pv2.center) -- (pv2s.north west);
\draw[-,thick] (pv2.center)+(.3,0) -- (pv2s.north east);
\draw[-,thick] (pv2.center)+(0,-.3) -- (pv2s.south west);
\end{tikzpicture}}
\caption{Partial volume effect in a $T_1$ parameter map. Left: true $T_1$ parameter map. Right: low resolution reconstruction.
}
\label{fig_pve}
\end{figure}

In general, qMRI techniques, particularly MRF-based methods \cite{Pierre2016,Zhao2016,Zhao2017,Doneva2017, Mazor2018, CSMRF,mcgivney2014}, assume that a voxel contains at most one type of tissue, e.g. white matter (WM), grey matter (GM), etc. This assumption is not suitable in practice. Consequently, voxels containing multiple tissue types may be assigned with incorrect parameters. This problem is known as the Partial Volume Effect (PVE) and appears in all medical imaging modalities with limited spatial resolution \cite{Tohka2014}. An example of PVE is given in Fig.~\ref{fig_pve}. The left image shows a spatial distribution of $T_1$ in a simulated brain. The right image shows a reconstruction using voxels four times bigger and assuming a single tissue per voxel. All low resolution voxels at the edge between tissues contain partial volumes, which implies a wrong estimate (single wrong value of $T_1$ rather than multiple values).

The PVE has been analysed in the supplementary material of \cite{MRF}. In this work, using a least-squares method, the signal is decomposed as a weighted sum of at most three distinct signals, each representing a different tissue. 
Although this method was shown to be robust to noise for long sequences, since it necessitates both information about the spatial distribution of the PV voxels and the true components of the original signal (which are unknown in practice), it is not adapted to handle \textit{in vivo} data. An extension of this approach has been proposed in \cite{Deshmane2017}, where the tissue parameters are learnt using a clustering approach on the parameter maps, obtained by the match filter. Then the data is matched with a PV dictionary varying the tissue proportion. This method has shown good results considering long sequences, but it is limited to the dictionary approximation and the accuracy of the parameters estimated during the first matching. However, on the other hand, all the reconstructions are performed with high aliased images requiring more acquisitions for accurate results.
Additionally, for short sequences, the noise in the measurements and the sampling of the manifold of fingerprints describing the signal can significantly affect the estimations. More recently, a Bayesian method was proposed in \cite{bayesianMRF}, to tackle the PVE in MRF. The authors show that their approach estimates the parameters of the PV voxels. However, due to the high aliasing effect encountered with undersampled noisy data, this estimation comes at the cost of an increased acquisition time with respect to traditional MRF based reconstructions (\textit{i.e.} three times longer sequences than traditional MRF). 
Furthermore, to obtain accurate results, this method relies on a high sampling of the fingerprint manifold, resulting in a high computational cost (in terms of both reconstruction time and memory requirement).

In this paper, we propose to tackle the PVE in MRF by reformulating the problem as a non-convex constrained least-squares minimisation problem. In our approach, we assume that the number of independent tissues in the imaged volume is upper bounded, and that there exists at least a region of the total volume with only pure voxels for each tissue. 
To solve the resulting non-convex constrained minimisation problem, we develop a greedy approximate projected gradient descent method, dubbed GAP-MRF. It can be seen as a generalisation of BLIP method for PVE. It consists in a projected gradient descent algorithm, where the projection is computed inexactly, through a memory efficient greedy approach. The proposed method is also generalised to compensate for phase errors in the model, due to timing or coil sensitivity errors, using an alternating minimisation approach \cite{Bertsekas99,Tseng2001, Bolte2014, Frankel2015, Repetti2016}. Through simulations on a simulated PV phantom, we show that our approach outperforms state-of-the-art methods. Our method is afterward validated on the EUROSPIN phantom and on \textit{in vivo} MRF datasets.

The remainder of the paper is organised as follows: In Section~\ref{Sec:MRF} we introduce the notation used throughout the paper, and we define the MRF inverse problem introducing the proposed PV model. 
In Section~\ref{Sec:GAP} we give the proposed algorithm to solve the PV problem. Finally, in Sections~\ref{Sec:simulations} and \ref{section_realdata}, we investigate the behaviour of the proposed method on simulated data and show the results on the \textit{in vivo} datasets, respectively. We conclude in Section~\ref{Sec:Conclusions}.

\section{Notation and MRF problem description}\label{Sec:MRF}

\subsection{Notation}

In this section, we introduce the notation we will use in the remainder of the paper. We refer the reader to \cite{Bauschke2011, Rockafellar2009} for additional details about optimisation.
To have a compact notation when selecting a specific row $n \in \{1, \ldots, N\}$ of a matrix $\mathbf{M}\in\mathbb{C}^{N\times L}$, we use the notation $\mathbf{M}_{n,:}=\left(\mathbf{M}_{n,l}\right)_{1\leq l\leq L}$.
Similarly, to select a specific column $l \in \{1, \ldots, L\}$ of this matrix, we use $\mathbf{M}_{:,l}=\left(\mathbf{M}_{n,l}\right)_{1\leq n\leq N}$.
More generally, this notation is also used to select subparts of tensors.
The operator $\text{real}(\cdot)$ gives the real part of its complex argument, 
the operator $\text{Diag}(\cdot)$ builds a diagonal matrix whose diagonal elements are given by its argument,
and $(\cdot)^\dagger$ gives its adjoint. The adjoint of a linear operator $g \colon \mathbb{C}^L \to \mathbb{C}^N$ is denoted by $g^\dagger$. 
The cardinality of a countable set $ \mathcal{T} $ is given by $\text{card}(\mathcal{T})$. 
The $\ell_p$ norm ($p\in ]0,+\infty]$) is denoted by $\|\cdot\|_p$. The $\ell_0$ pseudo-norm \cite{CS1}, counting the non-zero entries of its argument, is defined as 
    $(\forall \mathbf{x} \in \mathbb{R}^N) \quad
    \|\mathbf{x}\|_0 = \sum_{n=1}^N \left(\mathbf{x}_{n}\right)^0$,
with the convention $0^0=0$.
By abuse of notation, the $\ell_p$ norms and the $\ell_0$ pseudo-norm will be used for tensors by reshaping them into vectors. 
Finally, the projection of a vector $\overline{\mathbf{x}}\in \mathbb{C}^N$ onto a non-empty closed subset $\mathcal{S}$ of $\mathbb{C}^N$ is given by $\mathcal{P}_{\mathcal{S}} (\overline{\mathbf{x}} )= \argmin_{\mathbf{\mathbf{x}\in \mathcal{S}}} \frac{1}{2}\|\mathbf{x}-{\overline{\mathbf{x}}}\|_2^2$ \cite{Bauschke2011}.
The same notation is used for projections of tensors.


\subsection{Inverse problem for single tissue recovery}\label{subsec:inverse}

In the context of MRF, the objective is to estimate the parameters of each voxel in the imaged volume from degraded undersampled measurements. Let $\mathbf{Y}\in\mathbb{C}^{Q\times L\times C}$ be the measurement matrix, where $L$ is the excitation sequence length, $C$ is the number of coils and $Q$ is the number of measurements at each excitation and each coil. Let $\mathbf{M}\in \mathbb{C}^{N\times L}$ be the response of the imaged volume of interest with $N$ voxels. For every $(l,c) \in \{1,\ldots,L\}\times \{1,\ldots,C\}$, the corresponding observation $\mathbf{Y}_{:,l,c} \in \mathbb{C}^{Q}$ is given by
\begin{equation}
\mathbf{Y}_{:,l,c}=\boldsymbol{\Omega}_{:,:,l}\mathbf{F}\mathbf{S}_{:,:,c}\mathbf{M}_{:,l}+\boldsymbol{\eta}_{:,l,c},
\label{eq_mrfmodel}
\end{equation}
where $\boldsymbol{\Omega}\in\{1,0\}^{Q\times N\times L}$ is the concatenation of $L$ selection matrices, $\mathbf{F}\in \mathbb{C}^{N\times N}$ is the 2-dimensional discrete Fourier transform, $\mathbf{S}\in\mathbb{C}^{N\times N\times C}$ is the concatenation of $C$ spatial sensitivity coil diagonal matrices, and $\boldsymbol{\eta}\in \mathbb{C}^{Q\times L \times C}$ is a realisation of a random i.i.d. Gaussian noise. Let ${h:\mathbb{C}^{N\times L}\rightarrow \mathbb{C}^{Q\times L \times C}}$ be the linear mapping defining the complete acquisition process such that $\mathbf{Y}=h\left(\mathbf{M}\right) + \boldsymbol{\eta}$.

For each voxel $n\in \{1, \ldots, N\}$, the magnetisation response $\mathbf{M}_{n,:}$ is modelled through the smooth non-linear mapping $B:\mathcal{M} \rightarrow \mathbb{C}^{1\times L}$ (commonly Bloch equations or Extended Phase Graphs (EPG) model \cite{weigel2015}) scaled by the unknown proton density $\rho_n\in \mathbb{R}_+$, $\mathbf{M}_{n,:}=\rho_n B(\hat{\boldsymbol{\theta}}_{n,:},\boldsymbol{\Gamma})$,
where $\boldsymbol{\Gamma}\in \mathbb{R}^{A\times 1}$ represents the concatenation of $A$ known acquisition parameters (\textit{e.g.}, flip angles $\alpha$, repetition times TR) chosen such that $\mathbf{M}_{n,:}$ is only sensitive to the $P$ parameters $\hat{\boldsymbol{\theta}}_{n,:}\in \mathcal{M}$ under investigation, where $\mathcal{M}\subset \mathbb{R}^{1 \times P}$ denotes the subset of feasible parameters. In the remainder, we fix $P=2$ and choose $\mathcal{M}$ corresponding to $T_1$ and $T_2$ relaxation times.

\subsection{Proposed partial volume model}
\label{Sec:Prop_model}

The model described in the previous section considers that each voxel contains at most one element. PV voxels are introduced due to the spatial discretisation in the acquisition process. The magnetisation sequence can be described as $\mathbf{M} = \mathbf{X}\boldsymbol{\Phi}$, where $\mathbf{X} \in \mathbb{R}_+^{N\times D}$ is a sparse mixing matrix (each line of $\mathbf{X}$ represents the proton densities associated with a specific voxel, and would contain more than a nonzero value only for voxels with partial volumes), and $\boldsymbol{\Phi}\in\mathbb{C}^{D\times L}$ is the over-complete dictionary of fingerprints, introduced in \cite{MRF}, as a discrete sampling of the low dimensional manifold $B$. $\boldsymbol{\Phi}$ is constructed from $D$ samples of $\mathcal{M}$, stored in a matrix $\boldsymbol{\theta}\in \mathbb{R}^{D\times P}$. 
Due to the smoothness of $B$, $\boldsymbol{\Phi}$ is highly coherent. Consequently, the estimation of $\mathbf{X}$ from highly undersampled noisy data is expected to fail without additional priors. Leveraging CS theory \cite{CS1,CS2,CS3,CS4}, the sparsest matrix $\mathbf{X}$, fitting the measurement model, can be found by solving:
\begin{equation}
\underset{\mathbf{X} \in \mathbb{R}_{+}^{N\times D}}{\text{minimise}}\,\, \|\mathbf{X}\|_0 \mbox{ subject to } \|\mathbf{Y}-h(\mathbf{X}\boldsymbol{\Phi})\|_2 \leq \epsilon,
\label{eq_nonconvexproblem}
\end{equation}
where $\epsilon>0$ is a bound chosen according to the acquisition noise level. Since this function is non-convex and non-differentiable, problem~\eqref{eq_nonconvexproblem} is difficult to solve in practice, in particular in the context of high dimensional problems (usually, $D \sim 10^6$ and $L \sim 10^3$). Thus, the non-convexity of the $\ell_0$ pseudo-norm is often relaxed by the use of the $\ell_1$-norm \cite{Chen_2001}. Nevertheless, $\boldsymbol{\Phi}$ being highly coherent, this convex relaxation cannot be used to correctly estimate the coefficients of $\mathbf{X}$ \cite{CS5}. 
%
To overcome these difficulties, similarly to the BLIP approach, we propose to 
\begin{equation}
\underset{\mathbf{M}\in \mathcal{B}_{\mathcal{S}_+}\left(\boldsymbol{\Phi}\right)}{\text{minimise}}\,\, \frac{1}{2}\|\mathbf{Y}-h(\mathbf{M})\|_2^2
\label{eq_pvablipproblem}
\end{equation}
where
\begin{align}
&	\mathcal{B}_{\mathcal{S}_+}\left(\boldsymbol{\Phi}\right) 
	= \left\{\mathbf{M}\in \mathbb{C}^{N\times L} \,|\, \mathbf{M}=\mathbf{X} \boldsymbol{\Phi}\mbox{ with } \mathbf{X}\in \mathcal{S}_+ \right\},\label{eq_setbs+}	\\
&	\mathcal{S}_+
	=\overset{4}{\underset{s=1}{\cap}} \mathcal{S}_s,
	\label{eq_sets+}
\end{align}
and, for every $s\in \{1, \ldots,4\}$, $\mathcal{S}_s$ is a closed non-empty subset of $\mathbb{R}^{N \times D}$, used to impose feasibility constraints on $\mathbf{X}$. These sets are defined below.



\subsubsection{Positivity constraint}
Since the proton densities of the imaged volume must be non-negative, we can restrict our solution to be in the positive orthant:
\begin{equation}
\mathcal{S}_1=\mathbb{R}_+^{N\times D}.
\label{eq_positivity}
\end{equation}

\subsubsection{Constraint on the number of tissues}\label{subsec:Kconstraint}
Commonly MRF aims to obtain quantitative values of a small set of tissues. In practice, only $T\ll D$ elements of the dictionary $\boldsymbol{\Phi}$ are necessary to characterise $\mathbf{M}$. While $T$ is unknown, we have a reasonable estimate for it. We propose to introduce a loose upper bound $K$, such that $T \le K \le D$, to limit the number of active dictionary elements. 
Let us define a set $\mathcal{D}_\mathbf{X}$ that is formed by the column indices of $\mathbf{X}$ with non-zero coefficients. To avoid noisy voxels, only rows with proton density greater than $\xi>0$ (chosen according to the noise level) will be considered. Formally, this set  is defined as ${\mathcal{D}_\mathbf{X}=\left\{ d \in \{1,\ldots,D\} \, | \, (\exists n \in \mathcal{G}_\mathbf{X}) \;  \mathbf{X}_{n,d} \neq 0 \right\}}$, where $\mathcal{G}_\mathbf{X}=\{n\in\{1,\ldots,N\} \, | \, \|\mathbf{X}_{n,:}\|_1>\xi\}$. The set $\mathcal{D}_\mathbf{X}$ indicates the columns of $\mathbf{X}$ contributing to the magnetisation sequence. We can limit the number of used elements of the dictionary by upper bounding the cardinality of this set by~$K$: 
\begin{equation}
	\mathcal{S}_2=\left\{\mathbf{X}\in \mathbb{R}^{N \times D} \, | \, \text{Card}\!\left(\mathcal{D}_\mathbf{X}\right) \leq K \right\}.
	\label{eq_K}
\end{equation}

\subsubsection{Constraint on the manifold neighbourhoods}
The tissues of interest are unique and need to be sufficiently different to be distinguished. To incorporate this prior information in the reconstruction process, we define the neighbour set associated to each element $d\in \{1,\ldots,D\}$ of the dictionary as:
\begin{multline}
\mathcal{N}_v(d)=\{d'\in \{1,\ldots, D\} \backslash \{d\} | \\ 
\left(\forall p=\{1,\ldots,P\}\right) |\boldsymbol{\theta}_{d',p}-\boldsymbol{\theta}_{d,p}|<\upsilon\boldsymbol{\theta}_{d,p} \},
\label{eq_neighbours}
\end{multline}
where $\upsilon>0$. We define a set of all possible $\mathbf{X}$ such that, the parameters of each element in $\mathcal{D}_\mathbf{X}$ are sufficiently far from each other. Precisely, we constrict all the neighbour columns of each element in $\mathcal{D}_\mathbf{X}$ to be the null element $\boldsymbol{0}$ of $\mathbb{R}^N$:
\begin{equation}
	\mathcal{S}_3=\big\{\mathbf{X}\in \mathbb{R}^{N\times D} \, | \, \big(\forall d' \in \underset{d \in \mathcal{D}_\mathbf{X}}{\cup} \mathcal{N}_v(d) \big) \, \mathbf{X}_{:,d'}= \boldsymbol{0} \big\}
\end{equation}

\subsubsection{Constraint on the pure voxels}
Due to the additive noise in model \eqref{eq_mrfmodel}, some elements of $\mathbf{X}$ corresponding to non-used dictionary elements take non-zero values. In order to avoid these noisy elements in the reconstructions, 
we impose that at least $\kappa>0$ rows (i.e. voxels) of $\mathbf{X}$ contain only one non-zero value for each active column of $\mathbf{X}$. These rows identify the \emph{pure voxels}. This constraint can be formulated as follows:
\begin{equation}
\mathcal{S}_4=\{\mathbf{X}\in \mathbb{R}^{N\times D} \, | \, \left(\forall d \in \mathcal{D}_{\mathbf{X}}\right) \, \|\left(\mathbf{X}_{n,d}\right)_{n\in \mathcal{V}_\mathbf{X}}\|_0\geq \kappa \}
\label{eq_pure}
\end{equation}
where $\mathcal{V}_\mathbf{X}=\left\{n\in \left\{1,\ldots,N\right\} \, | \, \|\mathbf{X}_{n,:}\|_0=1\right\}$.
\section{Greedy Approximate Projection for MRF}\label{Sec:GAP}
\subsection{Proposed iterative projected gradient descent algorithm}
To solve problem~\eqref{eq_pvablipproblem}, we use an iterative projected gradient descent method \cite{IPA}. 
At each iteration $i\in \mathbb{N}$, this method updates $ \mathbf{M}^{(i+1)} $ by computing a gradient step followed by a projection step:
\begin{equation}	\label{algo:grad_proj_mod}
\mathbf{M}^{(i+1)} = 
\mathcal{P}_{\mathcal{B}_{S_+}\left(\boldsymbol{\Phi}\right)}\left(
\mathbf{M}^{(i)}-\mu h^\dagger\big( h (\mathbf{M}^{(i)}) -\mathbf{Y} \big) \right),
\end{equation}
where $\mu >0$. In \cite{CSMRF}, it is shown that choosing $\mu\approx N/Q$ is theoretically justifiable. However, in order to ensure the stability of the iterative projected gradient descent algorithm and accelerate convergence, in \cite{IHT, CSMRF} the authors proposed to choose $\mu$ using a backtracking method. 
In order to handle efficiently the constraint $\mathcal{B}_{S_+}\left(\boldsymbol{\Phi}\right) $, we propose to compute inexactly the projection onto this set in \eqref{algo:grad_proj_mod}. 
The resulting method, named Greedy Approximate Projection for MRF (GAP-MRF), is described in Algorithm~\ref{alg_IPA_BT}. It can be noticed that the GAP-MRF method and BLIP are solving similar problems, using the same algorithmic structure. In this context, as in \cite{CSMRF}, a condition on both $L$ and the undersampling ratio $N/Q$ might be derived for recovery guarantee. However, the investigation of such condition is beyond the scope of this article.
\begin{algorithm}[!t]
	\caption{GAP-MRF global iterations}
    \label{alg_IPA_BT}
	{\begin{algorithmic}[1]
		\State 
        	\textbf{Input:} $\mathbf{Y} \in \mathbb{C}^{Q\times L \times C}$, 
        	$\zeta <1$, $\mathbf{M}^{(0)} \in \mathbb{C}^{N \times L} $
		\State \textbf{Iterations:} 
		\For{$i=0, 1,\ldots$}
		\State $\mu = 2N/Q$, $ \nu = 0 $
        \While{$\mu > \nu $}
		\State $\mu = \mu/2$ 
		\State \textbf{Gradient Step:}
		\State $\overline{\mathbf{M}}^{(i)}= \mathbf{M}^{(i)}-\mu h^\dagger\left(h\left(\mathbf{M}^{(i)}\right)-\mathbf{Y}\right)$
		\State \textbf{Projection Step:}
		\State \label{algo_IPA_BT:proj}
  				$\mathbf{M}^{(i+1)} \approx \mathcal{P}_{\mathcal{B}_{S_+} \left(\boldsymbol{\Phi}\right)} \left( \overline{\mathbf{M}}^{(i)} \right)$
        \State \textbf{Backtracking step}
        \State 
        $\nu =\zeta \dfrac{\|\mathbf{M}^{(i+1)}-\mathbf{M}^{(i)}\|_2^2}{\|h\left(\mathbf{M}^{(i+1)}-\mathbf{M}^{(i)}\right)\|_2^2}$
        \EndWhile
		\EndFor
	\end{algorithmic}}
\end{algorithm}

\subsection{Approximate projection}
\label{Ssec:Algo:approx_proj}
For every $\overline{\mathbf{M}}\in \mathbb{C}^{N\times L}$, we have:
\begin{align}
\mathbb{\mathcal{P}}_{\mathcal{B}_{\mathcal{S}_+}\left(\boldsymbol{\Phi}\right)}\left(\,\overline{\mathbf{M}}\right)
	&=\argmin_{\mathbf{M}\in\mathcal{B}_{\mathcal{S}_+}\left(\boldsymbol{\Phi}\right)} \frac{1}{2} \|\mathbf{M}-\overline{\mathbf{M}}\|_2^2	\nonumber	\\
	&=\argmin_{\mathbf{M}=\mathbf{X}\boldsymbol{\Phi},\, \mathbf{X}\in \mathcal{S}_+} \frac{1}{2} \|\mathbf{X}\boldsymbol{\Phi}-\overline{\mathbf{M}}\|_2^2	\nonumber	\\
	&= \big( \argmin_{\mathbf{X}\in \mathcal{S}_+} \frac{1}{2} \|\mathbf{X}\boldsymbol{\Phi}-\overline{\mathbf{M}}\|_2^2 \big) \boldsymbol{\Phi},
	\label{eq_trueprojection}
\end{align}
Note that $\mathcal{S}_2,\mathcal{S}_3$ and $\mathcal{S}_4$ can be handled through the definition of $\boldsymbol{\Phi}$. Let $\mathbf{M}=\mathbf{X}\boldsymbol{\Phi} \in \mathcal{B}_{\mathcal{S}_+}\! (\boldsymbol{\Phi})$ and $\overline{T}\in \{1,\ldots,K\}$ ($K$ is the upper bound defined in \eqref{eq_K}). 
Let $\mathbf{U}\in \mathbb{R}^{N\times \overline{T}}$ be a subpart of $\mathbf{X}$ with non-zero columns and $\boldsymbol{\Delta}\in\mathbb{C}^{\overline{T}\times L} $ the corresponding subpart of $\boldsymbol{\Phi}$ such that 
$\mathbf{M}=\mathbf{U}\boldsymbol{\Delta}$. Then we have 
\begin{equation}	\label{eq:Proj_XS_PhiS}
\mathcal{P}_{\mathcal{B}_{\mathcal{S}_+}\left(\boldsymbol{\Phi}\right)} (\overline{\mathbf{M}} )=
\big(\argmin_{\mathbf{U}\in\mathbb{R}_+^{N\times \overline{T}}} \frac{1}{2} \|\mathbf{U} \boldsymbol{\Delta} - \overline{\mathbf{M}}\|_2^2 \big)\boldsymbol{\Delta}.
\end{equation}	
In \eqref{eq:Proj_XS_PhiS}, the dictionary $\boldsymbol{\Delta}$ is defined as
\begin{equation}
	\boldsymbol{\Delta}=\argmin_{\overline{\boldsymbol{\Delta}}\in \mathcal{C}} \big( \min_{\overline{\mathbf{U}}\in \mathbb{R}_+^{N\times \overline{T}}} \frac{1}{2}\|\overline{\mathbf{U}}\,\overline{\boldsymbol{\Delta}}-\overline{\mathbf{M}}\|_2^2 \big),
	\label{eq_Delta}
\end{equation}
where $\mathcal{C}$ is the set given by
\begin{multline}
\mathcal{C} = \Big\{ 
\boldsymbol{\Delta} \in \mathbb{C}^{\overline{T} \times L} \, | \, 
(\exists \mathbf{X}\in \mathcal{S}_+) \; 
\mathbf{X} = \mathcal{Z} \big( \overline{\mathbf{U}} \big) \\
\text{with } \overline{\mathbf{U}} = \underset{\mathbf{U}\in\mathbb{R}_+^{N\times \overline{T}}}{\text{argmin}} \frac{1}{2}\|\mathbf{U}\boldsymbol{\Delta} - \overline{\mathbf{M}} \|_2^2
\Big\}.
\label{eq_dictionaries}
\end{multline}
with $\mathcal{Z} \colon \mathbb{R}_+^{N\times \overline{T}} \to \mathbb{R}_+^{N \times D}$ defined such that $\mathcal{Z}(\mathbf{U})\boldsymbol{\Phi}=\mathbf{U}\boldsymbol{\Delta}$. 

As mentioned earlier, $\boldsymbol{\Phi}$ is an over-complete dictionary which makes the exact projection practically impossible to compute. To overcome this difficulty, we propose a greedy approach to approximate the projection by finding a reduced dictionary $\widetilde{\boldsymbol{\Delta}}\in \mathbb{C}^{\overline{T}\times L}$ and its corresponding mixing matrix $\widetilde{\mathbf{U}}\in \mathbb{C}^{N\times \overline{T}}$, with $\overline{T}\leq K$, such that $\mathbf{U}\boldsymbol{\Delta} \approx \widetilde{\mathbf{U}}\widetilde{\boldsymbol{\Delta}}$. Then the projection in step~\ref{algo_IPA_BT:proj} of Algorithm~\ref{alg_IPA_BT} can be approximated as 
\begin{equation}
\mathbb{\mathcal{P}}_{\mathcal{B}_{\mathcal{S}_+}\left(\boldsymbol{\Phi}\right)} (\overline{\mathbf{M}})\approx \big( \argmin_{\widetilde{\mathbf{U}} \in \mathbb{R}_+^{N\times \overline{T}}} \frac{1}{2} \| \widetilde{\mathbf{U}} \widetilde{\boldsymbol{\Delta}} - \overline{\mathbf{M}} \|_2^2 \big) \widetilde{\boldsymbol{\Delta}}.
\label{eq_pvaprojection}
\end{equation}
As mentioned in \cite{CSMRF}, it is a common practice to allow the proton density to be complex-valued in order to absorb phase terms correcting for timing and coil sensitivity errors. We incorporate a vector $\boldsymbol{\lambda} \in \mathbb{C}^{N}$ to compensate for these errors. 
Let $\widetilde{\mathcal{B}}_{\mathcal{S}_+}(\boldsymbol{\Phi})$ be the set of magnetisation sequences of the form $\mathbf{M} = \text{Diag} (\boldsymbol{\lambda})\mathbf{X} \boldsymbol{\Phi}$ such that $\mathbf{X}\in \mathcal{S}_+$ and $ \boldsymbol{\lambda}\in\mathbb{C}^{N} $ satisfies $ (\forall n \in \{1, \ldots, N\}) \, |\boldsymbol{\lambda}_{n}|=1$.
The approximate projection with the phase compensation is given by:
\begin{equation}
\mathbb{\mathcal{P}}_{\widetilde{\mathcal{B}}_{\mathcal{S}_+}\left(\boldsymbol{\Phi}\right)} (\overline{\mathbf{M}} ) \approx \text{Diag}(\boldsymbol{\lambda}) \widetilde{\mathbf{U}}\widetilde{\boldsymbol{\Delta}},
\end{equation}
where 
$(\boldsymbol{\lambda},\widetilde{\mathbf{U}})$ are obtained by solving: 
\begin{multline}
\underset{\boldsymbol{\lambda}\in \mathbb{C}^{N}, \widetilde{\mathbf{U}}\in\mathbb{R}^{N\times \overline{T}}_+}{{\text{minimise}}} 
\frac{1}{2}\|\text{Diag}(\boldsymbol{\lambda}) \widetilde{\mathbf{U}} \widetilde{\boldsymbol{\Delta}} - \overline{\mathbf{M}} \|_2^2 \\
\text{subject to } (\forall n \in \{1, \ldots, N\}) \, |\boldsymbol{\lambda}_{n}|=1.
\label{eq_complexapvprojection}
\end{multline}
It is worth mentioning that in \eqref{eq_pvaprojection} and \eqref{eq_complexapvprojection}, all the rows of $\widetilde{\mathbf{U}}$ can be computed independently in parallel. 

On the one hand, forward-backward based algorithms \cite{Chen97, Tseng2000, Combettes2005} can be used to solve problem~\eqref{eq_pvaprojection} (in particular, in our simulations, we use the built-in \textsc{Matlab} function of non-negative least-squares, that is an implementation of \cite{Lawson1938}).
On the other hand, to solve problem~\eqref{eq_complexapvprojection}, to jointly estimate $\boldsymbol{\Lambda}$ and $\mathbf{U}$, block coordinate approaches must be considered (e.g. Gauss-Seidel approaches \cite{Bertsekas99}, alternating forward-backward methods \cite{Tseng2001, Bolte2014, Frankel2015, Repetti2016}). 
Note that in comparison with the traditional MRF methods which densely sample the manifold, our approach reduces the memory requirements, by using the dictionary $\widetilde{\boldsymbol{\Delta}}$ containing at most $K$ elements, without the inaccuracies related to the manifold discretisation.


\subsection{Greedy dictionary estimation}\label{ssec:gready}

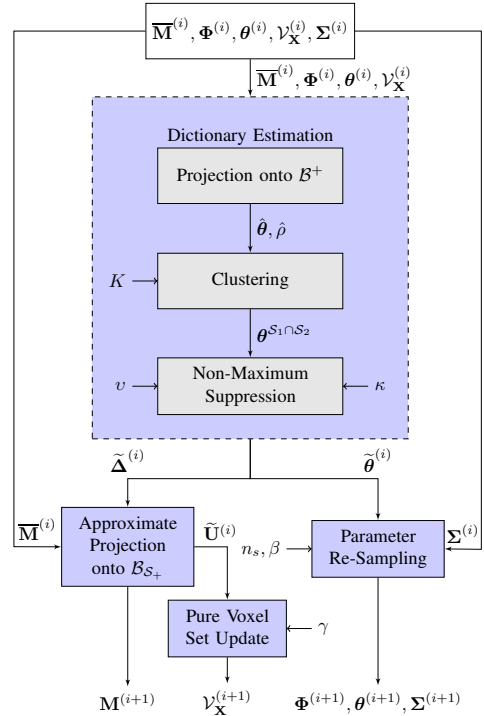
\begin{figure}[!t]
\scalebox{.7}{
\tikzstyle{block} = [draw, fill=blue!20, rectangle, 
    minimum height=3em, minimum width=4em]
    \tikzstyle{block4} = [draw, fill=gray!20, rectangle, 
    minimum height=3em, minimum width=10em]
    \tikzstyle{block2} = [draw, fill=white, rectangle, 
    minimum height=3em, minimum width=4em]
\tikzstyle{bigblock} = [draw,dashed, fill=blue!20, 
    minimum height=6.5cm, minimum width=6cm]
\tikzstyle{sum} = [draw, fill=blue!20, circle, node distance=1cm]
\tikzstyle{input} = [coordinate]
\tikzstyle{output} = [coordinate]
\tikzstyle{dot} = [coordinate]
\tikzstyle{pinstyle} = [pin edge={to-,thin,black}]
\tikzstyle{pinstyle2} = [pin edge={to-,thin,black},pin distance = 1cm]
\tikzstyle{pinstyle3} = [pin edge={->,thin,black},pin distance = 1.9cm]
\tikzstyle{pinstyle4} = [pin edge={->,thin,black},pin distance = 0.5cm]
\tikzstyle{pinstyle5} = [pin edge={->,thin,black},pin distance = 2.05cm]
\begin{tikzpicture}[auto, node distance=2cm,>=latex']
    \node [block2] (variables) {   $\overline{\mathbf{M}}^{(i)},\boldsymbol{\Phi}^{(i)},\boldsymbol{\theta}^{(i)},\mathcal{V}_{\mathbf{X}}^{(i)},\boldsymbol{\Sigma}^{(i)}$};
    \node [bigblock, below of=variables, node distance=4.5cm](big){};
    \node [block4, below of=variables,
            node distance=2.75cm] (projection) {Projection onto $\mathcal{B}^+$};
     \node [block4, below of=projection,pin={[pinstyle]left:$K$},
            node distance=2cm] (clustering) {Clustering};
            \node [block4, below of=clustering,pin={[pinstyle]left:$\upsilon$},pin={[pinstyle]right:$\kappa$},
            node distance=2cm] (Filter) {\begin{tabular}{c}Non-Maximum\\Suppression\end{tabular}};
            \node [block, below left = 1.75cm and -0.7cm of Filter,,pin={[pinstyle3]below:$\mathbf{M}^{(i+1)}$} ] (nnlsq) {\begin{tabular}{c}Approximate\\Projection\\onto $\mathcal{B}_{\mathcal{S}_+}$\end{tabular}};
             \node [block, below right = 2cm and -0.6cm of Filter,pin={[pinstyle]left:$n_s,\beta$},,pin={[pinstyle5]below:$\boldsymbol{\Phi}^{(i+1)},\boldsymbol{\theta}^{(i+1)},\boldsymbol{\Sigma}^{(i+1)}$}] (Resample) {\begin{tabular}{c}Parameter\\Re-Sampling\end{tabular}};
            \node [block, below right =  0.25cm and -0.5cm of nnlsq,pin={[pinstyle]right:$\gamma$},pin={[pinstyle4]below:$\mathcal{V}_{\mathbf{X}}^{(i+1)}$}] (Pure Voxel Set) {\begin{tabular}{c}Pure Voxel\\Set Update\end{tabular}};
    \node [draw=none, below of=variables,
            node distance=2cm] (dot) {Dictionary Estimation};
    	
    
    \node [output, below of=nnlsq,
            node distance=3cm] (output) {$\mathbf{M}^{(i+1)}$};
    
            
    \draw [->] (projection.south) -- node {$\hat{\boldsymbol{\theta}},\hat{\rho}$}(clustering.north);
     \draw [->] (clustering.south) -- node {$\boldsymbol{\theta}^{\mathcal{S}_1\cap\mathcal{S}_2}$}(Filter.north);
      \draw [->] (nnlsq.east) -- +(0.5,0) node [above] {$\widetilde{\mathbf{U}}^{(i)}$} -- +(0.5,0) -| (Pure Voxel Set.north);
      \draw [->] (big.south) |- +(0,-0.75) -|  node [above] {$\widetilde{\boldsymbol{\Delta}}^{(i)}$} (nnlsq.north);
      \draw [->] (big.south) |- +(0,-0.75) -|  node [above] {$\widetilde{\boldsymbol{\theta}}^{(i)}$} (Resample.north);
      \draw [->] (variables.east) -| +(2.4,-9.85)  -- node [above]{$\boldsymbol{\Sigma}^{(i)}$} (Resample.east);
      \draw [->] (variables.west) -| +(-2.5,-9.8)  -- node [above]{$\overline{\mathbf{M}}^{(i)}$} (nnlsq.west);
      \draw[->] (variables.south) -- node [right]{$\overline{\mathbf{M}}^{(i)},\boldsymbol{\Phi}^{(i)},\boldsymbol{\theta}^{(i)},\mathcal{V}_{\mathbf{X}}^{(i)}$} (big.north);
       
\end{tikzpicture}}

\caption{Greedy approximate projection diagram. The blue boxes represent the main steps in the approximate projection, the gray boxes represent the intermediate steps for the dictionary estimation and the arrows show the input and output variables.}
\label{fig_projectiondiagram}
\end{figure}

The GAP-MRF algorithm takes advantage of the dictionary coherence and 
the constraints imposed on $\mathbf{X}$ (described in Section~\ref{Sec:Prop_model}) to approximate the projection onto $\mathcal{B}_{\mathcal{S}_+}\left(\boldsymbol{\Phi}\right)$ in line~\ref{algo_IPA_BT:proj} of Algorithm~\ref{alg_IPA_BT}. 
As described in Section~\ref{Ssec:Algo:approx_proj}, this projection can be approximated at each iteration $i\in \mathbb{N}$, by solving~\eqref{eq_pvaprojection}, which necessitates to estimate the dictionary $\widetilde{\boldsymbol{\Delta}}^{(i)}$. We propose to estimate it using a greedy approach, leveraging both the knowledge of $\overline{\mathbf{M}}^{(i)}$ and the properties of the sets $\mathcal{S}_{2}$, $\mathcal{S}_{3}$ and $\mathcal{S}_{4}$ (note that the constraint $\mathcal{S}_1$ is handled directly in \eqref{eq_pvaprojection}). The proposed approach is described in details in this section.

The process to obtain $\widetilde{\boldsymbol{\Delta}}^{(i)}$ consists in three main steps leveraging the set of pure voxels. The first step consists in approximating the parameters of the pure voxels ($\mathcal{S}_4$ constraint) using the projection onto the set $\mathcal{B}^+$ defined as:
\begin{multline}
\mathcal{B}^+=
\big\{\mathbf{M}\in \mathbb{C}^{N\times L}\,|\, 
(\forall n\in\{1,\ldots,N\}) \; 
\mathbf{M}_{n,:} = \rho \mathbf{m} ,\\
\; \text{with }
\rho\in\mathbb{R}_+ 
\text{ and }
\mathbf{m} \in B\left(\mathcal{M},\boldsymbol{\Gamma} \right) 
\big\}.
\label{eq_blochset}
\end{multline}
The objective of the second step is to find $K$ regions of interest ($\mathcal{S}_2$ constraint) of the manifold by exploiting its smoothness. Finally, in the third step, the parameters that are too close to each other are discarded ($\mathcal{S}_3$ constraint) by using a Non-Maximum Suppression based method \citep{Canny1986}. This method acts on the number of voxels that corresponds to each parameter and keeps only the elements which have enough pure voxels to satisfy the $\mathcal{S}_4$ constraint. This process is summarised in the dictionary estimation step on Fig.~\ref{fig_projectiondiagram}. The remaining blue blocks in the diagram are used to update the variables in the greedy approximate projection. More precisely, we compute the mixing matrix $\widetilde{\mathbf{U}}^{(i)}$ and the magnetisation sequence $\mathbf{M}^{(i+1)}$ using equation \eqref{eq_pvaprojection} with the resulting dictionary $\widetilde{\boldsymbol{\Delta}}^{(i)}$. Then, we update the pure voxel set $\mathcal{V}_{\mathbf{X}}$ using the mixing matrix $\widetilde{\mathbf{U}}^{(i)}$. Finally, the dictionary $\boldsymbol{\Phi}$ is refined by randomly sampling around the parameters $\widetilde{\boldsymbol{\theta}}^{(i)}$. The complete method is described in Algorithm~\ref{alg_cluster} and explained in the following paragraphs.

\begin{algorithm}[!t]
	\caption{Greedy Approximate Projection}\label{alg_cluster}
	\begin{algorithmic}[1]
		\State \textbf{Input:} $\overline{\mathbf{M}}^{(i)},\boldsymbol{\Phi}^{(i)},\boldsymbol{\theta}^{(i)},\mathcal{V}_\mathbf{X}^{(i)},\boldsymbol{\Sigma}^{(i)},K,\boldsymbol{\Gamma},\kappa,\upsilon,\gamma,\beta,\xi,n_s$
		\State \colorbox{blue!20}{\textbf{Dictionary Estimation:}}
        \Indent
        \State \colorbox{gray!20}{\textbf{Projection onto $\mathcal{B}^+$}}
        \Indent
		\For{$n=1,2,...,N$}\label{alg:pb+_ini}
		\State $\displaystyle\hat{d}_n=\underset{d}{\argmax} \,\,\, \text{real} ( \overline{\mathbf{M}}^{(i)}_{n,:} \boldsymbol{\Phi}^{\dagger(i)}_{d,:} ) / \|\boldsymbol{\Phi}^{(i)}_{d,:}\|_2$
		\State $\displaystyle \hat{\rho}_n=\max (\text{real} ( \overline{\mathbf{M}}^{(i)}_{n,:} \boldsymbol{\Phi}^{\dagger(i)}_{\hat{d}_n,:} ) / \| \boldsymbol{\Phi}^{(i)}_{\hat{d}_n,:} \|_2^2,0 )$
		\State $\displaystyle\hat{\boldsymbol{\theta}}_{n,:}=\boldsymbol{\theta}^{(i)}_{\hat{d}_n,:}$
		\EndFor\label{alg:pb+_end}
        \EndIndent
		\State \colorbox{gray!20}{\textbf{Clustering}}
        \Indent
		\State $\displaystyle \mathcal{I}=\{n\in\mathcal{V}^{(i)}_\mathbf{X} \,|\, \hat{\rho}_n>\xi\}$
        
		\State $\displaystyle [\boldsymbol{\theta}^{\mathcal{S}_1\cap\mathcal{S}_2},\mathbf{c} ] = \text{k-means} (\hat{\boldsymbol{\theta}}_{\mathcal{I},:},K )$
        \EndIndent
        \State \colorbox{gray!20}{\textbf{Non-Maximum Suppression}}
        \Indent
        \State $\displaystyle\widetilde{\boldsymbol{\theta}}^{(i)}=\text{NonMaximumSuppression} (\boldsymbol{\theta}^{\mathcal{S}_1\cap\mathcal{S}_2},\mathbf{c},\upsilon,\kappa )$
        
		\State $\displaystyle\widetilde{\boldsymbol{\Delta}}^{(i)}=B(\widetilde{\boldsymbol{\theta}}^{(i)},\boldsymbol{\Gamma})$
        \EndIndent
\EndIndent
\State \colorbox{blue!20}{\textbf{Approximate Projection onto $\displaystyle\mathcal{B}_{\mathcal{S}_+}$}}
\Indent
        \State $\displaystyle\widetilde{\mathbf{U}}^{(i)}=\underset{\overline{\mathbf{U}}\in \mathbb{R}_+^{N\times T}}{\argmin} \,\, \dfrac{1}{2}\|\overline{\mathbf{U}}\widetilde{\boldsymbol{\Delta}}^{(i)}-\overline{\mathbf{M}}^{(i)}\|^2_2$
       \State $\displaystyle\mathbf{M}^{(i+1)}=\widetilde{\mathbf{U}}^{(i)}\widetilde{\boldsymbol{\Delta}}^{(i)}$
        \EndIndent
        \State \colorbox{blue!20}{\textbf{Pure Voxel Set Update}}
        \Indent
        \State $\mathcal{G}_{\mathbf{X}}^{(i)}=\{n\in \{1,\ldots,N\} \,|\, \|\widetilde{\mathbf{U}}^{(i)}_{n,:}\|_1>\xi \}$
		\State $\mathcal{V}_\mathbf{X}^{(i+1)}=\{n\in\mathcal{G}_{\mathbf{X}}^{(i)}\,|\,
		\max (\widetilde{\mathbf{U}}^{(i)}_{n,:} )\ge \gamma \|\widetilde{\mathbf{U}}^{(i)}_{n,:}\|_1
		 \}$\label{algo:pure}
         \EndIndent
         \State \colorbox{blue!20}{\textbf{Parameter Re-sampling}}
        \Indent
        \State $\displaystyle\boldsymbol{\theta}^{(i+1)}=\text{ParameterReSampling}(\widetilde{\boldsymbol{\theta}}^{(i)},\boldsymbol{\Sigma}^{(i)},n_s)$
         
         \State $\displaystyle\boldsymbol{\Phi}^{(i+1)}=B(\boldsymbol{\theta}^{(i+1)},\boldsymbol{\Gamma})$
        \State \label{algo:stepsigma} $\displaystyle\boldsymbol{\Sigma}^{(i+1)}=\boldsymbol{\Sigma}^{(i)}\beta$
         \EndIndent
		\State \textbf{Output:} $\boldsymbol{\theta}^{(i+1)} ,\boldsymbol{\Phi}^{(i+1)},\boldsymbol{\Sigma}^{(i+1)},\mathcal{V}_\mathbf{X}^{(i+1)}$ and $\mathbf{M}^{(i+1)}$
	\end{algorithmic}
\end{algorithm}
\subsubsection{Projection onto $\mathcal{B}^+$}
At iteration $i\in \mathbb{N}$, we have:
\begin{equation}
\displaystyle
\mathcal{P}_{\mathcal{B}_{\mathcal{S}_+ (\boldsymbol{\Phi})}}
\big( \overline{\mathbf{M}}_{\mathcal{V}_{\mathbf{X}}^{(i)},:} \big)
=
\mathcal{P}_{\mathcal{B}^+ ( \boldsymbol{\Phi} )} \big( \overline{\mathbf{M}}_{\mathcal{V}_{\mathbf{X}}^{(i)},:} \big),
 	\label{eq_purevoxels}
\end{equation}
where $\mathcal{B}^+$ is the set defined in equation~\eqref{eq_blochset}, and $\displaystyle \overline{\mathbf{M}}_{\mathcal{V}_{\mathbf{X}}^{(i)},:} = ( \overline{\mathbf{M}}_{n,:} )_{n \in \mathcal{V}_{\mathbf{X}}^{(i)}} $, $\mathcal{V}_{\mathbf{X}}^{(i)}$ corresponding to an estimate of the pure voxel positions in $\mathbf{X}^{(i)}$ at iteration $i$ (the true set $\mathcal{V}_{\mathbf{X}}$ corresponding to the pure voxels of the original $\mathbf{X}$ being unknown). At the first iteration, we choose $ \mathcal{V}_{\mathbf{X}}^{(0)} = \{1, \ldots,N \} $, and it is updated during the greedy process (see Algorithm~\ref{alg_cluster}, step \ref{algo:pure}).

From \eqref{eq_purevoxels}, we can estimate the parameters $\hat{\boldsymbol{\theta}}$ and the proton density $\hat{\rho}$ of the voxels in $\mathcal{V}_{\mathbf{X}}^{(i)}$ using the projection onto $\mathcal{B}^{+}$ with a dictionary $\boldsymbol{\Phi}^{(i)}$ (see steps \ref{alg:pb+_ini}-\ref{alg:pb+_end} of Algorithm \ref{alg_cluster}). $\boldsymbol{\Phi}^{(i)}$ is an adaptive dictionary that is refined at each iteration to reduce the computational cost, the simulations suggest that the accuracy of the reconstructions is preserved. Since there are at least $\kappa$ pure voxels for each active element in $\boldsymbol{\Phi}$ and the value of the proton density is at least $\xi$, we expect that the voxel parameters in $\mathcal{V}^{(i)}_\mathbf{X}$ with $\hat{\rho}>\xi$ will form clusters around the true values of the dictionary elements, an example can be seen in Fig.~\ref{fig:examples}-(a). 

\subsubsection{Clustering} In order to find $K$ centers approximating the parameters of interest, we propose to use the k-means algorithm \cite{Lloyd1982}. 
The objective of k-means is to find $K$ centers that minimise the squared distance from all points to its closest center. The centers obtained by solving the k-means problem $\boldsymbol{\theta}^{\mathcal{S}_1\cap\mathcal{S}_2}\in \mathbb{R}^{K\times P}$ can be used to compute a dictionary $\boldsymbol{\Delta}^{\mathcal{S}_1\cap\mathcal{S}_2}\in \mathbb{C}^{K \times L}$. By solving equation \eqref{eq_pvaprojection} with $\boldsymbol{\Delta}^{\mathcal{S}_1\cap\mathcal{S}_2}$, we would obtain a $\mathbf{U}^{\mathcal{S}_1\cap\mathcal{S}_2}\in \mathbb{R}^{N\times K}$ such that $\mathcal{Z}\left(\mathbf{U}^{\mathcal{S}_1\cap\mathcal{S}_2}\right)\in \mathcal{S}_1\cap \mathcal{S}_2$.

\subsubsection{Non-maximum suppression} 
The k-means algorithm also provides a label to each voxel corresponding to the matched center. We define $\mathbf{c}\in \mathbb{R}^{K \times 1}$ to be the vector containing the number of voxels associated with each center. Inspired by the Non-Maximum Suppression method in \cite{Canny1986}, we use the number of pure voxels assigned to each center to remove the neighbours defined in equation \eqref{eq_neighbours}. We first take the parameters of the highest value of $\mathbf{c}$, and we add all the $\mathbf{c}$ values of the neighbours to the maximum value of $\mathbf{c}$ if it is greater than $\kappa$ we keep the parameters, if not we discard them and set the corresponding values of $\mathbf{c}$ to 0 (see Fig.~\ref{fig:examples}-(b)). We repeat the process until all values of $\mathbf{c}$ are 0. Finally, we use the resulting parameters $\widetilde{\boldsymbol{\theta}}^{(i)}\in \mathbb{R}^{\overline{T}\times P}$ to construct $\widetilde{\boldsymbol{\Delta}}^{(i)}\in \mathbb{C}^{\overline{T}\times L}$.

\subsubsection{Inexact projection onto $\mathcal{B}_{\mathcal{S}_+}$} Once the dictionary $\widetilde{\boldsymbol{\Delta}}^{(i)}$ is approximated, computing the three steps described above, the magnetisation sequence $\mathbf{M}^{(i+1)}$ can be updated. To this aim, we use equation \eqref{eq_pvaprojection}, where the minimisation problem is solved using \textsc{Matlab} built-in function for non-negative least-squares problems \cite{Lawson1938}.

\subsubsection{Pure voxel set update}
In order to avoid noisy voxels, we re-define the set $\mathcal{G}_\mathbf{X}$, introduced in Section~\ref{subsec:Kconstraint}, for $\widetilde{\mathbf{U}}^{(i)}$. Note that $\mathcal{Z}\big(\widetilde{\mathbf{U}}^{(i)}\big)$ is a matrix of the size of $\mathbf{X}$ filling the missing values of $\widetilde{\mathbf{U}}^{(i)}$ with zeros, and thus we can re-define the set $\mathcal{G}_{\mathbf{X}}^{(i)}$ in terms of $\widetilde{\mathbf{U}}^{(i)}$ as:
\begin{equation}
 \mathcal{G}_{\mathbf{X}}^{(i)}= \{ n\in \{1,\ldots,N\} \,|\, \|\widetilde{\mathbf{U}}^{(i)}_{n,:}\|_1>\xi  \}.
\end{equation}
Then, we update the pure voxel set as:
\begin{equation}
	\mathcal{V}_\mathbf{X}^{(i+1)}= \{ n\in\mathcal{G}_{\mathbf{X}}^{(i)}\,|\,
		\max( \widetilde{\mathbf{U}}^{(i)}_{n,:} ) \ge \gamma \|\widetilde{\mathbf{U}}^{(i)}_{n,:}\|_1
		 \},
		 \label{eq_gamma}
\end{equation}
where $0<\gamma<1$ is a relaxation factor used to compensate both for the noise and for the fact that the true dictionary elements are not guaranteed to be present. Note that the parameter $\gamma$ is defined as a proportion of the total proton density in the voxels, and it is used as a threshold to determine if a voxel is pure or not.

\subsubsection{Parameter re-sampling} 
\begin{figure}[!t]
\centering
\begingroup
    \setlength{\tabcolsep}{0mm}
\renewcommand{\arraystretch}{0}
\footnotesize
\begin{tabular}{ccc} 
\includegraphics[width=0.33\linewidth]{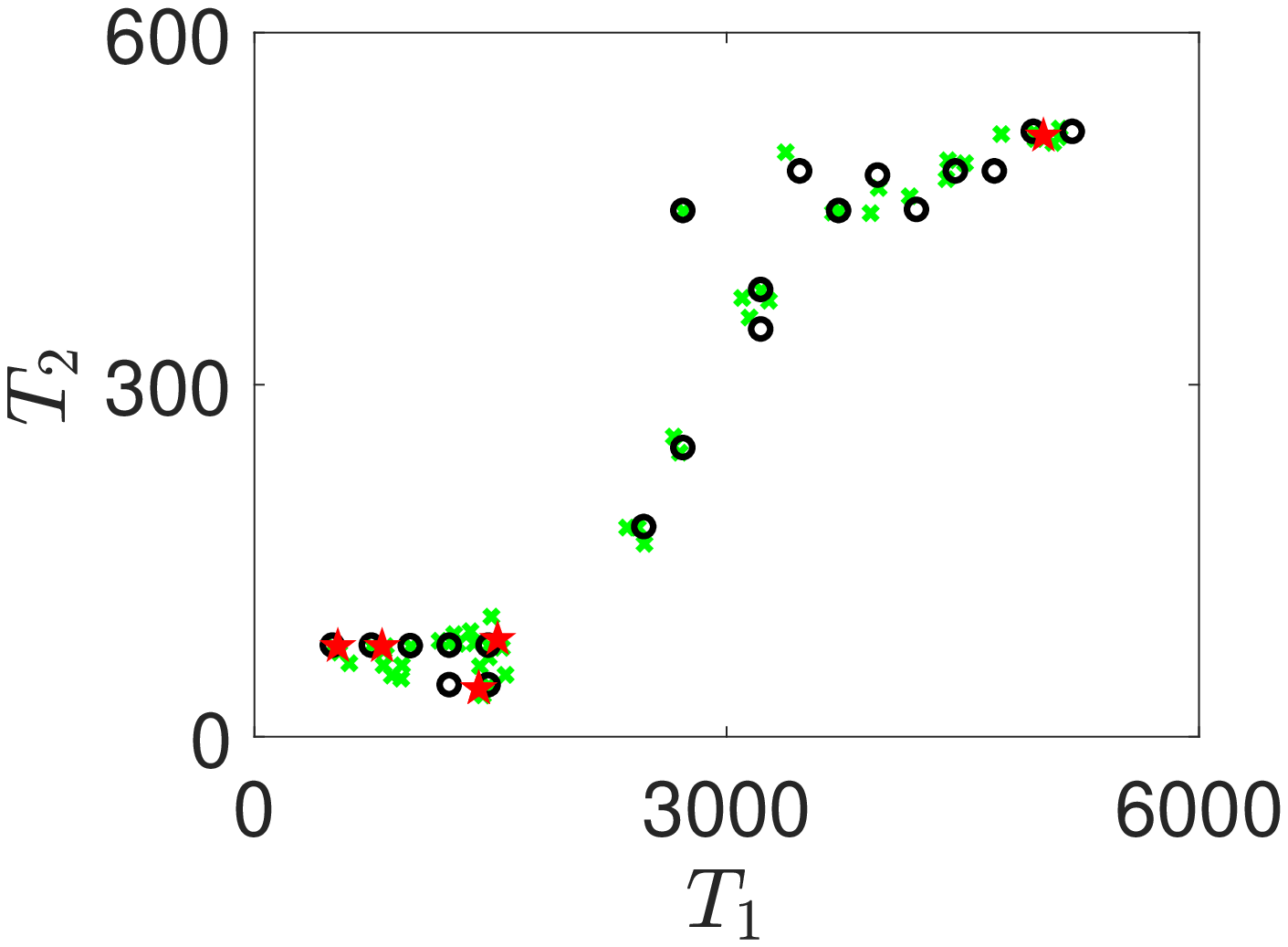}
& \includegraphics[width=0.33\linewidth]{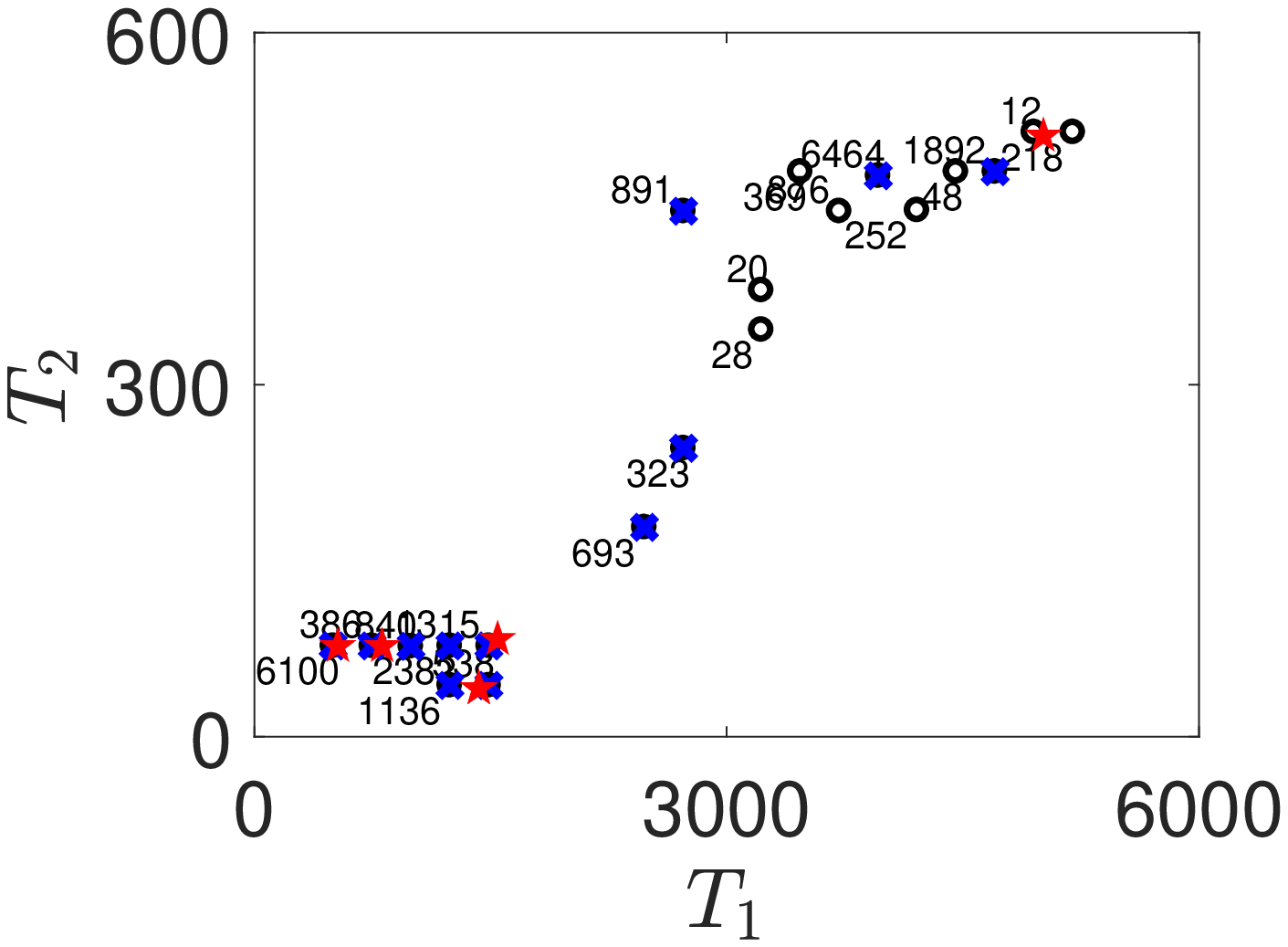} &
\includegraphics[width=0.33\linewidth]{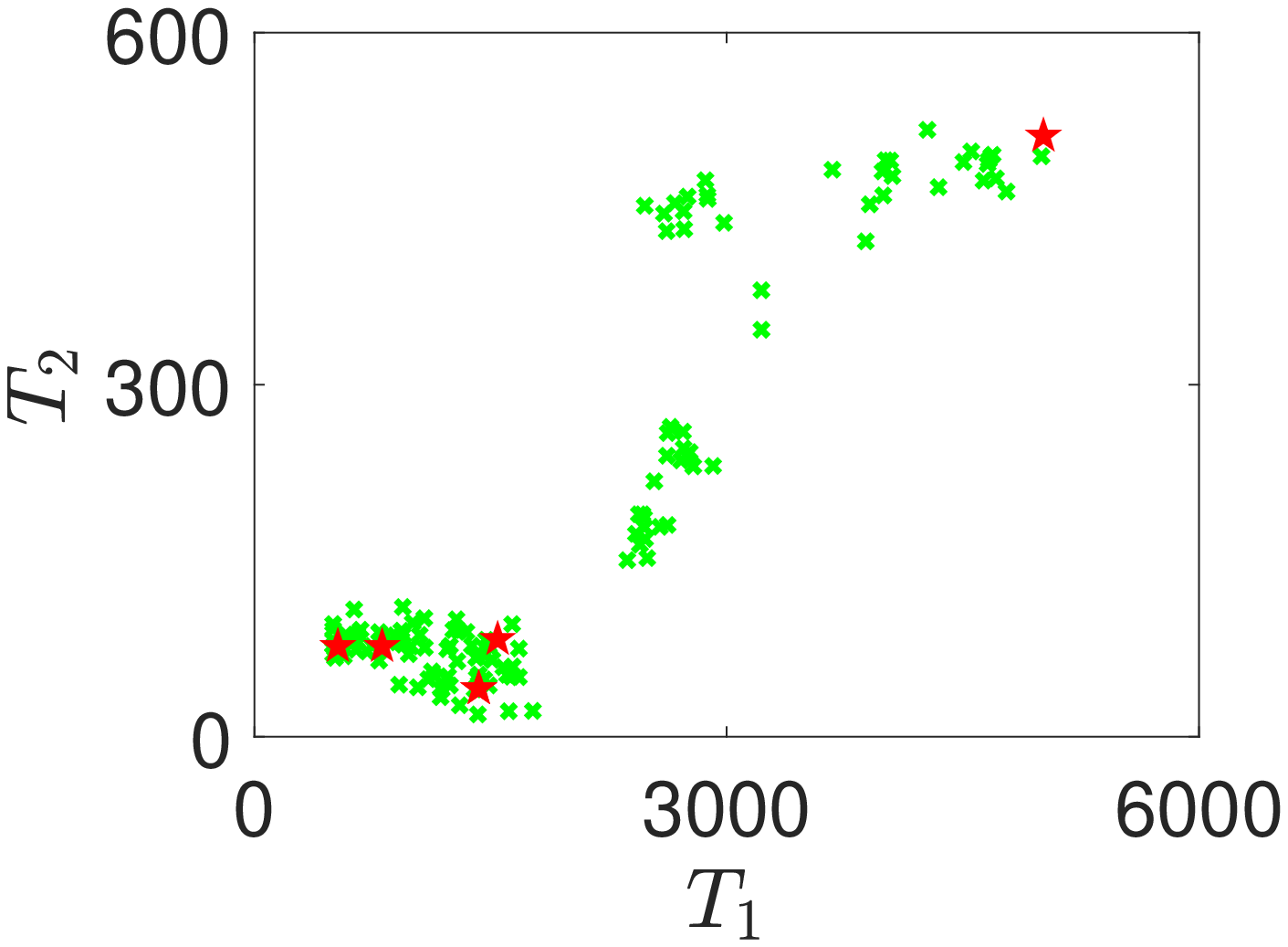}
\\
\end{tabular}
\endgroup
\caption{Examples of the clustering, non-maximum suppression and parameter re-sampling. For all examples the red stars represent the true phantom parameters. (Left) Clustering. The parameters of the voxels in $\mathcal{V}_{\mathbf{X}}^{(i)}$ which its corresponding proton density is greater than $\xi$ (green crosses) are the input of the k-means algorithm and the output are the centers (black circles). (Center) Non-maximum suppression. The centers obtained by the k-means (black circles) and the filtered centers are output of the Non-Maximum Suppression (blue crosses). (Right) Parameter re-sampling. The parameters of the dictionary $\boldsymbol{\Phi}^{(i+1)}$ are obtained by randomly sampling around the parameters obtained by the Non-Maximum Suppression (green crosses).}
\label{fig:examples}
\end{figure}

We update $\boldsymbol{\Phi}^{(i)}$ to refine the manifold elements of interest. For this process, we produce $n_s$ random samples around the elements in $\widetilde{\boldsymbol{\theta}}^{(i)}$ using a Gaussian distribution with a diagonal covariance matrix $\boldsymbol{\Sigma}^{(i)}$ (see Fig.~\ref{fig:examples}-(c)). The values of the covariance matrix $\boldsymbol{\Sigma}^{(i)}$ are reduced by a factor $0<\beta<1$ at each iteration. When the values of $\boldsymbol{\Sigma}^{(i)}$ are sufficiently small, the dictionary $\widetilde{\boldsymbol{\Delta}}$ will not change anymore and after a fixed number of iterations the sequences generated by Algorithm~\ref{alg_IPA_BT} will stabilise. Since the samples are randomly Gaussian distributed, the parameter values are not limited to a given resolution.

\begin{table}\footnotesize
\begin{tabular}{@{}p{1.9cm}@{}p{6.8cm}@{}}
\hline
$N$ & 
    Number of voxels in the volume of interest\\
$Q$ & 
    Number of measurements per acquisition and per coil\\
$L$ & 
    Number of acquisitions\\
$\mathbf{Y}\in \mathbb{C}^{Q\times L\times C}$ & 
    Measurement matrix \\ 
$\mathbf{M}\in \mathbb{C}^{N\times L}$ & 
    Magnetisation Response of the volume of interest, introduced in \eqref{eq_mrfmodel} \\
$h$ & 
    Linear operator from $\mathbb{C}^{N\times L}$ to $\mathbb{C}^{Q\times L \times C}$ defining the acquisition process \\
$P$ & 
    Number of the tissue parameters \\
$\mathcal{M}\subset \mathbb{R}^{1 \times P}$ & 
    Subset describing the feasible parameter space \\
$B$ & 
    Non-linear smooth operator from $\mathcal{M} $ to $ \mathbb{C}^{1\times L}$ describing the magnetic resonance experiment, introduced in Section~\ref{subsec:inverse}\\ 
$\boldsymbol{\Phi} \in \mathbb{C}^{D\times L}$ & 
    Discretisation of $B$ with $D$ elements\\
$\mathcal{B}^+$ & 
    Set describing $\mathbf{M}$ as a magnetisation response of pure voxels\\
$\mathbf{X}\in \mathbb{R}_{+}^{N\times D}$ & 
    Mixing matrix used to describe the PVE, introduced in Section~\ref{Sec:Prop_model}\\
$\mathcal{B}_{\mathcal{S}_+}(\boldsymbol{\Phi})$ & 
    Set describing all possible $\mathbf{M}$ satisfying the proposed PV model, defined in \eqref{eq_setbs+}\\
$\mathcal{S}_+$ & 
    Set describing the intersection of the sets $\mathcal{S}_1$, $\mathcal{S}_2$, $\mathcal{S}_3$ and $\mathcal{S}_4$, defined in \eqref{eq_sets+} \\
$\mathcal{S}_1$ & 
    Positive orthant, defined in \eqref{eq_positivity} \\
$\mathcal{S}_2$ &
    Set describing the maximum number of active dictionary elements, defined in \eqref{eq_K}  with: \newline
    \begin{tabular}{@{}p{0.7cm}@{}p{6.0cm}@{}}
    $\mathcal{G}_{\mathbf{X}}$ & Set indicating the columns of $\mathbf{X}$ contributing to the magnetisation response \\
    $\xi$ & Minimum voxel proton density\\
    $\mathcal{D}_{\mathbf{X}}$ & Set giving the voxels with significant contribution to the magnetisation response\\
    $K$ & Maximum number of active dictionary elements
    \end{tabular}\\ 
$\mathcal{S}_{3}$ &
    Set describing the constraint on the distinct active dictionary elements, defined in \eqref{eq_neighbours} with: \newline
    \begin{tabular}{@{}p{0.7cm}@{}p{6.0cm}@{}}
    $\upsilon$ & Constant used to define neighbourhoods for the dictionary elements in the parameter space\\
    $\mathcal{N}_{\upsilon}$ & Set describing the neighbour dictionary elements for given dictionary element $d$\\
    \end{tabular} \\
$\mathcal{S}_{4}$ &
    Set giving the minimum number of pure voxels per active dictionary element, defined in \eqref{eq_pure} with: \newline
    \begin{tabular}{@{}p{0.7cm}@{}p{6.0cm}@{}}
    $\mathcal{V}_{\mathbf{X}}$ & Set describing the pure voxels in $\mathbf{X}$ \\
    $\kappa$ & Minimum number of pure voxels per active dictionary element\\
    \end{tabular} \\
$\mu$ & 
    Step size in Algorithm~\ref{alg_IPA_BT}\\
$\mathbf{U}\in \mathbb{R}^{N \times \overline{T}}$ & 
    Sub-matrix of $\mathbf{X}$ (see eq.~\eqref{eq:Proj_XS_PhiS}) \\
$\mathbf{\boldsymbol{\Delta}}\in \mathbb{C}^{\overline{T} \times L}$ & 
    Sub-matrix of $\boldsymbol{\Phi}$ (see eq.~\eqref{eq_Delta})    \\
$\mathcal{Z}$ & 
    Operator from $\mathbb{R}_+^{N\times \overline{T}} $ to $ \mathbb{R}_+^{N \times D} $ that maps a matrix $\mathbf{U}$ to the correspronding $\mathbf{X}$\\
$\boldsymbol{\lambda}\in \mathbb{C}^{N}$ & 
    Vetor used to compensate for the complex phase errors in the model\\
$\gamma$ & 
    Tolerance parameter for a pure voxels (see \eqref{eq_gamma})\\
$\tau, \tau_k, \tau_\upsilon, \tau_\kappa$ & 
    Tolerance parameters for the initialisation process (see Section~\ref{sec:Initialisation})\\
\hline
\end{tabular}
\caption{Table of symbols.}
\end{table}

\section{Choice of the parameters and initialisation}\label{sec:Initialisation}
Since $\mathcal{S}_+$ is a non-convex set, the choice of the initialisation is important. If the initial magnetisation sequence or the dictionary are not close to the desired values, the greedy approximate projection can fail. In this section, we will describe the initialisation for our algorithm.

\subsection{Choice of the parameters}
The choice of $\xi$, setting the minimum proton density, is related to the background noise, the ideal $\xi$ is a value between the background noise and the signal in the volume of interest. If $\xi$ is too small, empty voxels will affect the clustering process. If it is too big, the tissue voxels will not be considered in the clustering process.

As mentioned before, the dictionary $\boldsymbol{\Phi}^{(i)}$ is updated through the iterations to reduce the complexity of the algorithm. We fix $\boldsymbol{\Phi}^{(0)}$ to all possible combinations of 20 values of $T_1$ and 20 values of $T_2$, equally spaced in $\mathcal{M}$. 

Concerning the number of random samples $n_s$, on the one hand if we choose it too big, we increase the complexity of our pure voxel projection. On the other hand if we set $n_s$ too small, more iterations will be needed to find the elements of interest. In all our simulations (simulated and \textit{in vivo} data) we fix $n_s=10$. 

For the diagonal elements of the covariance matrix (i.e. $\boldsymbol{\Sigma}_{1,1}^{(0)}$ and $\boldsymbol{\Sigma}_{2,2}^{(0)}$) associated to the resampling of the dictionary, if they are chosen too big, the parameter sampling will be far from the parameters of interest, increasing the number of iterations required to find them. If they are too small, the algorithm may not find the parameter of interest. $\boldsymbol{\Sigma}^{(0)}$ should be chosen based on the parameter separation of $\boldsymbol{\Phi}^{(0)}$. In all the reconstructions we fix $\boldsymbol{\Sigma}_{1,1}^{(0)}=40$ and $\boldsymbol{\Sigma}_{2,2}^{(0)}=10$. 

Similarly, for the decreasing parameter $\beta$ of the covariance matrix (see step \ref{algo:stepsigma} in Algorithm~\ref{alg_cluster}), if it is chosen too big, the algorithm will need more iterations to find the correct elements while if it is too small the algorithm may not explore the true parameters. We fix $\beta=0.9$ in the considered scenarios. 

The choice of the pure voxel tolerance $\gamma$ is related to the noise and the accuracy of the dictionary during the iterations of the algorithm. If it is too big, the elements of interest could be eliminated through the iterations since pure voxels may be considered as PV voxels, if it is too small, the PV voxels may be considered as pure affecting the clustering process. We found in our simulations that $\gamma=0.85$ is a suitable choice.

The choice of the different parameters $K$, $\upsilon$ and $\kappa$ has been investigated during preliminary work. In particular, we observed a significant increase in the residual $\|\mathbf{Y}-h(\mathbf{M})\|_2$ when $K$ is not sufficiently large. For $\upsilon$ and $\kappa$, we see a significant increase in the residual when they are chosen too large (i.e. merging proton density maps of the true tissues), and an increase of noisy proton density maps when they are chosen too low. 

We propose to automatically choose $K$, $\upsilon$ and $\kappa$ by analysing the residual. Precisely, we choose a tolerance value on the residual, denoted by $\tau>0$. This value, indicates the minimum contribution of an element of the dictionary in the residual. If $\tau$ is chosen too big, our solution will contain noisy elements. While if it is chosen too small our elements of interest will be removed from the reconstruction.

\subsection{Initialisation}
The global GAP-MRF method, including the initialisation process, is described in Algorithm~\ref{alg_initialisation}. It describes the process to choose the parameters $K$, $\upsilon$ and $\kappa$. Firstly, the estimation of $K$ is described in steps \ref{algo:eKi}-\ref{algo:eKe}. Fixing all the other parameters, $K$ is estimated by running multiple times the GAP-MRF iterations given in Algorithm 1. We assume that we have a suitable estimate of $K$ when the stopping criteria given in step \ref{algo:eKe} of Algorithm~\ref{alg_initialisation} is reached. The same process is adopted for the estimation of $\upsilon$ described (steps \ref{algo:eUi}-\ref{algo:eUe}) and $\kappa$ (steps \ref{algo:ekappai}-\ref{algo:ekappae}). For these two estimates, we allow for a small tolerance ($\tau_\upsilon>0$ and $\tau_\kappa>0$, respectively), for robustness purposes. 
Note that each new run of Algorithm~\ref{alg_IPA_BT} uses the previous estimated of $\mathbf{M}$, $\boldsymbol{\Delta}$ and $\boldsymbol{\theta}$, in order to accelerate the global method.

 \begin{algorithm}[!t]
	\caption{GAP-MRF global method}
    \label{alg_initialisation}
	{\begin{algorithmic}[1]
		\State 
        	\textbf{Input:} $\mathbf{Y}$, $\boldsymbol{\Gamma}$,   $\boldsymbol{\Phi}$, $\boldsymbol{\theta}$, $\xi$, $\tau$, $\boldsymbol{\Sigma}^{(0)}_{1,1}=40$,  $\boldsymbol{\Sigma}^{(0)}_{2,2}=10$, $\zeta=0.99$, $\mathcal{V}_\mathbf{X}^{(0)}=\{1,\ldots,N\}$,  $\beta=0.9$, $n_s=10$, $\mathbf{M}^{(0)}=\boldsymbol{0}$, $(\tau_K, \tau_\upsilon, \tau_\kappa) = (10, 0.02, 10)$
        	
        \vspace*{0.2cm}
        
		\State  \label{algo:eKi}
		        \textbf{Estimation of $K$:}
		\State  $\quad$ 
		        \textbf{Input:} $(\gamma, K, \upsilon, \kappa)=(0,0,0,0)$, $\widetilde{\boldsymbol{\Delta}}^{(0)}=\{\}$, $\widetilde{\boldsymbol{\theta}}^{(0)}=\{\}$, $j=0$.
        \State $\quad$ \textbf{Do}
            \State  $\quad\quad$ 
                    $\boldsymbol{\Phi}^{(0)}=\left[\boldsymbol{\Phi} ,\widetilde{\boldsymbol{\Delta}}^{(j)}\right]$, 
                    $\boldsymbol{\theta}^{(0)}=\left[(\boldsymbol{\theta})^{T} ,(\widetilde{\boldsymbol{\theta}}^{(j)})^T\right]^T$
            \State  $\quad\quad$ 
                    $K=K+ \tau_{K}$
            \State  $\quad\quad$ 
                    $\left[\mathbf{M}^{(j+1)},\widetilde{\boldsymbol{\Delta}}^{(j+1)},\widetilde{\boldsymbol{\theta}}^{(j+1)}\right] = \text{Algorithm1} (\mathbf{Y},\zeta,\mathbf{M}^{(j)} )$
            \State  $\quad\quad$ 
                    j=j+1
		\State  $\quad$ 
		        \textbf{while} $\|\mathbf{Y}-h(\mathbf{M}^{(j)})\|_2-\|\mathbf{Y}-h(\mathbf{M}^{(j-1)})\|_2> \tau$.
		\State  \label{algo:eKe} $\quad$ 
		        \textbf{Output:} $K^\star = K$, $j=j-1$
        	
        \vspace*{0.2cm}
        
		\State  \label{algo:eUi}
		        \textbf{Estimation of $\upsilon$:}
		\State $\quad$ \textbf{Input:} $(\gamma, K, \upsilon, \kappa)=(0.85, K^\star, \tau_{\upsilon}, 0)$ 
        \State $\quad$ \textbf{Do}
            \State  $\quad\quad$
                    $\boldsymbol{\Phi}^{(0)}=\widetilde{\boldsymbol{\Delta}}^{(j)}$, 
                    $\boldsymbol{\theta}^{(0)}=\widetilde{\boldsymbol{\theta}}^{(j)}$
            \State  $\quad\quad$
                    $\upsilon=\upsilon+ \tau_{\upsilon}$
            \State  $\quad\quad$
                    $\left[\mathbf{M}^{(j+1)},\widetilde{\boldsymbol{\Delta}}^{(j+1)},\widetilde{\boldsymbol{\theta}}^{(j+1)}\right] = \text{Algorithm1}\left(\mathbf{Y},\zeta,\mathbf{M}^{(j)}\right)$
            \State  $\quad\quad$
                    j=j+1;
		\State  $\quad$ 
		        \textbf{while} $\|\mathbf{Y}-h(\mathbf{M}^{(j)})\|_2-\|\mathbf{Y}-h(\mathbf{M}^{(j-1)})\|_2> \tau$.
		\State  \label{algo:eUe} $\quad$ 
		        \textbf{Output:} $\upsilon^\star = \upsilon- 2\tau_{\upsilon}$, $j=j-1$
        	
        \vspace*{0.2cm}
        
		\State  \label{algo:ekappai}
		        \textbf{Estimation of $\kappa$:}
		\State $\quad$ \textbf{Input:} $(\gamma, K, \upsilon, \kappa)= (0.85, K^\star, \upsilon^\star, 10)$ 
        \State $\quad$ \textbf{Do}
            \State  $\quad\quad$
                    $\boldsymbol{\Phi}^{(0)}=\widetilde{\boldsymbol{\Delta}}^{(j)}$, 
                    $\boldsymbol{\theta}^{(0)}=\widetilde{\boldsymbol{\theta}}^{(j)}$
            \State  $\quad\quad$
                    $\kappa=\kappa+ \tau_{\kappa}$
            \State  $\quad\quad$
                    $\left[\mathbf{M}^{(j+1)},\widetilde{\boldsymbol{\Delta}}^{(j+1)},\widetilde{\boldsymbol{\theta}}^{(j+1)}\right] = \text{Algorithm1}\left(\mathbf{Y},\zeta,\mathbf{M}^{(j)}\right)$
            \State  $\quad\quad$
                    j=j+1;
		\State  $\quad$ 
		        \textbf{while} $\|\mathbf{Y}-h(\mathbf{M}^{(j)})\|_2-\|\mathbf{Y}-h(\mathbf{M}^{(j-1)})\|_2> \tau$.
		\State  \label{algo:ekappae} $\quad$ 
		        \textbf{Output:} $\kappa^\star = \kappa-2\tau_{\kappa}$ 
        	
        \vspace*{0.2cm}
        
		\State \textbf{GAP-MRF Global Iterations:}
		\State  $\quad$ 
		        \textbf{Input:} $(\gamma, K, \upsilon, \kappa)= (0.85, K^\star, \upsilon^\star, \kappa^\star)$, $\boldsymbol{\Phi}^{(0)}=\widetilde{\boldsymbol{\Delta}}^{(j-1)}$, $\boldsymbol{\theta}^{(0)}=\widetilde{\boldsymbol{\theta}}^{(j-1)}$
		\State  $\quad$ 
		        $\left[ \overline{\mathbf{M}}, \overline{\boldsymbol{\Delta}}, \overline{\boldsymbol{\theta}}\right] = \text{Algorithm1}\left( \mathbf{Y}, \zeta, \mathbf{M}^{(j-1)} \right)$
		\State  $\quad$ 
		        $\displaystyle \overline{\mathbf{U}} = \argmin_{\mathbf{U}\in\mathbb{R}_+^{N\times \overline{T}}} \frac{1}{2} \| \mathbf{U} \overline{\boldsymbol{\Delta}} - \overline{\mathbf{M}} \|_2^2$
		\State  $\quad$ 
		        \textbf{Output:} $\overline{\mathbf{M}}$, $\overline{\boldsymbol{\Delta}}$, $\overline{\boldsymbol{\theta}}$ and $\overline{\mathbf{U}}$
\end{algorithmic}}
\end{algorithm}

\section{Simulations and results}\label{Sec:simulations}

In this section, we present the procedure used to evaluate the reconstruction with simulated data using a simulated PV phantom. For the sake of simplicity, we consider a particular case of model~\eqref{eq_mrfmodel}, with only one coil (i.e. $C=1$) and the corresponding sensitivity map $\mathbf{S}_{:,:,1}$ to be the identity matrix. 
An Echo-planar Imaging (EPI) undersampling scheme is used \cite{Rieger2017,Rieger2018}. The Bloch Equations are used for the non-linear mapping, with the random flip angles $\alpha$ and fixed repetition times $TR$ as described in \cite{CSMRF}. We compare the BLIP algorithm \cite{CSMRF} to the proposed GAP-MRF method, considering two different experiments. 
In the first experiment, we investigate the effect of measurement noise by varying the input SNR (iSNR in dB), defined as $\text{iSNR}= 20\log\left( \|h(\mathbf{M})\|_2/(\sqrt{Q LC}\sigma_{\mathbf{M}}\right))$, where $\sigma_{\mathbf{Y}}$ is the standard deviation of the noise. We vary the iSNR from $10$dB to $50$dB.
In the second experiment, we investigate the effect of the magnetisation sequence length $L \in [200, 600]$, affecting directly the acquisition time. 
In both the cases, we choose the undersampling ratio $N/Q=16$ to simulate the EPI \textit{in vivo} data in Section~\ref{section_realdata}. 

The BLIP algorithm and Algorithm~\ref{alg_IPA_BT} are stopped when the following stopping criterion is satisfied $|E^{(i+1)} - E^{(i)}| < 10^{-4} E^{(i+1)}$, where $ E^{(i)} = \|h \big(\mathbf{M}^{(i)} \big)-\mathbf{Y} \|^2_2 $, and $(\mathbf{M}^{(i)})_{i\in \mathbb{N}}$ is a sequence generated by the algorithms. In all simulations, GAP-MRF takes at most 120 iterations of Algorithm~\ref{alg_IPA_BT} to converge taking the initialisation into account. Both algorithms were implemented in \textsc{Matlab}. For the longest test, BLIP takes around 30 minutes and GAP-MRF takes around 3 hours using a computer with a 3rd generation Quad Core Intel i5 processor. The computational time can be significantly improved with a parallel implementation of both algorithms.

\subsection{Partial volume simulated phantom}
We create a simulated phantom according to \cite{brainphantom}, with five tissues: adipose, WM, GM, muscle and cerebrospinal fluid (CSF). More precisely, to introduce the PVE, we use blocks of $2 \times 2$ voxels to form a lower resolution phantom containing PV voxels. The resulting volume is resized to $256\times 256$ voxels. In the first column of Fig.~\ref{im_e2pdmaps} proton density maps and the voxel distribution of the simulated phantom are shown. Using this representation we can see the structure of the tissues of interest. Traditionally in qMRI, individual parameter maps are evaluated since only a tissue per voxel is considered but in a PV scenario this is not meaningful since several parameter maps would be needed and visually do not show the tissue structures. We also compute the dominant tissue (highest proton density in the voxel) parameter maps for a traditional evaluation. The phantom dominant tissue parameter maps can be seen in the first column of Fig.~\ref{fig:e2dominanttissues}. Note that for the construction of the phantom, we only consider in-plane PV, while in reality through-plane PV and in-plane PV occurs. Both kind of PV are modelled the same way and should not make any difference in the reconstructions.

\subsection{Evaluation}
In order to evaluate the algorithms, we use the Signal-to-Noise Ratio (SNR in dB) defined as $\mbox{SNR}(\mathbf{U}_{:,t},\widetilde{\mathbf{U}}_{:,t})=10\log\left( \sum_{n=1}^N \left(\mathbf{U}_{n,t}\right)^2/\sum_{n=1}^N (\mathbf{U}_{n,t}-\widetilde{\mathbf{U}}_{n,t} )^2 \right)$, where $t\in\{1,\ldots,T\}$ is the index of the evaluated tissue, $\mathbf{U}$ is the mixing matrix ground truth and $\widetilde{\mathbf{U}}$ is the estimation. Similarly for the magnetisation sequence SNR, we sum for all values in the matrix. To construct the matrix $\widetilde{\mathbf{U}}$, a tolerance of $15\%$ from the ground truth parameter values is used (i.e. for $T_1=530$ and $T_2=77$ milliseconds (ms) all the dictionary elements that fall for $T_1$ in the range of $[450.5-609.5]$ms and simultaneously for $T_2$ in the range of $[65.45-88.55]$ms are considered). In order to evaluate if the tissues are correctly identified, we define the success rate (SR) index as the proportion of voxels where the number of elements are correctly identified and its corresponding parameters fall within the $15\%$ of the true parameters. The same definition of SR is used for both pure and PV voxels (considering only the corresponding phantom voxels). Due to noise, there could be small values in $\widetilde{\mathbf{U}}$ that could significantly affect the SR. In consequence, we choose not to consider values that are smaller than $30$, given that the range of the proton densities is from $80$ to $400$.

\subsection{Experiment 1 - Impact of the iSNR}
In this experiment, we investigate the behaviour of both the BLIP and the GAP-MRF algorithms while changing the input noise. We fix the magnetisation sequence length $L=1000$. The dictionary for BLIP is defined as in \cite{CSMRF} with $D=16170$. The results correspond to an average (with standard deviation) over 10 runs of each choice of iSNR.

\begin{figure}[!t]
	\centering
    \begingroup
	\setlength{\tabcolsep}{0.2mm}
\renewcommand{\arraystretch}{.2}
    \begin{tabular}{cccc}
    
        \includegraphics[width=0.3\linewidth]{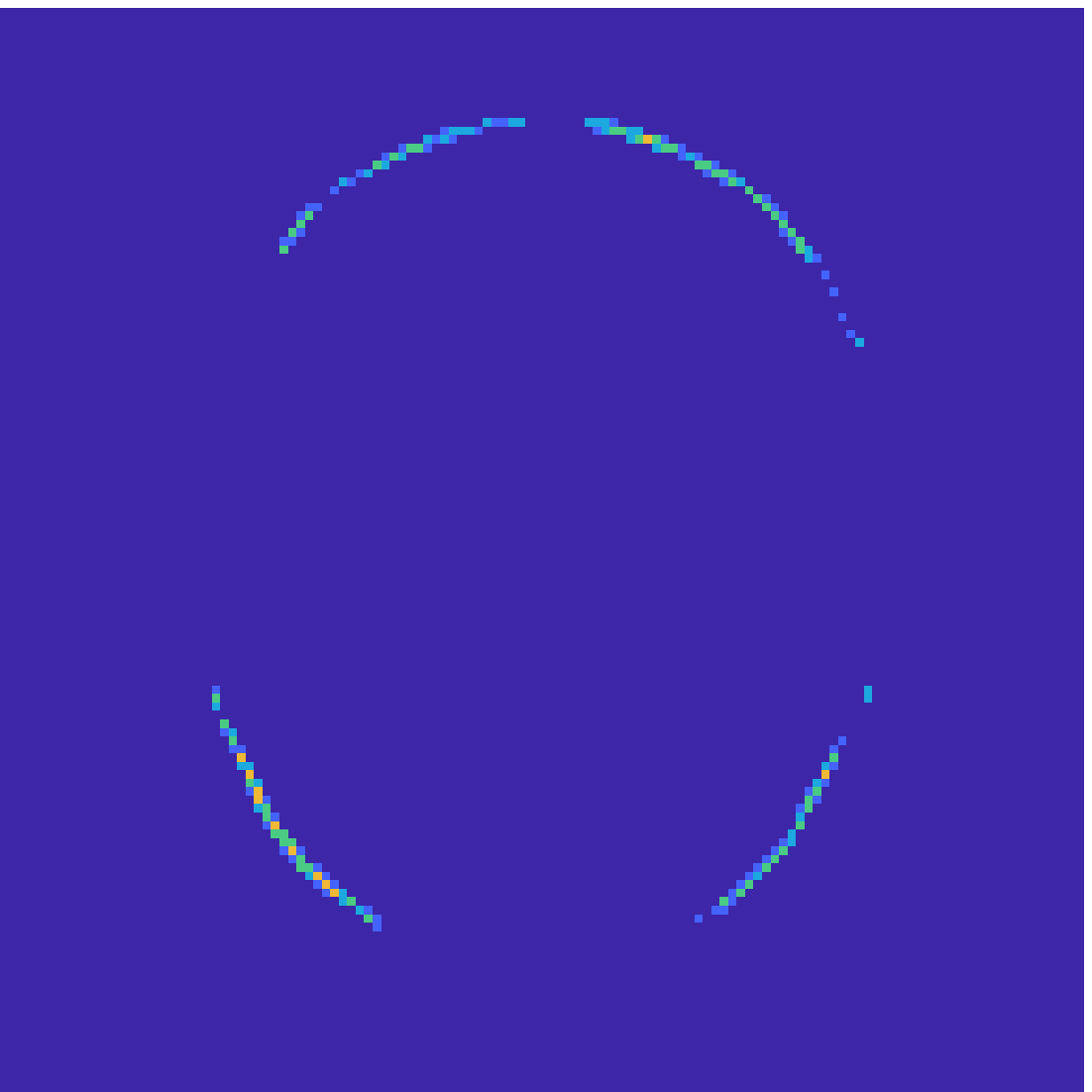}
        & \includegraphics[width=0.3\linewidth]{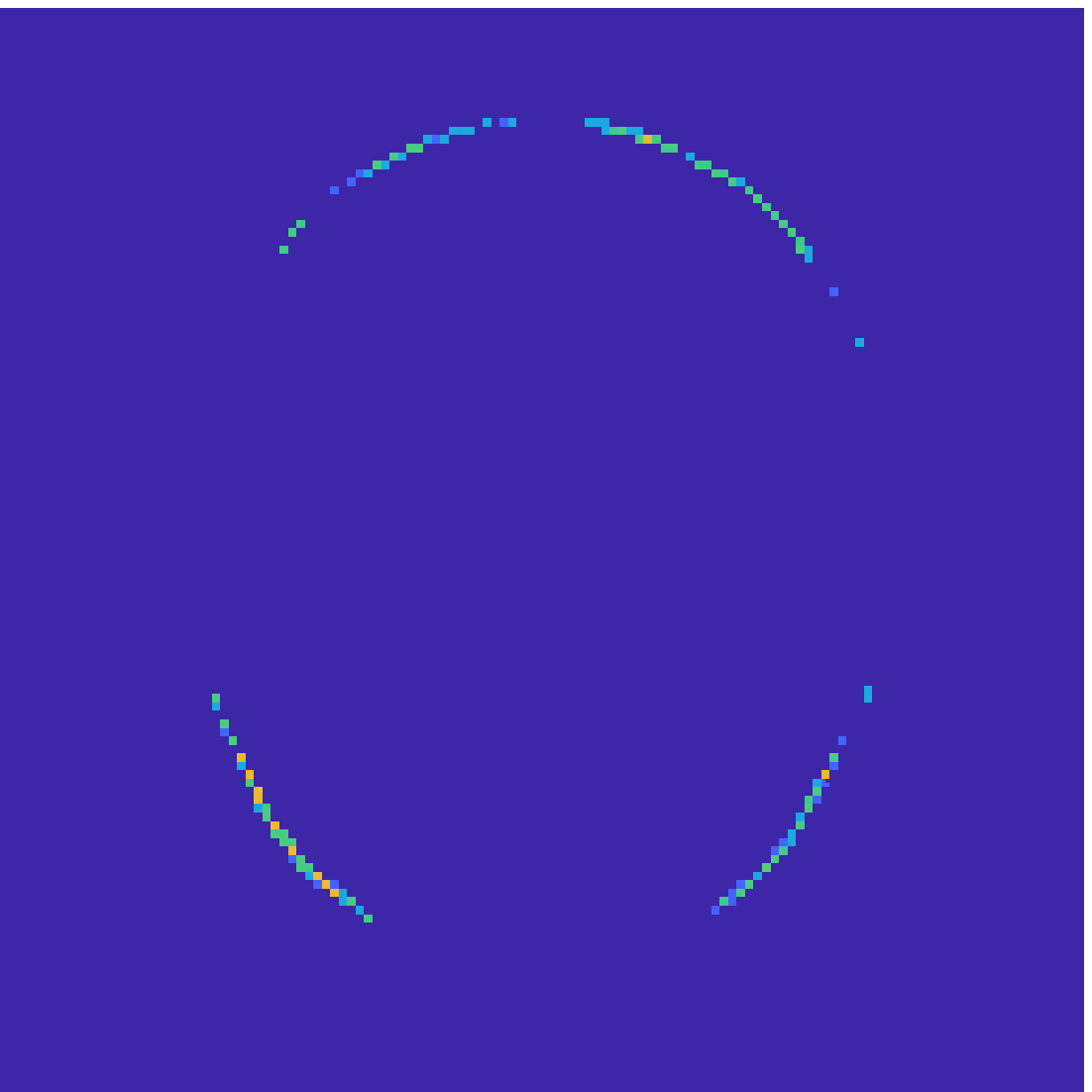}
        & \includegraphics[width=0.3\linewidth]{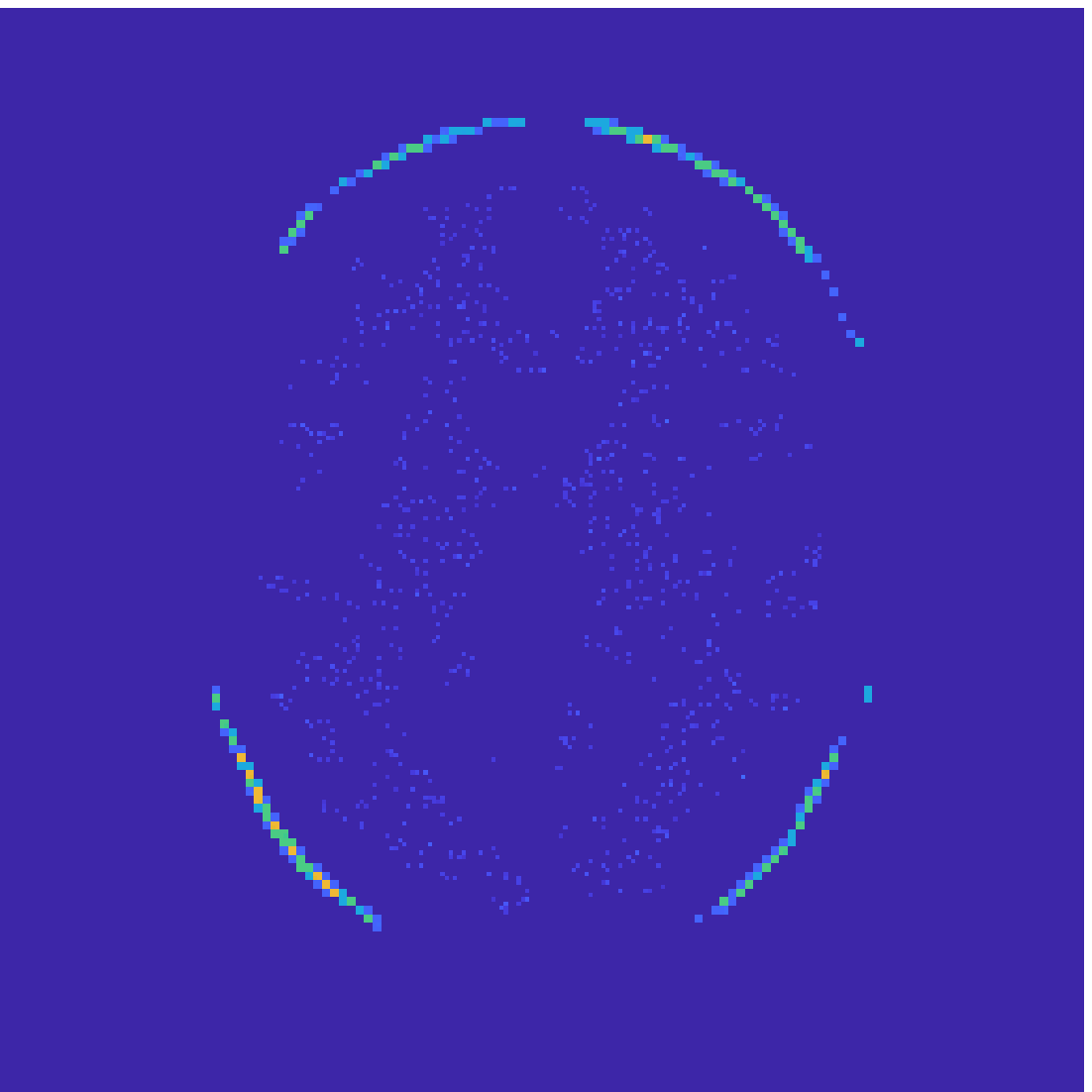} & \includegraphics[width=0.045\linewidth]{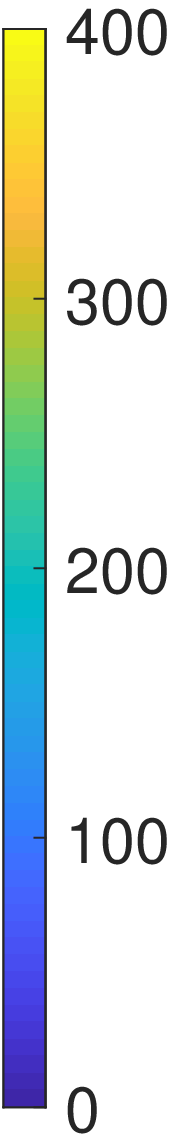}
        \\
        \includegraphics[width=0.3\linewidth]{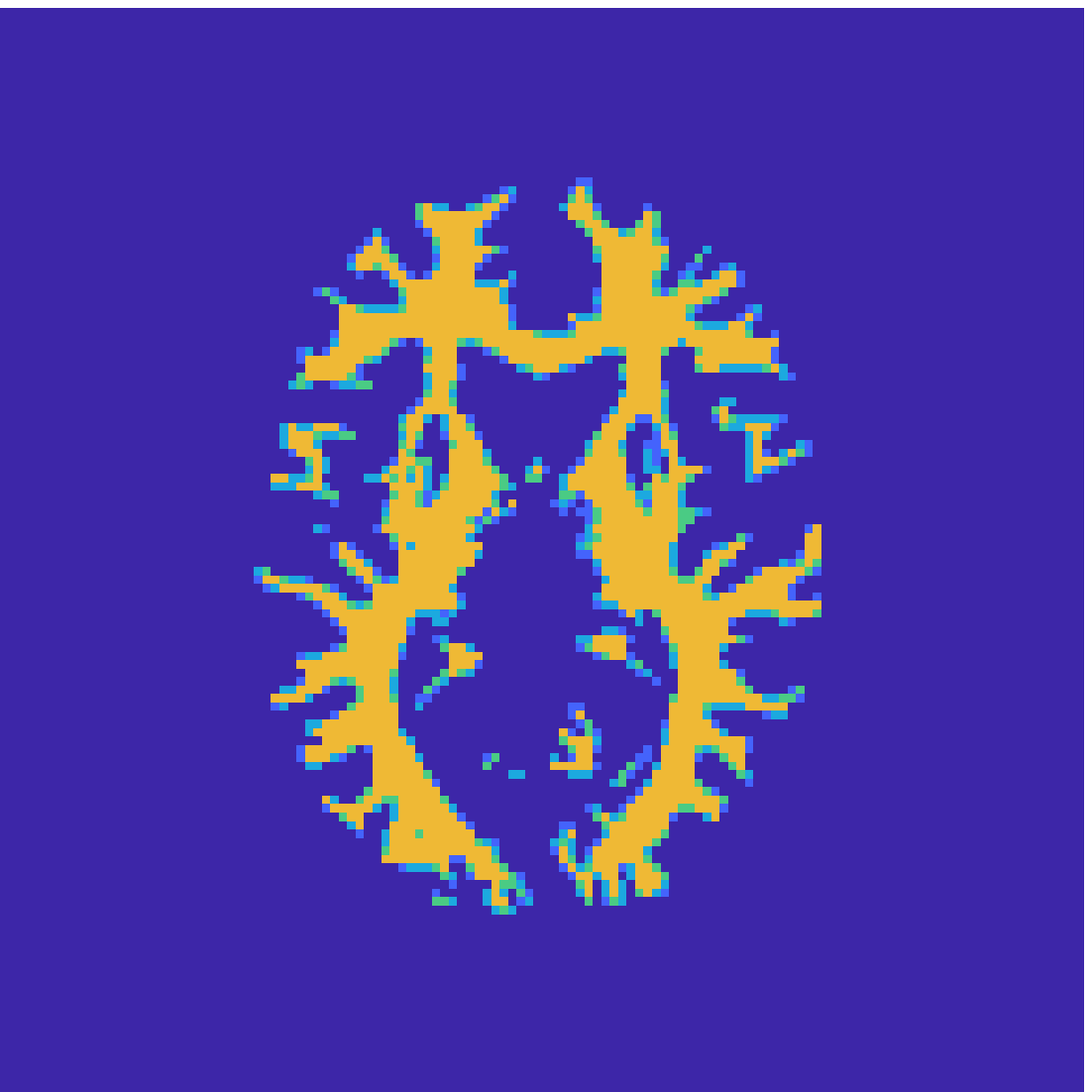}
        & \includegraphics[width=0.3\linewidth]{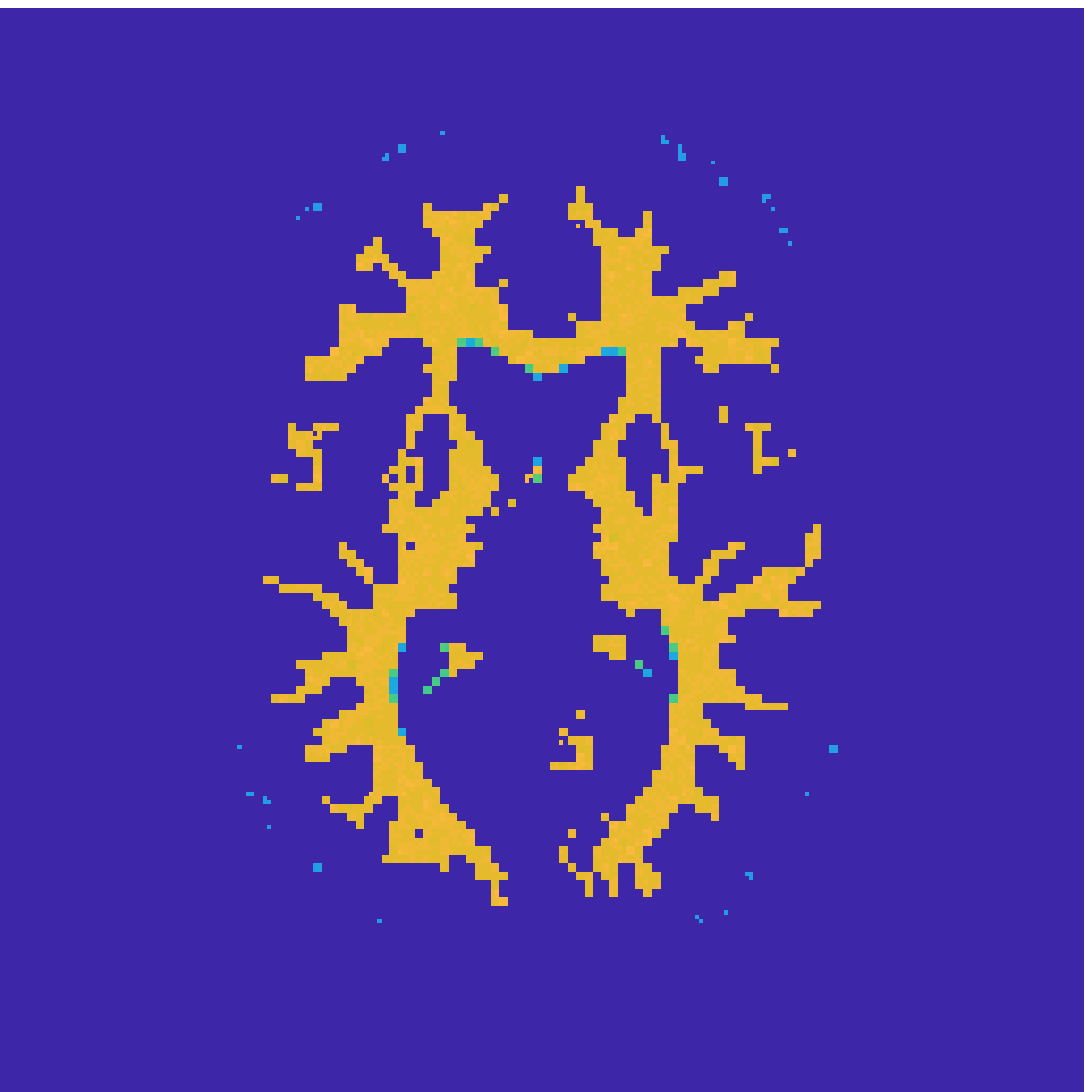}
        & \includegraphics[width=0.3\linewidth]{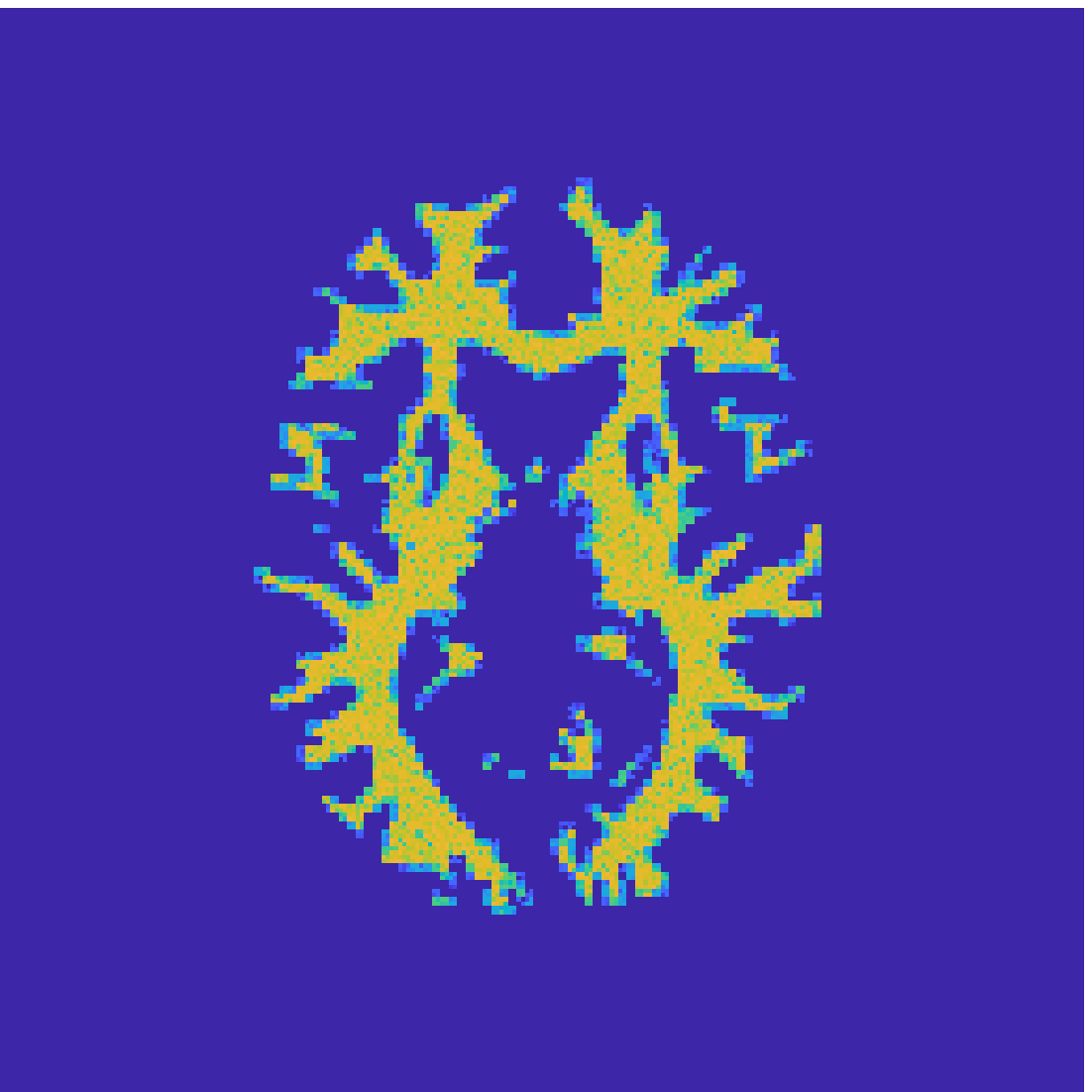} & \includegraphics[width=0.045\linewidth]{fig/colorbar6.eps}
        \\
        \includegraphics[width=0.3\linewidth]{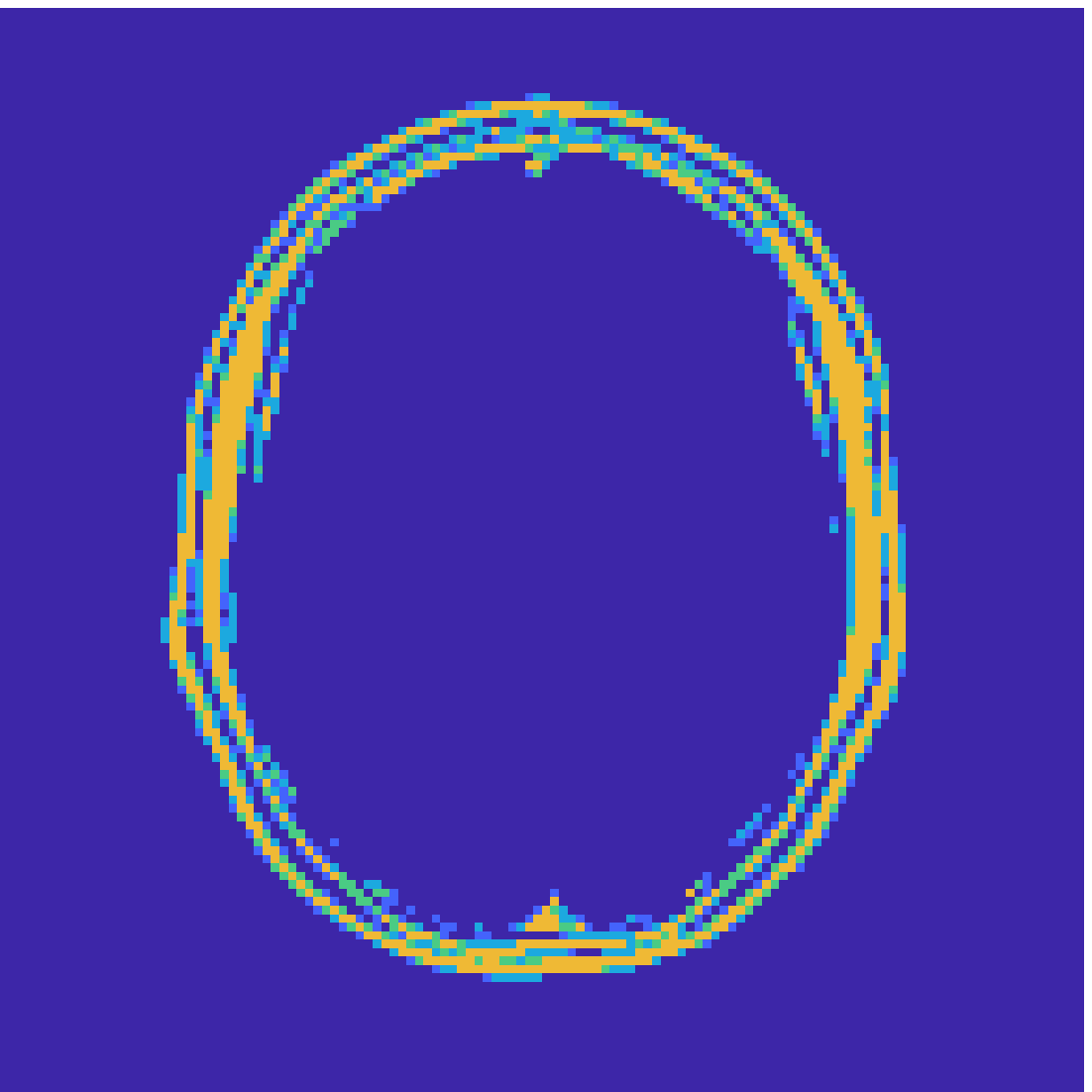}
        & \includegraphics[width=0.3\linewidth]{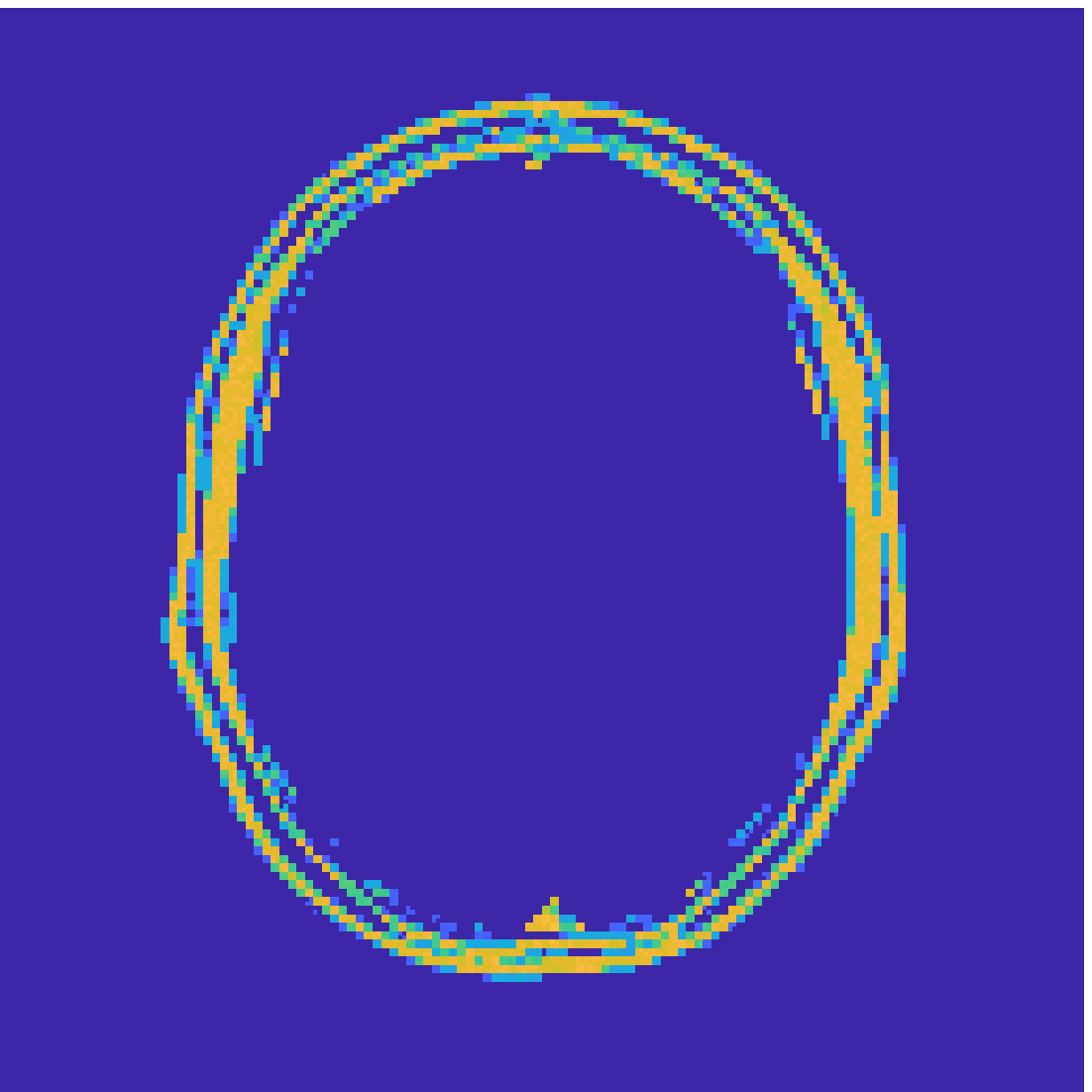}
        & \includegraphics[width=0.3\linewidth]{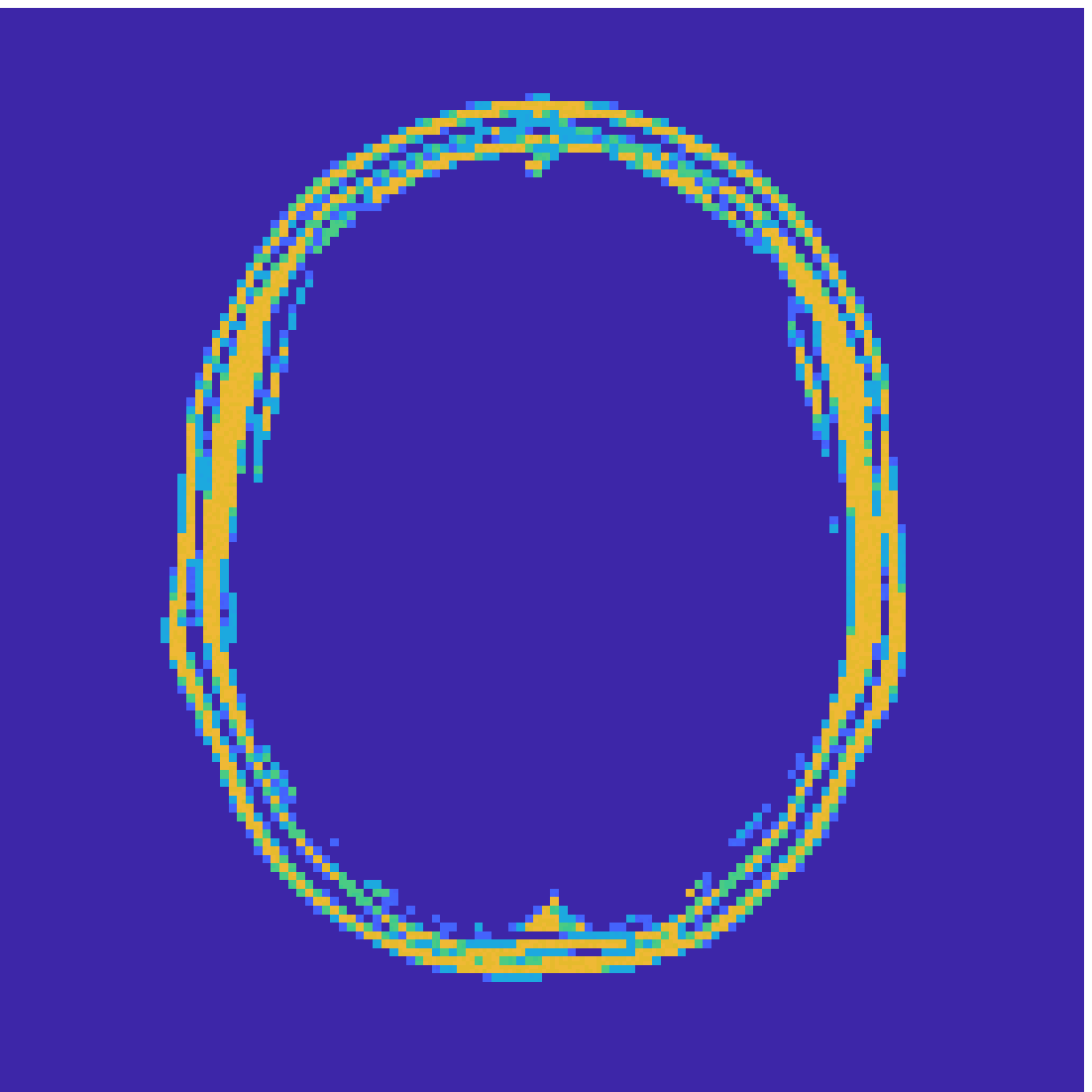} & \includegraphics[width=0.045\linewidth]{fig/colorbar6.eps}
        \\
        \includegraphics[width=0.3\linewidth]{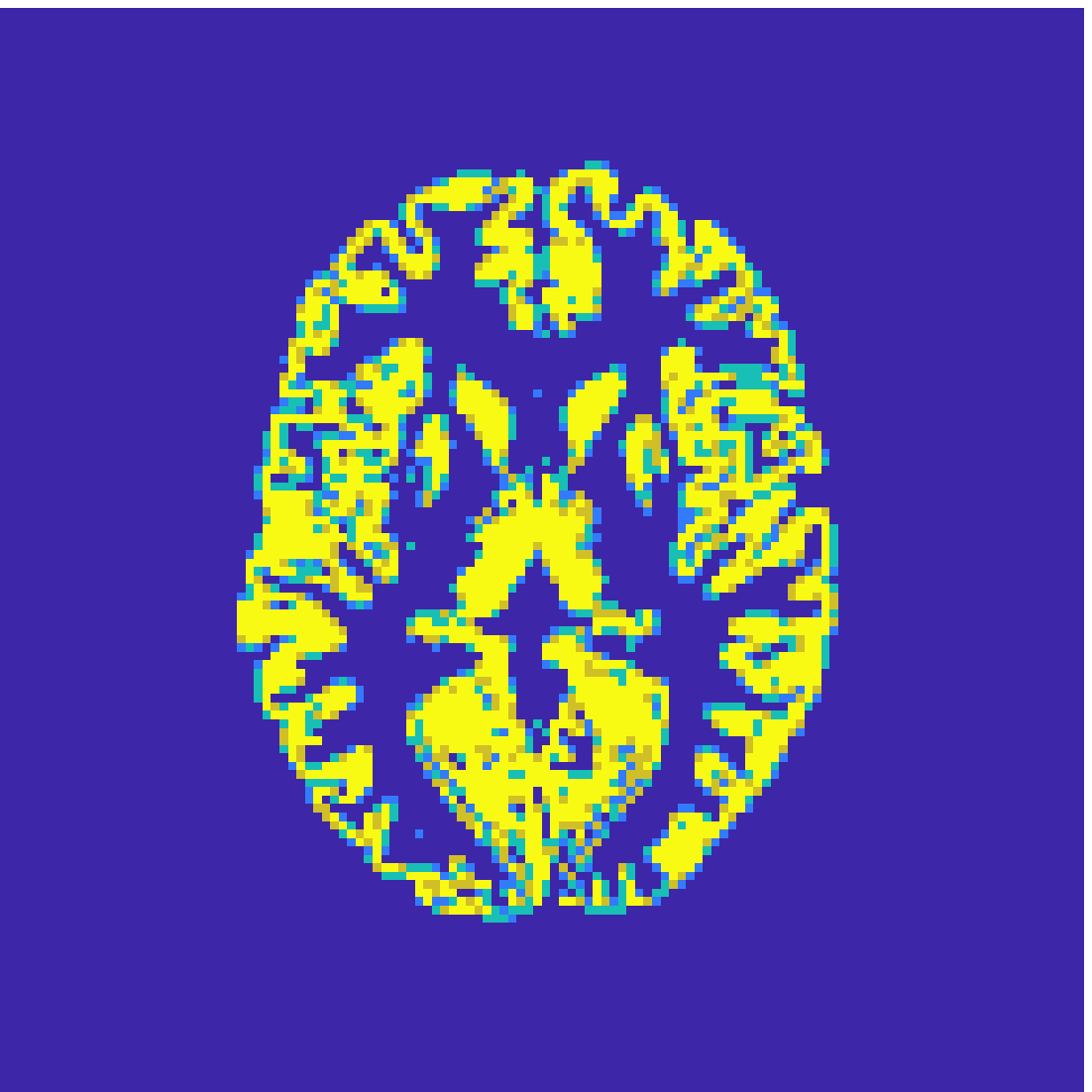}
        & \includegraphics[width=0.3\linewidth]{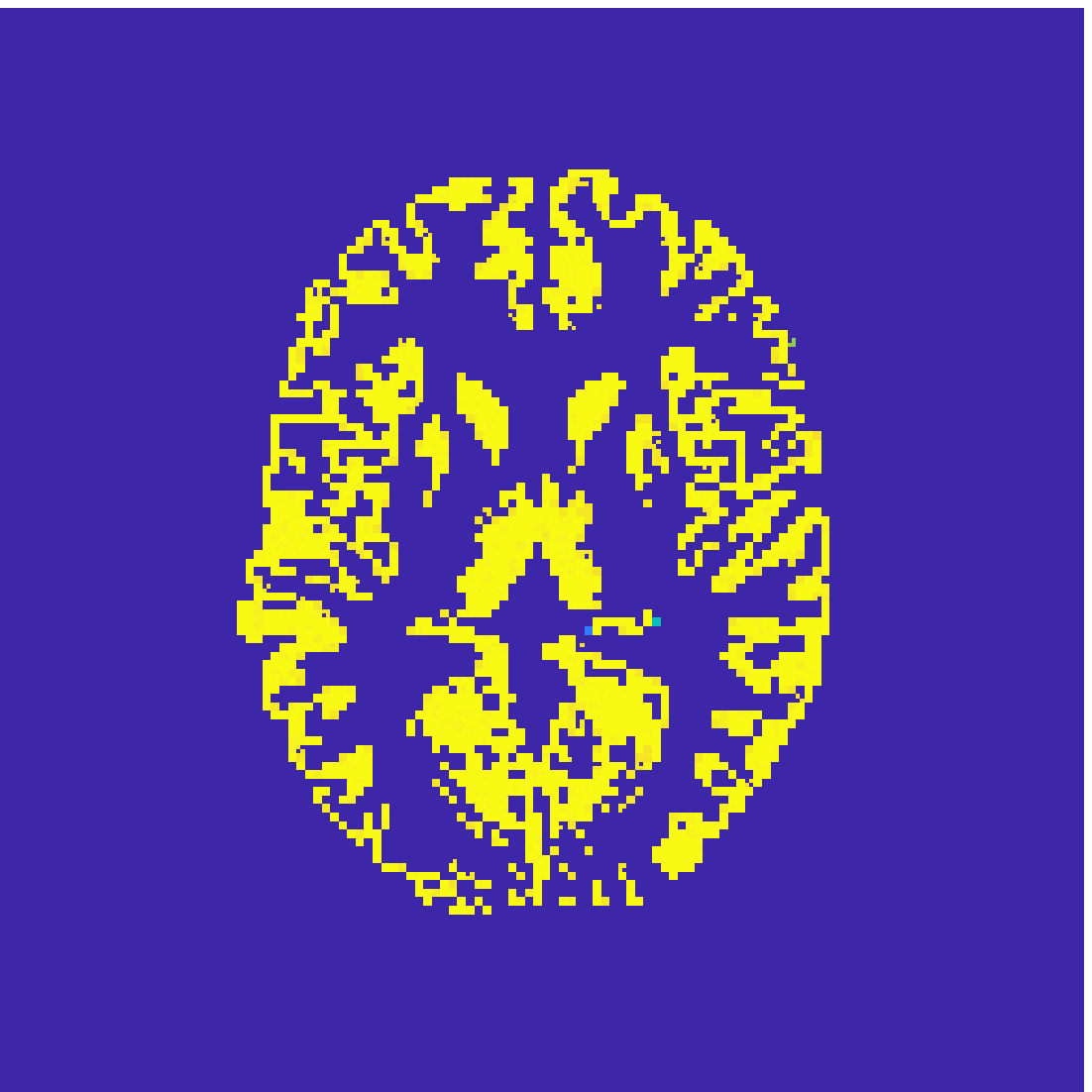}
        & \includegraphics[width=0.3\linewidth]{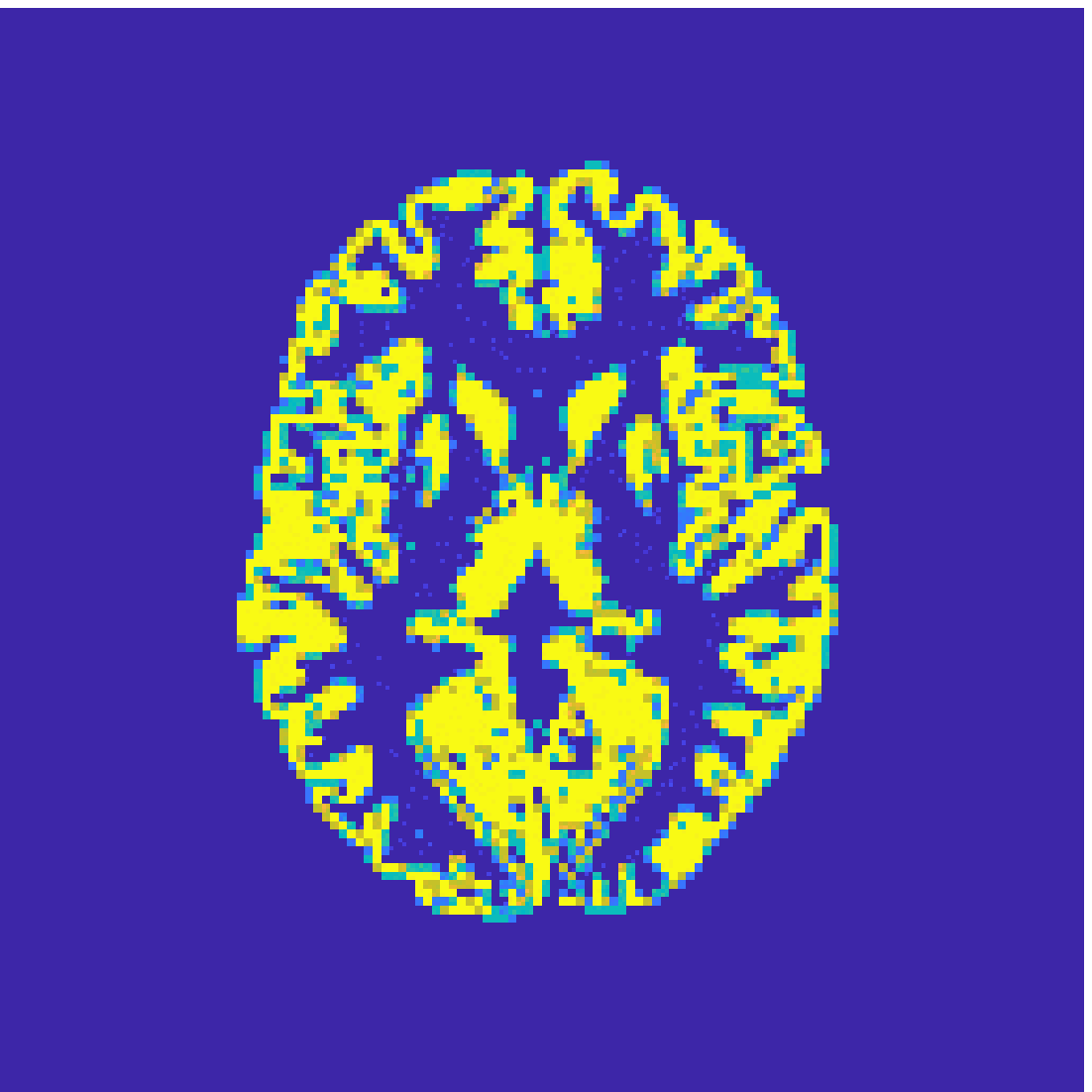} & \includegraphics[width=0.045\linewidth]{fig/colorbar6.eps}
        \\
        \includegraphics[width=0.3\linewidth]{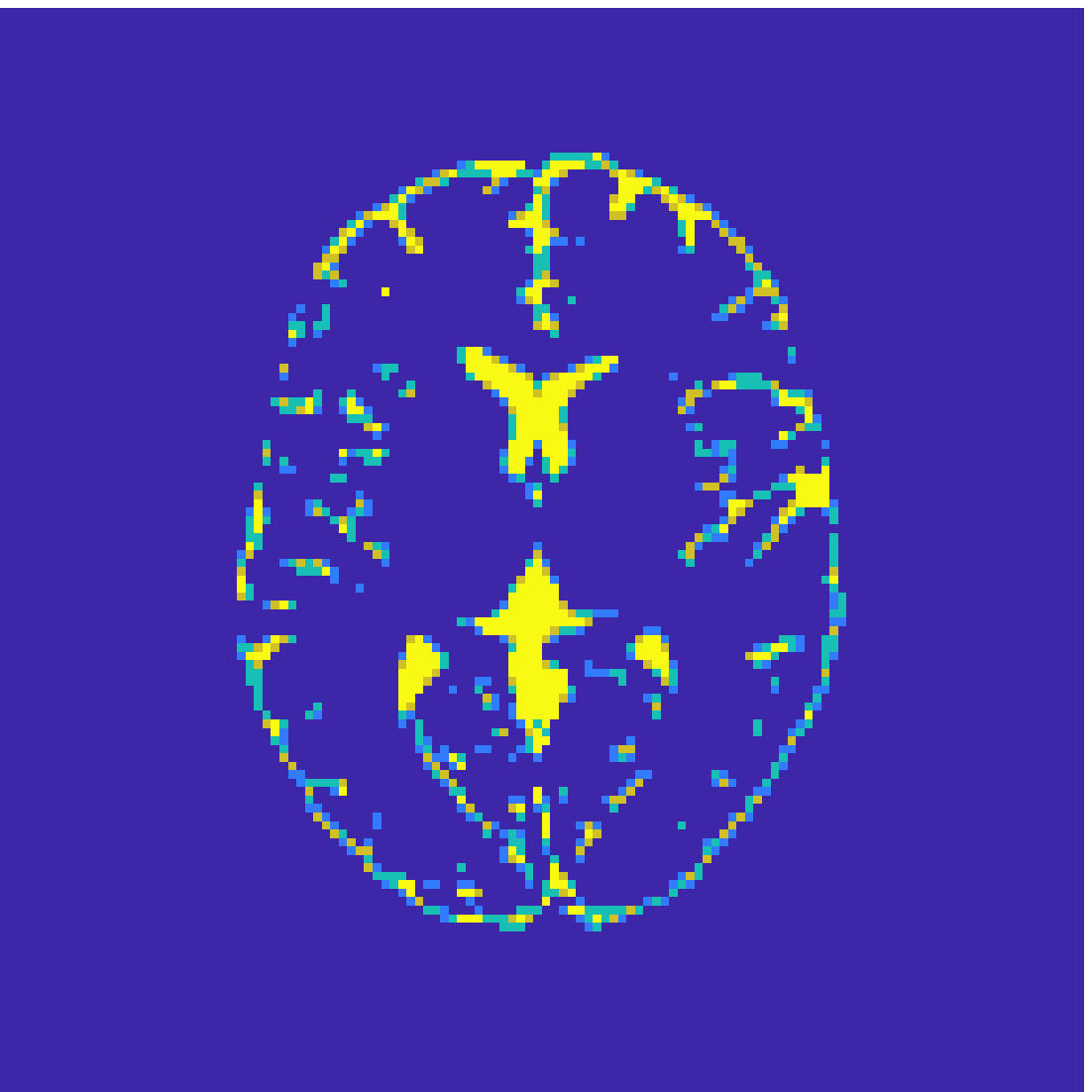}
        & \includegraphics[width=0.3\linewidth]{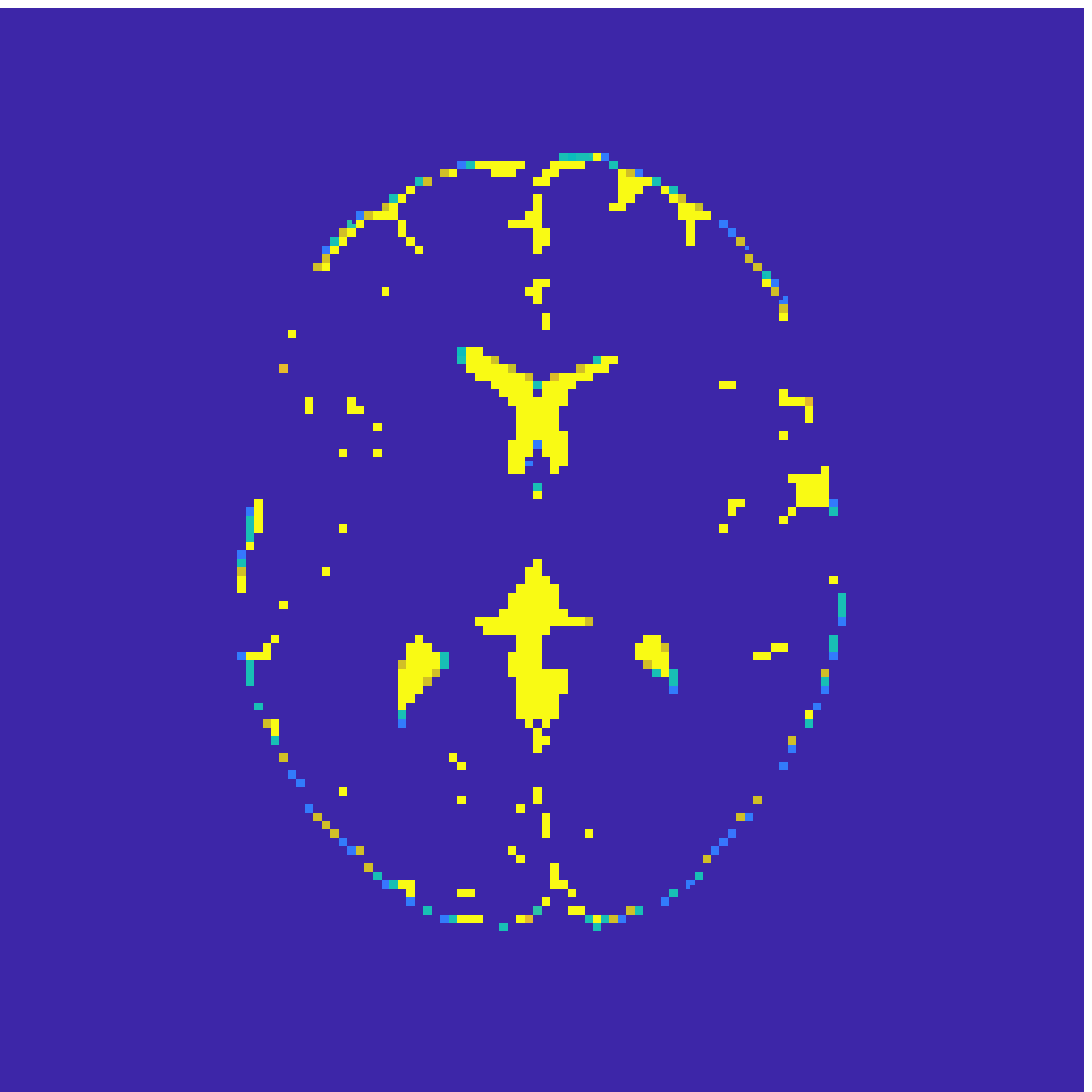}
        & \includegraphics[width=0.3\linewidth]{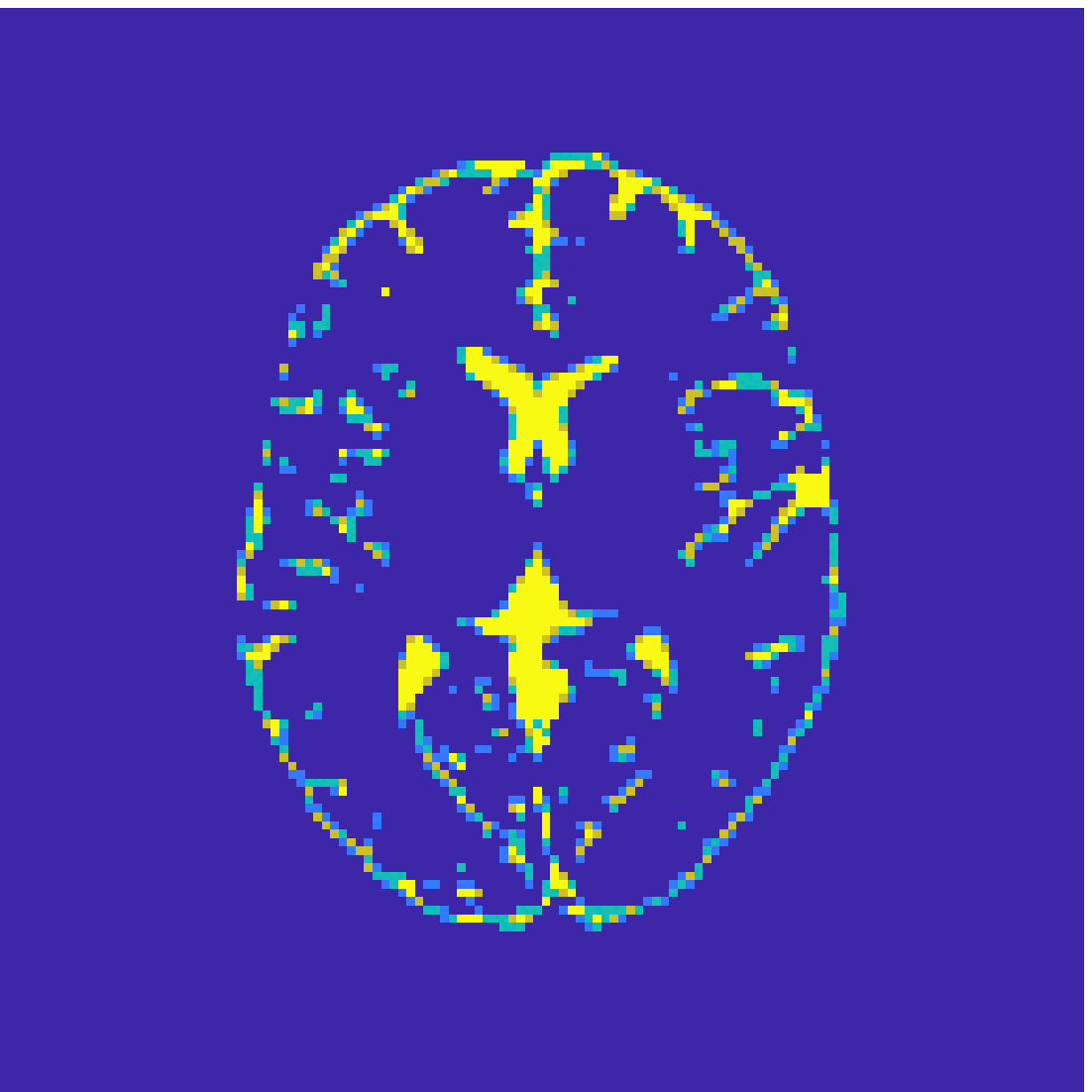} & \includegraphics[width=0.045\linewidth]{fig/colorbar6.eps}
        \\
        \includegraphics[width=0.3\linewidth]{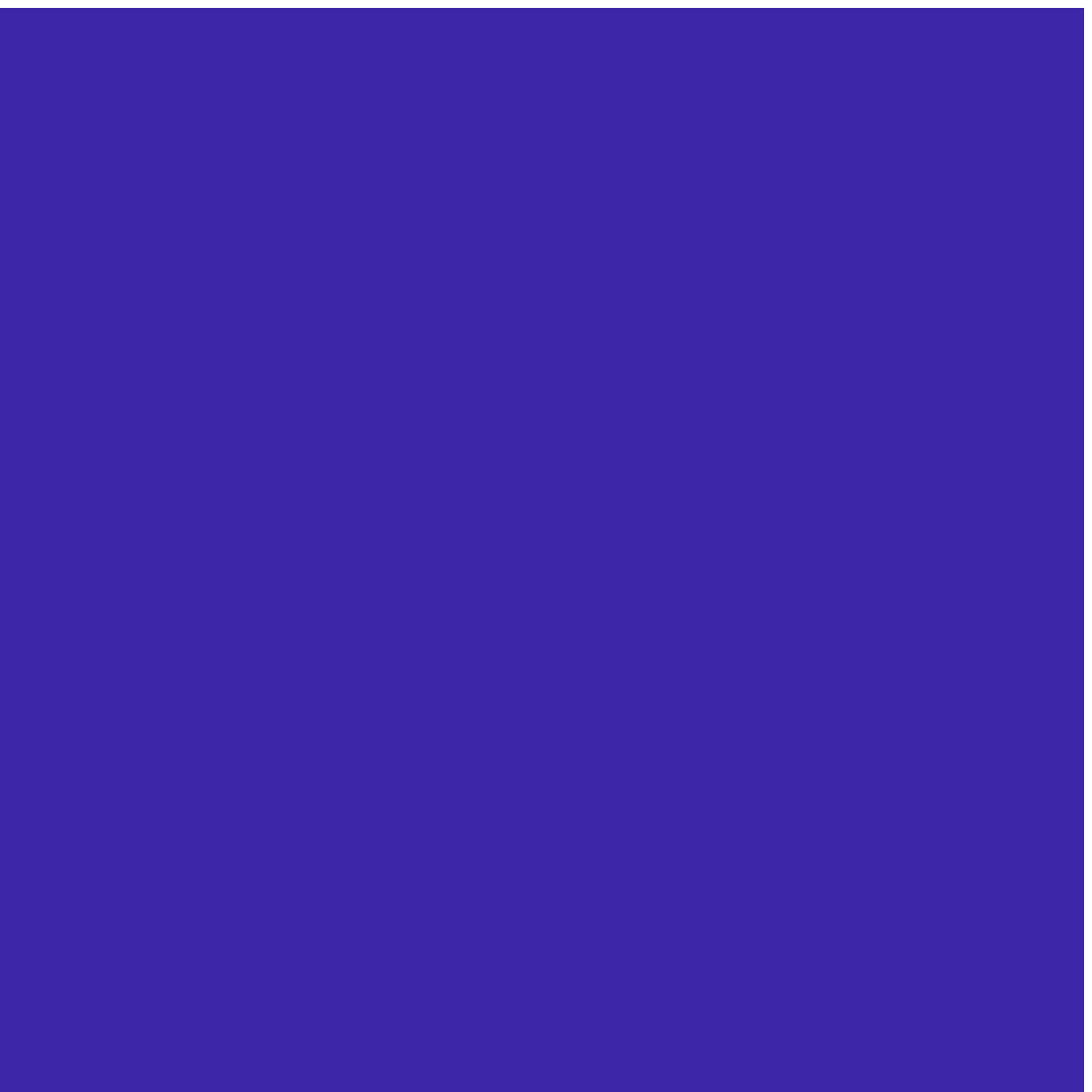}
        & \includegraphics[width=0.3\linewidth]{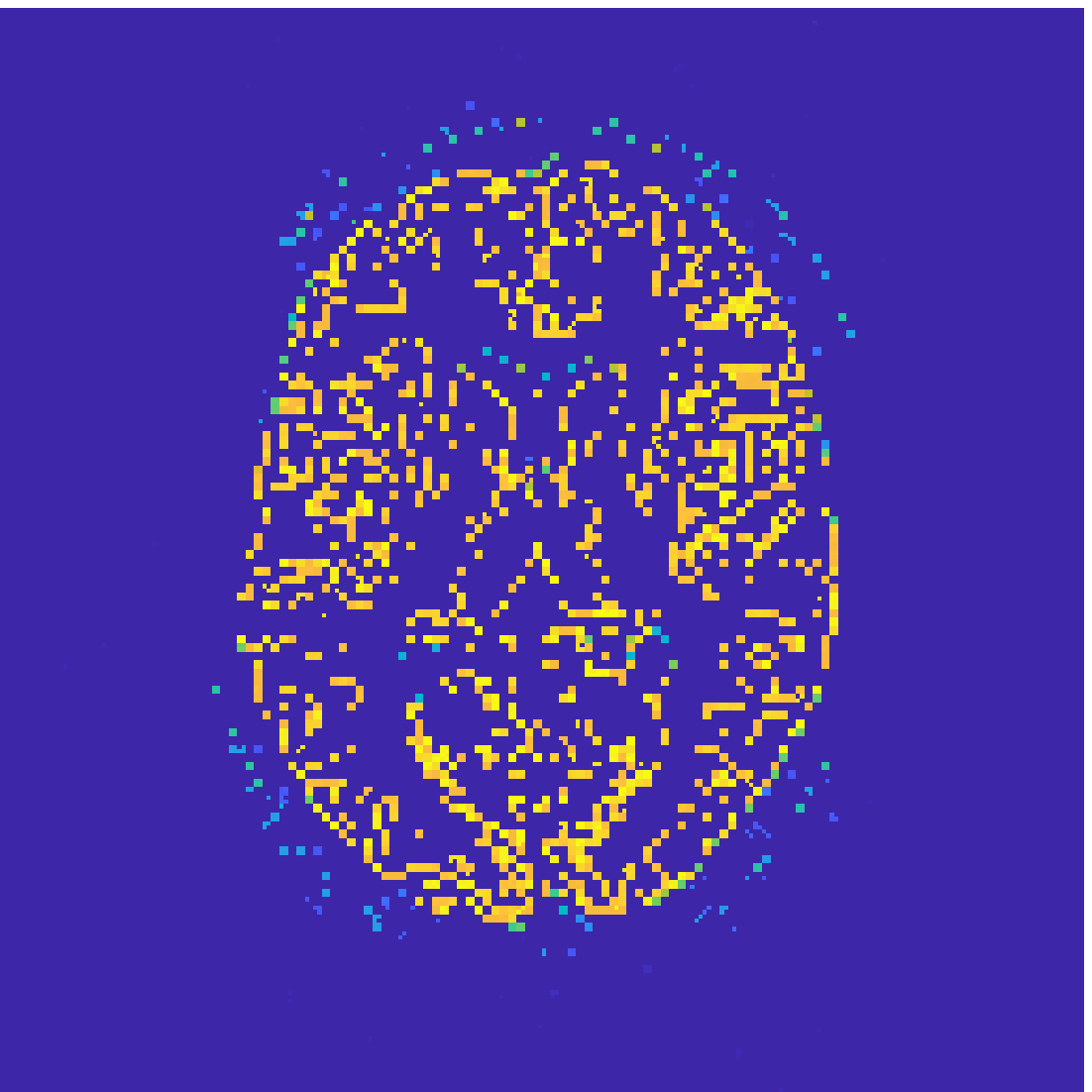}
        & \includegraphics[width=0.3\linewidth]{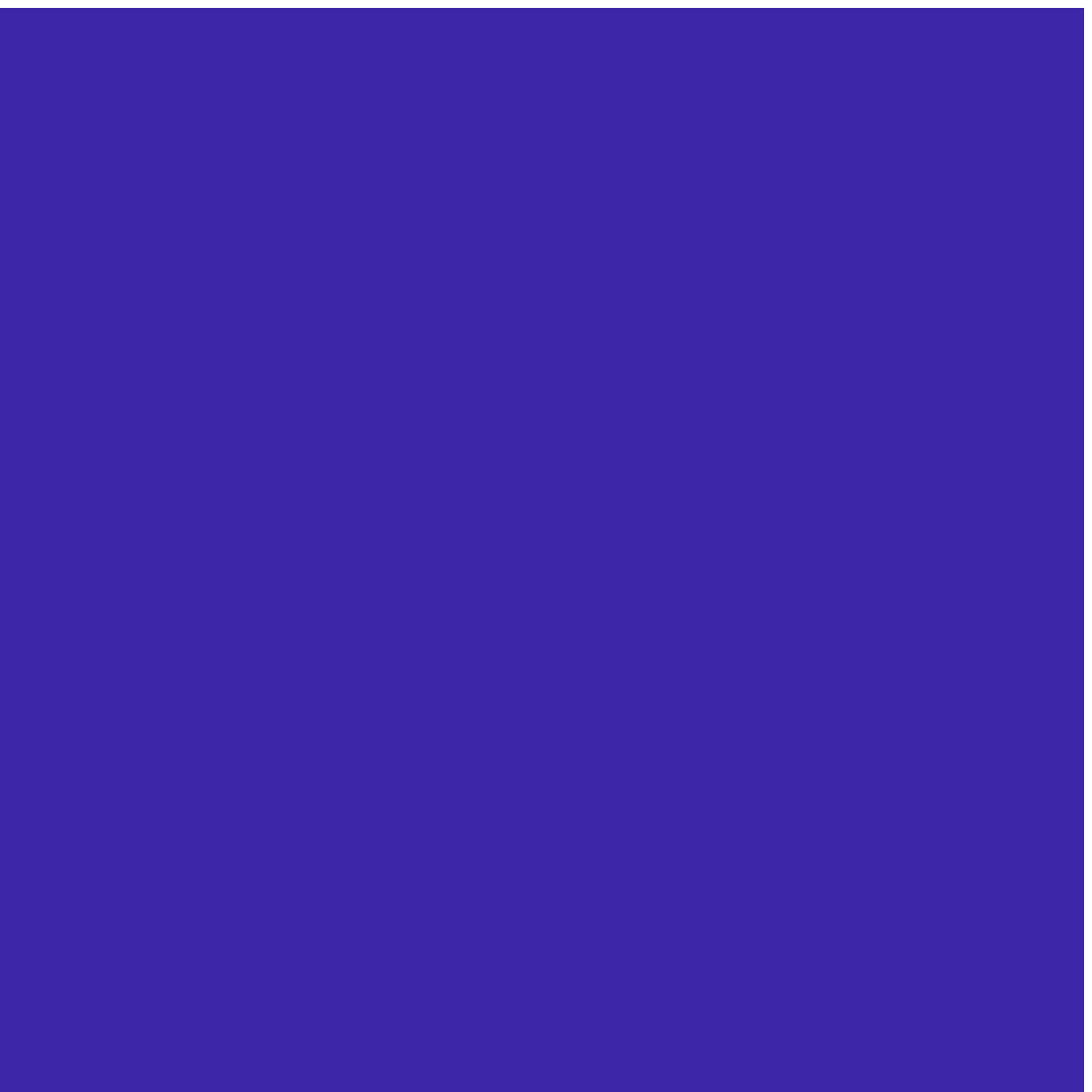} & \includegraphics[width=0.045\linewidth]{fig/colorbar6.eps}
        \\
        \includegraphics[width=0.3\linewidth]{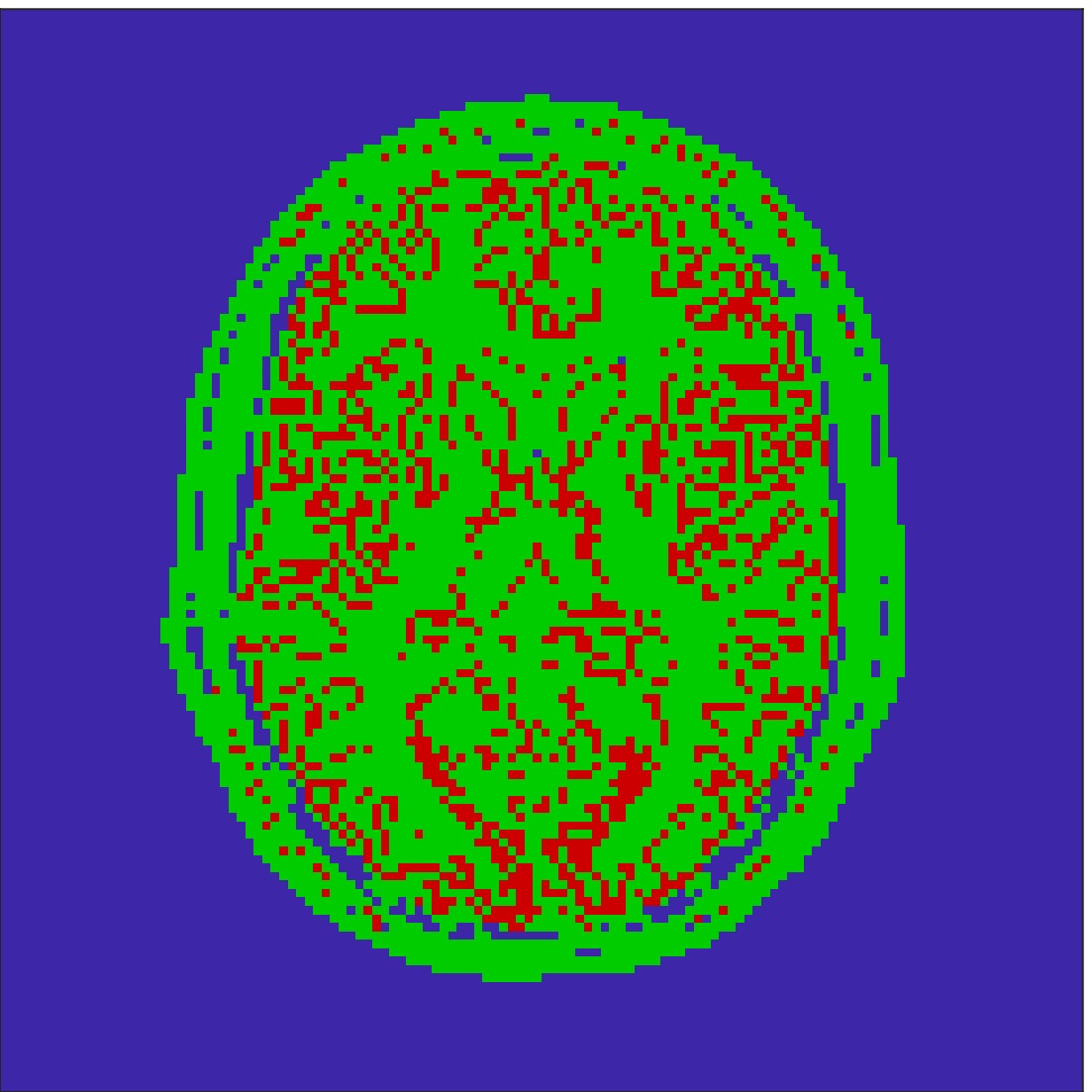}&&\\
    \end{tabular}
    \endgroup
    \caption{\label{im_e2pdmaps}
    Experiment 1 - 
    Example of the proton density maps with $L=1000$ and an iSNR of $30$dB. 
    From first to last column: Ground truth images, BLIP reconstructions, and GAP-MRF reconstructions. 
    From first to fifth row: Adipose, WM, muscle, GM and CSF. Sixth row: Proton density sum of all other matched elements that are not in the 15\% range of the ground truth elements. 
    The corresponding $T_1$ and $T_2$ values are given in Table~\ref{table:SNRexperiment}. 
    Last row: Voxel distribution map showing the pure voxels (green) and the PV voxels (red).
    }
\end{figure}

\begin{figure}[!t]
    \centering
    \begingroup
    \setlength{\tabcolsep}{0.2mm}
\renewcommand{\arraystretch}{.2}
    \begin{tabular}{cccc}
	\includegraphics[width=0.3\linewidth]{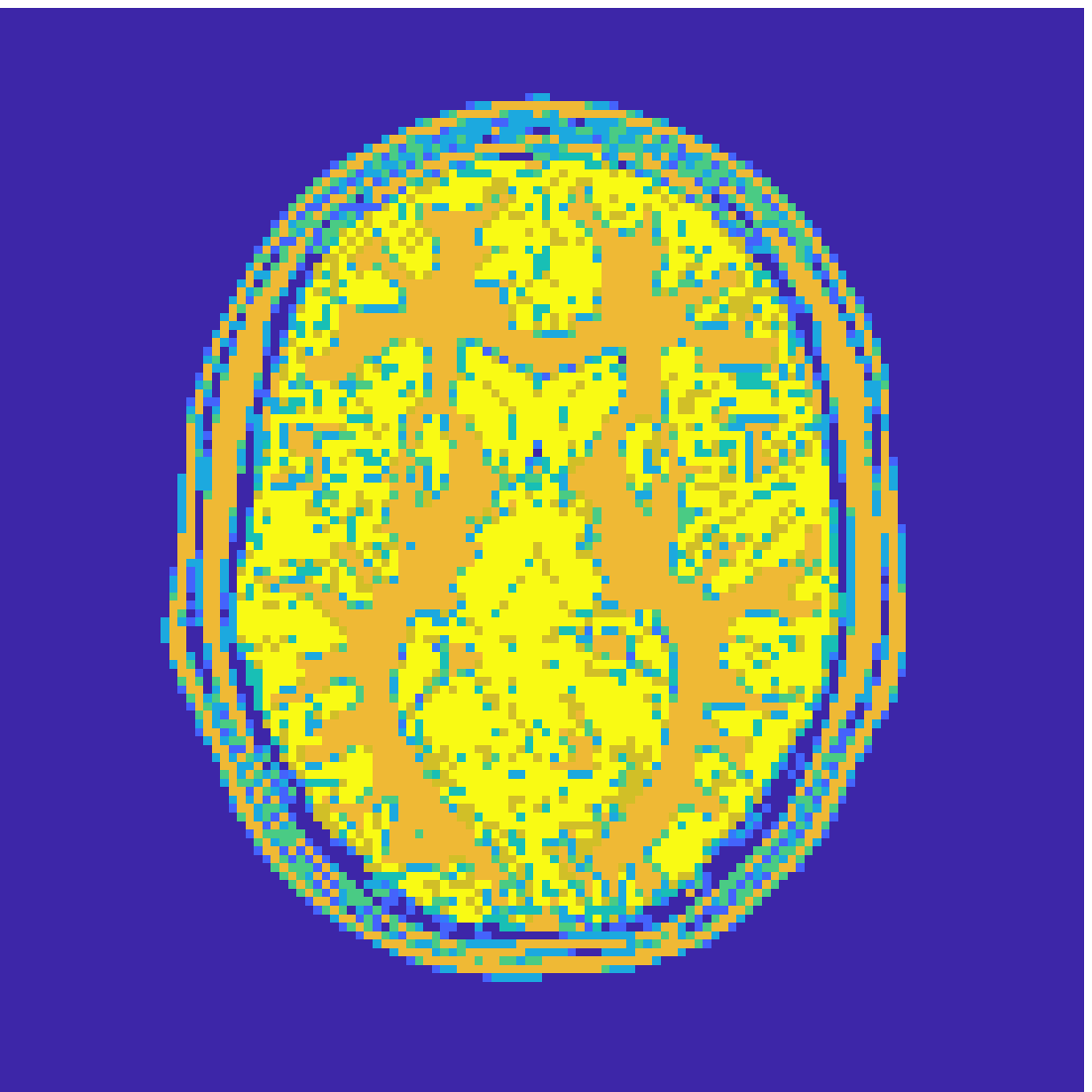}
&	\includegraphics[width=0.3\linewidth]{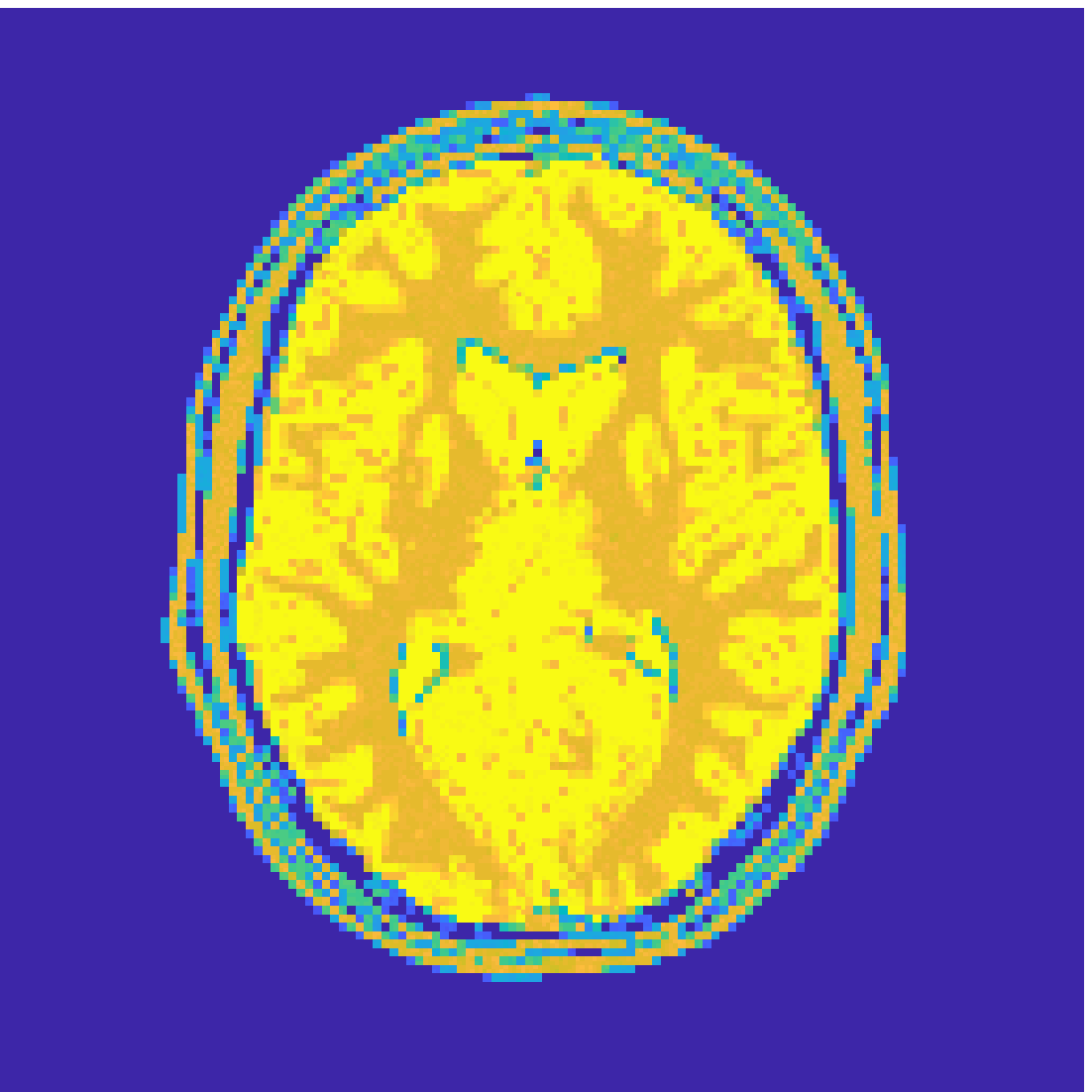}
&	\includegraphics[width=0.3\linewidth]{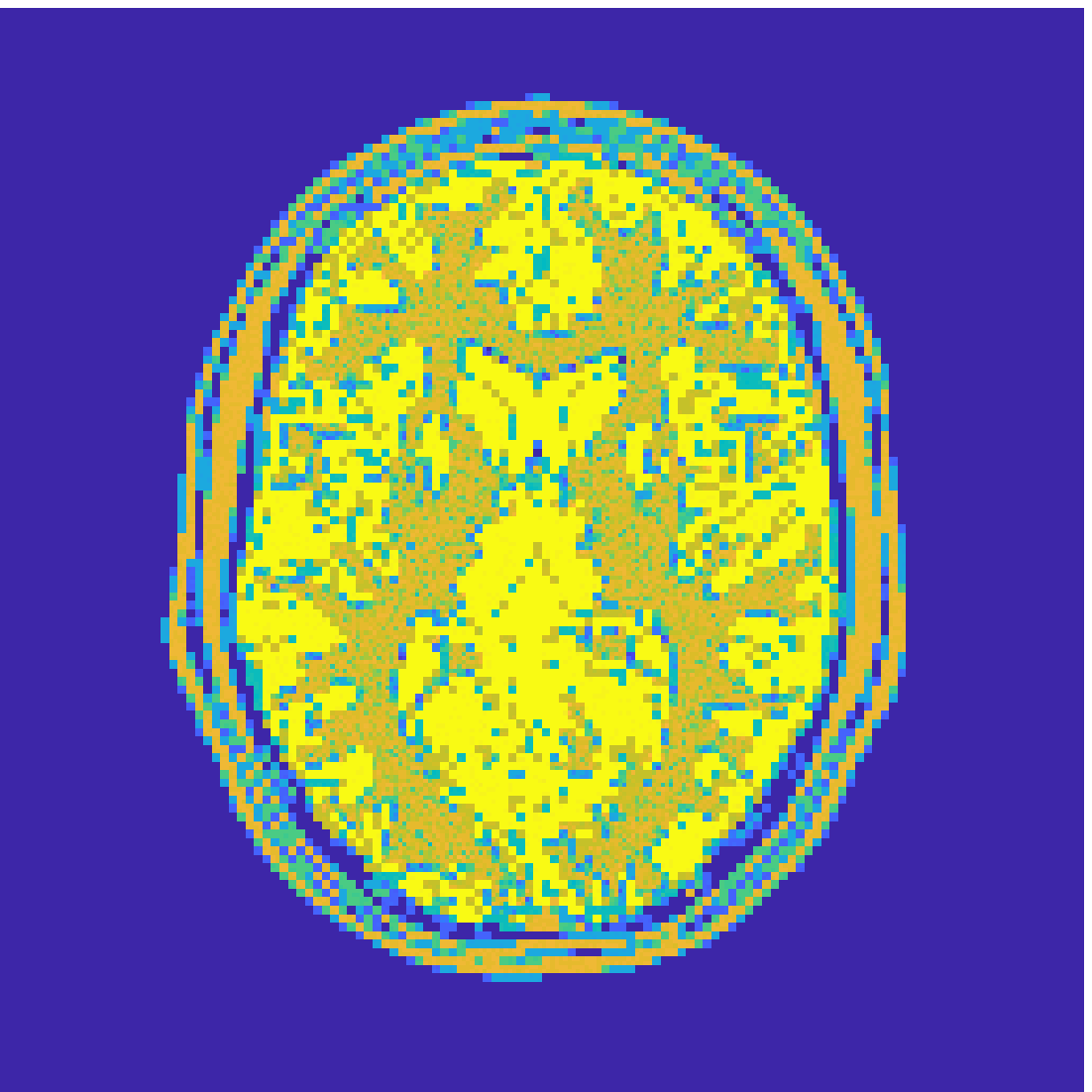}
&   \includegraphics[width=0.045\linewidth]{fig/colorbar6.eps}\\
    \includegraphics[width=0.3\linewidth]{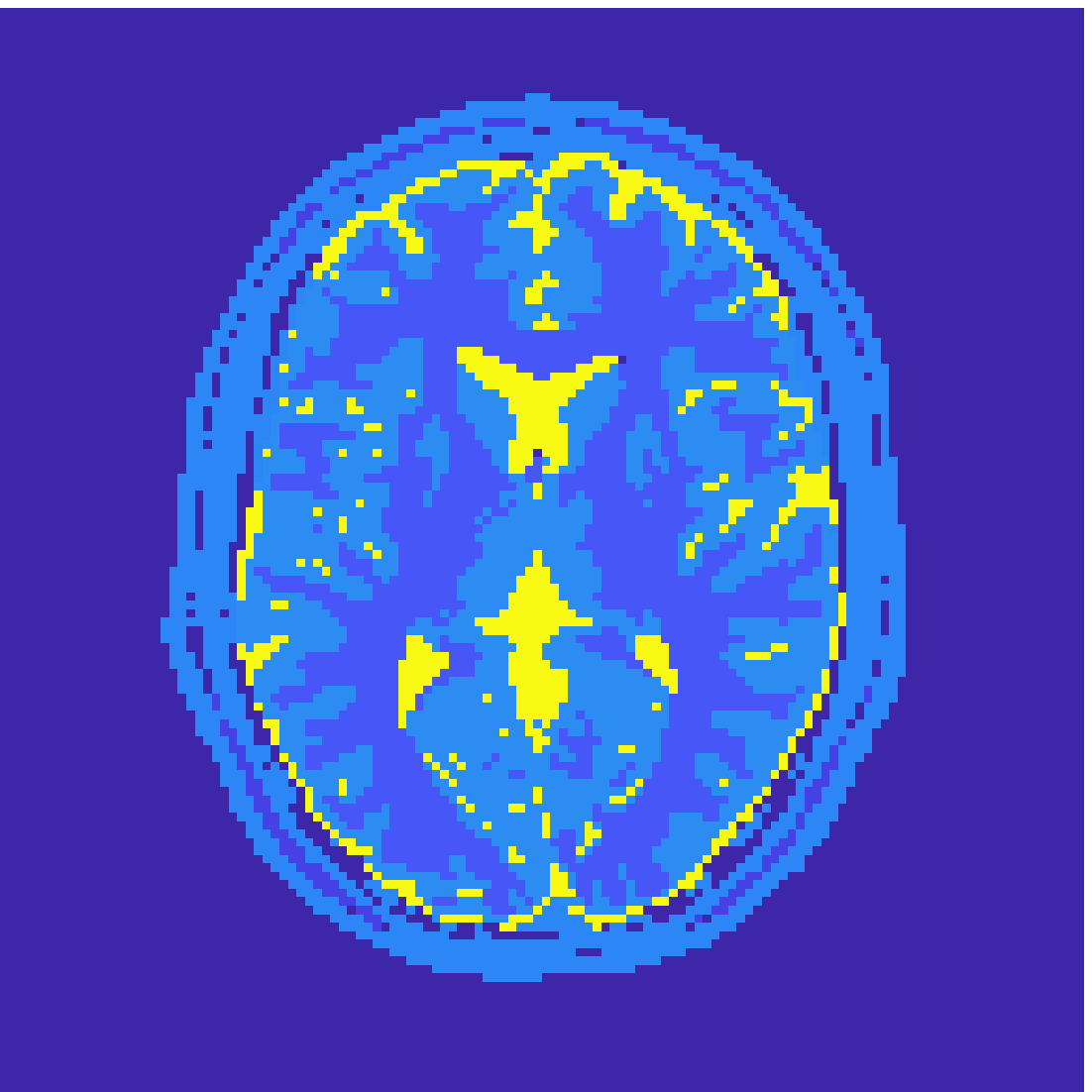}
&	\includegraphics[width=0.3\linewidth]{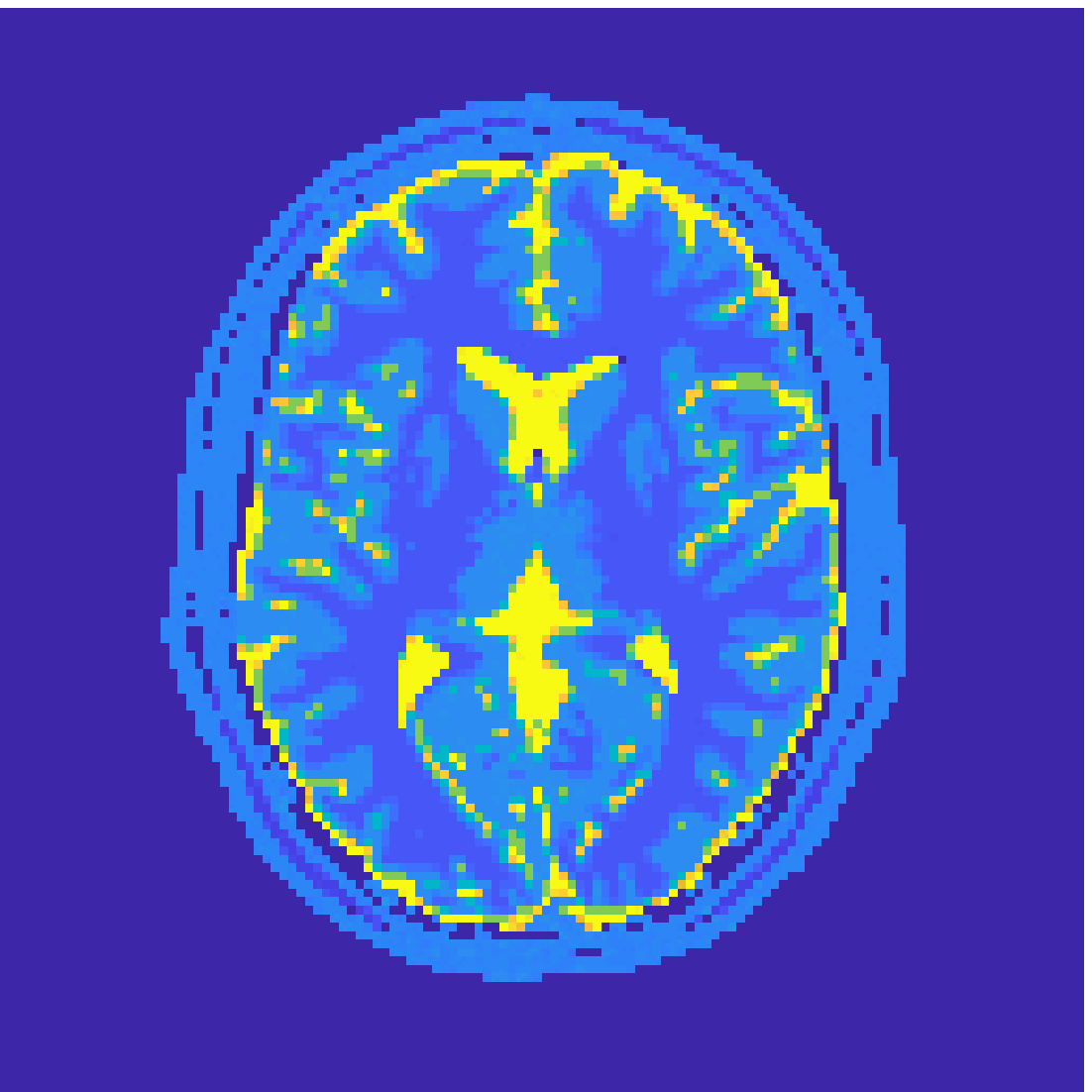}
&	\includegraphics[width=0.3\linewidth]{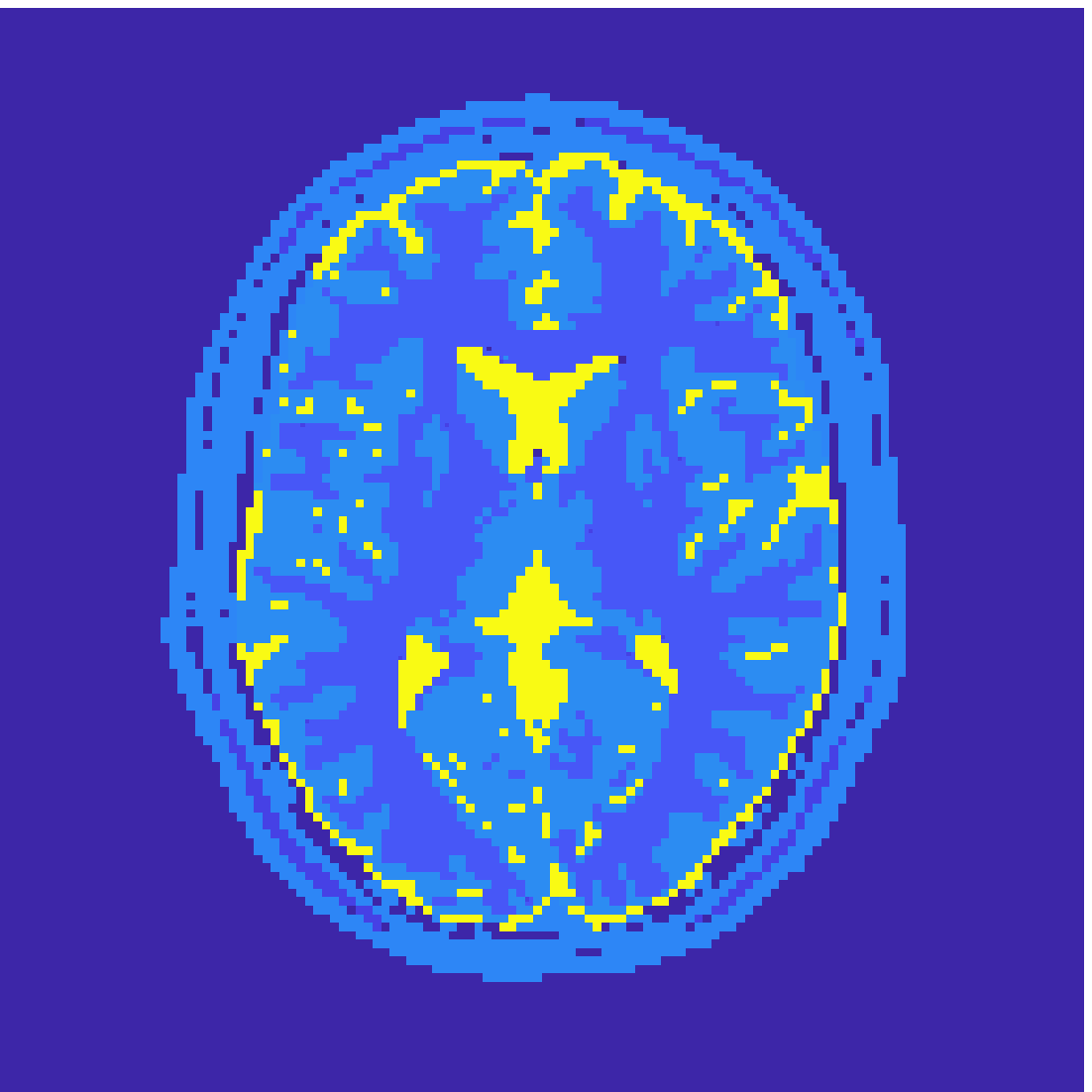}
&   \includegraphics[width=0.48cm]{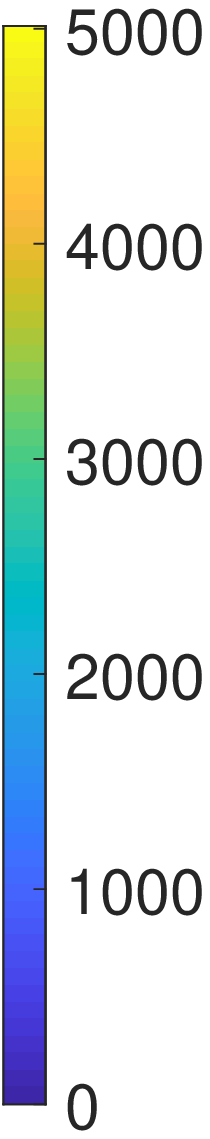}\\
    \includegraphics[width=0.3\linewidth]{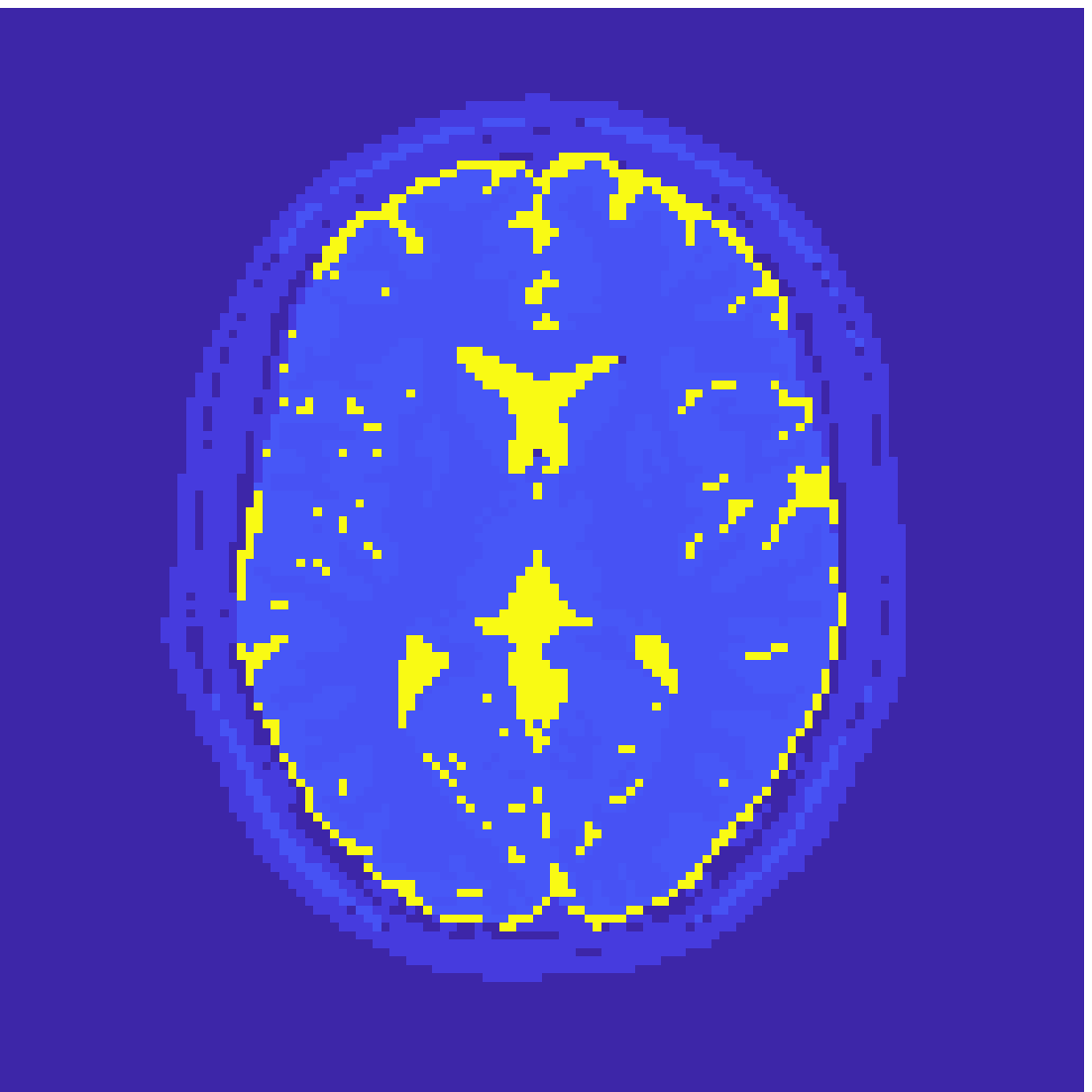}
&	\includegraphics[width=0.3\linewidth]{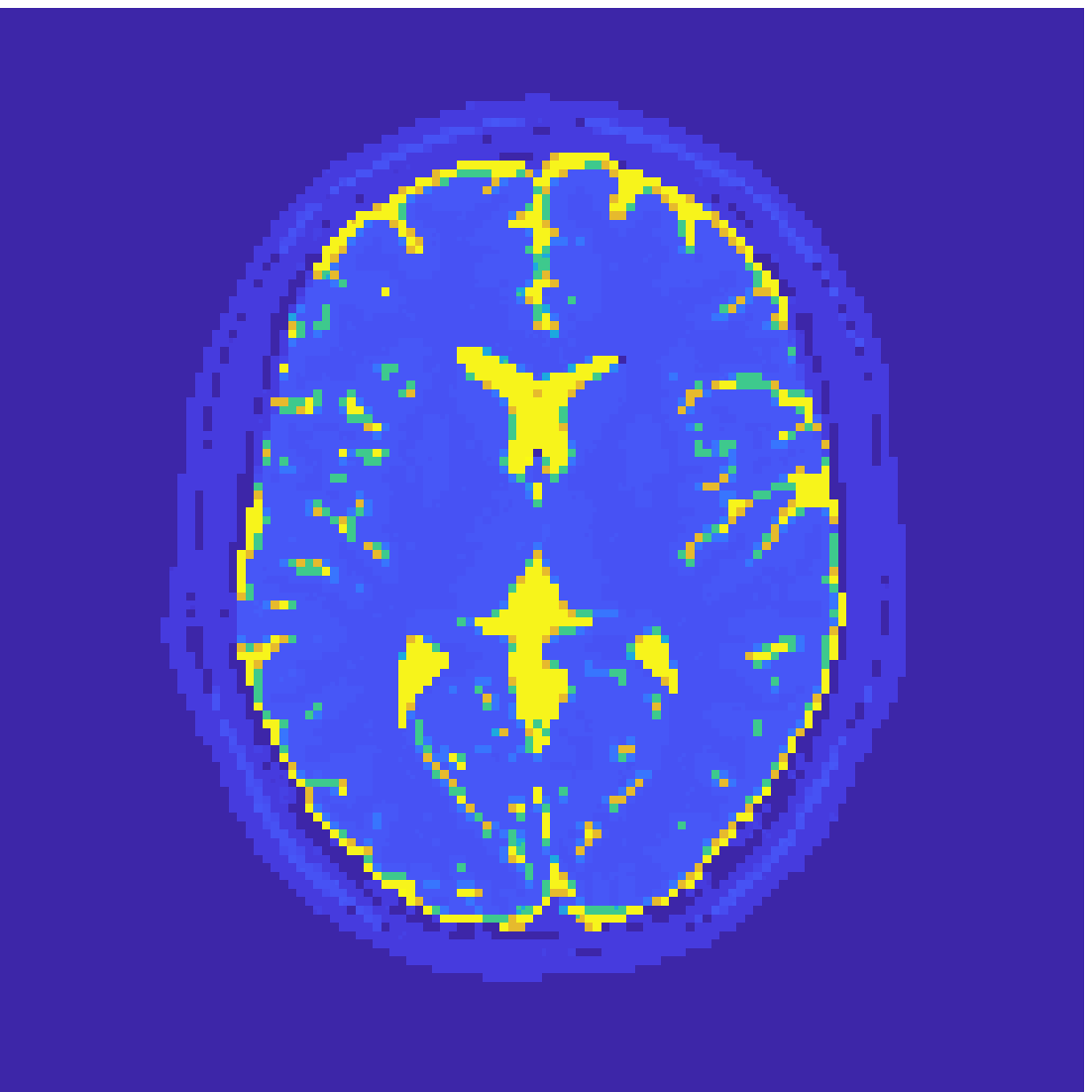}
&	\includegraphics[width=0.3\linewidth]{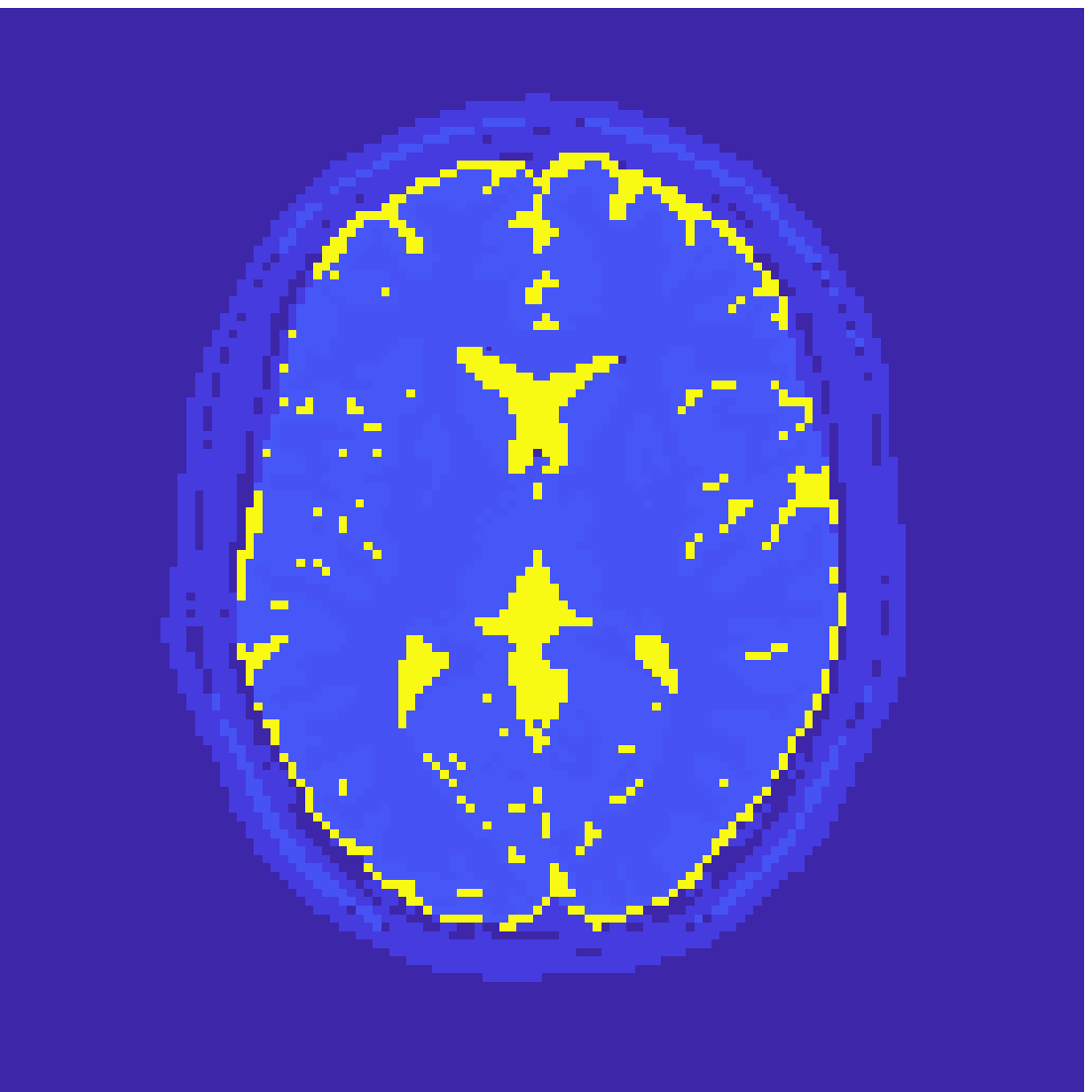}
&   \includegraphics[width=0.045\linewidth]{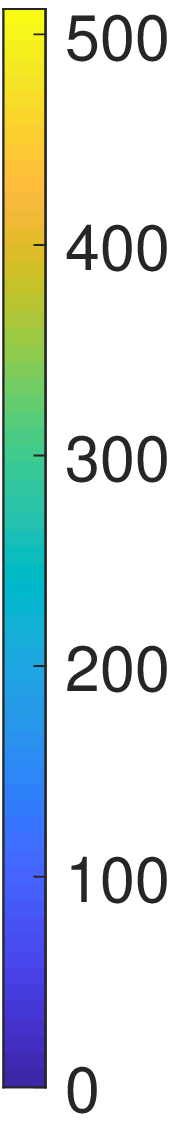}\\
\end{tabular}
\endgroup
    \caption{\label{fig:e2dominanttissues}
    Experiment 1 - 
    Example of the dominant tissue parameters with $L=1000$ and an iSNR of $30$dB. 
    From first to last column: Ground truth, BLIP reconstructions, and GAP-MRF reconstructions. 
    From first to last row: Proton density, $T_1$ and $T_2$ parameter maps.
    }
\end{figure}

\begin{figure}[!t]
	\centering
	\begingroup
\setlength{\tabcolsep}{0pt} 
	\begin{tabular}{ccc}
	\includegraphics[align=c,width=0.3\linewidth]{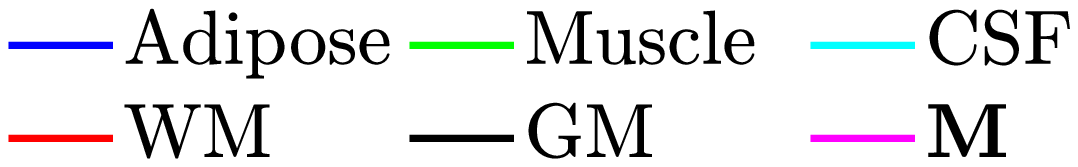}& \includegraphics[align=c,width=0.28\linewidth]{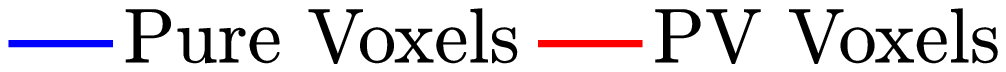}& \includegraphics[align=c,width=0.2\linewidth]{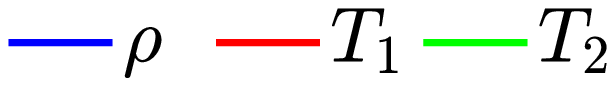}\\
	     \includegraphics[align=c,width=0.32\linewidth]{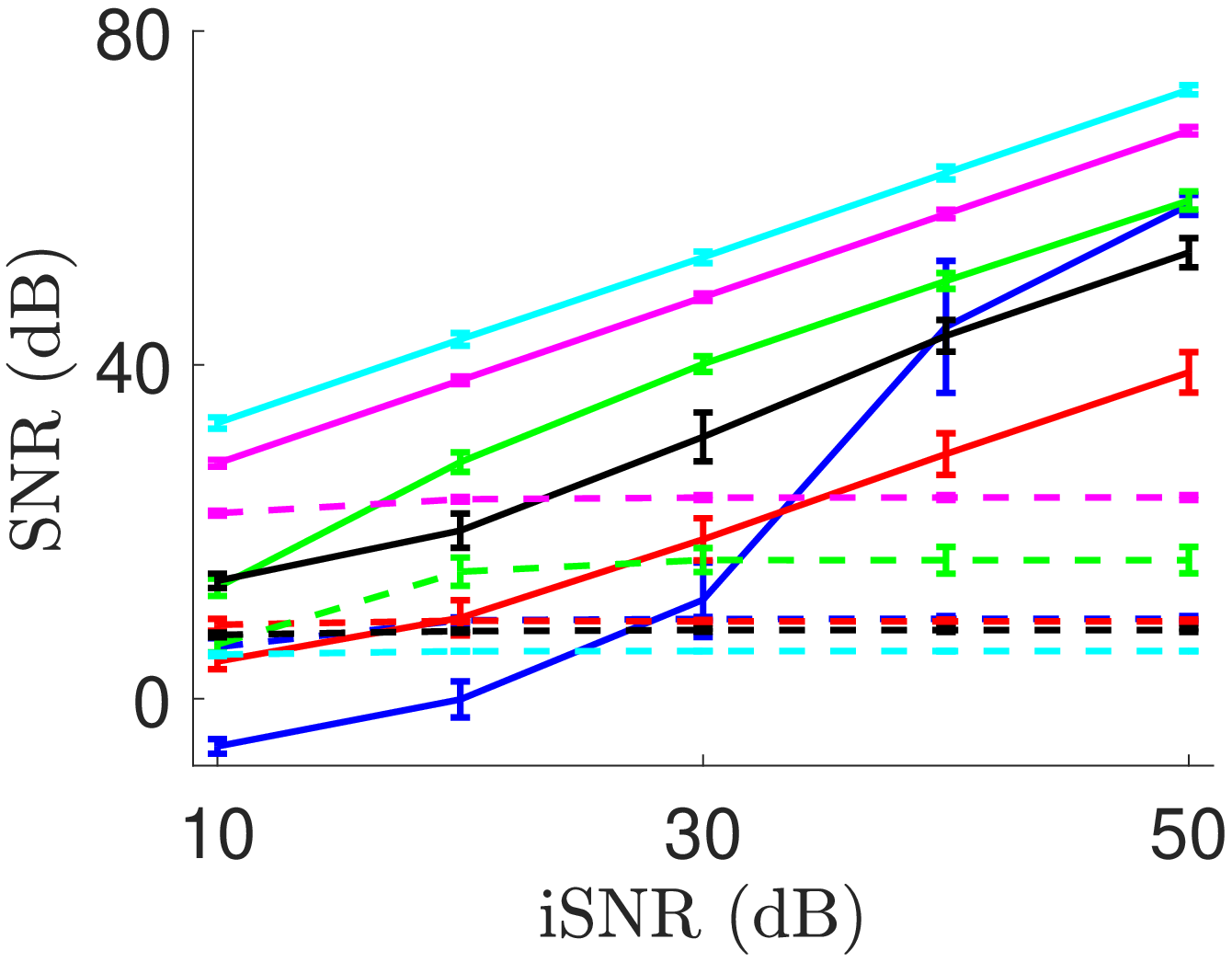}&\includegraphics[align=c,width=0.32\linewidth]{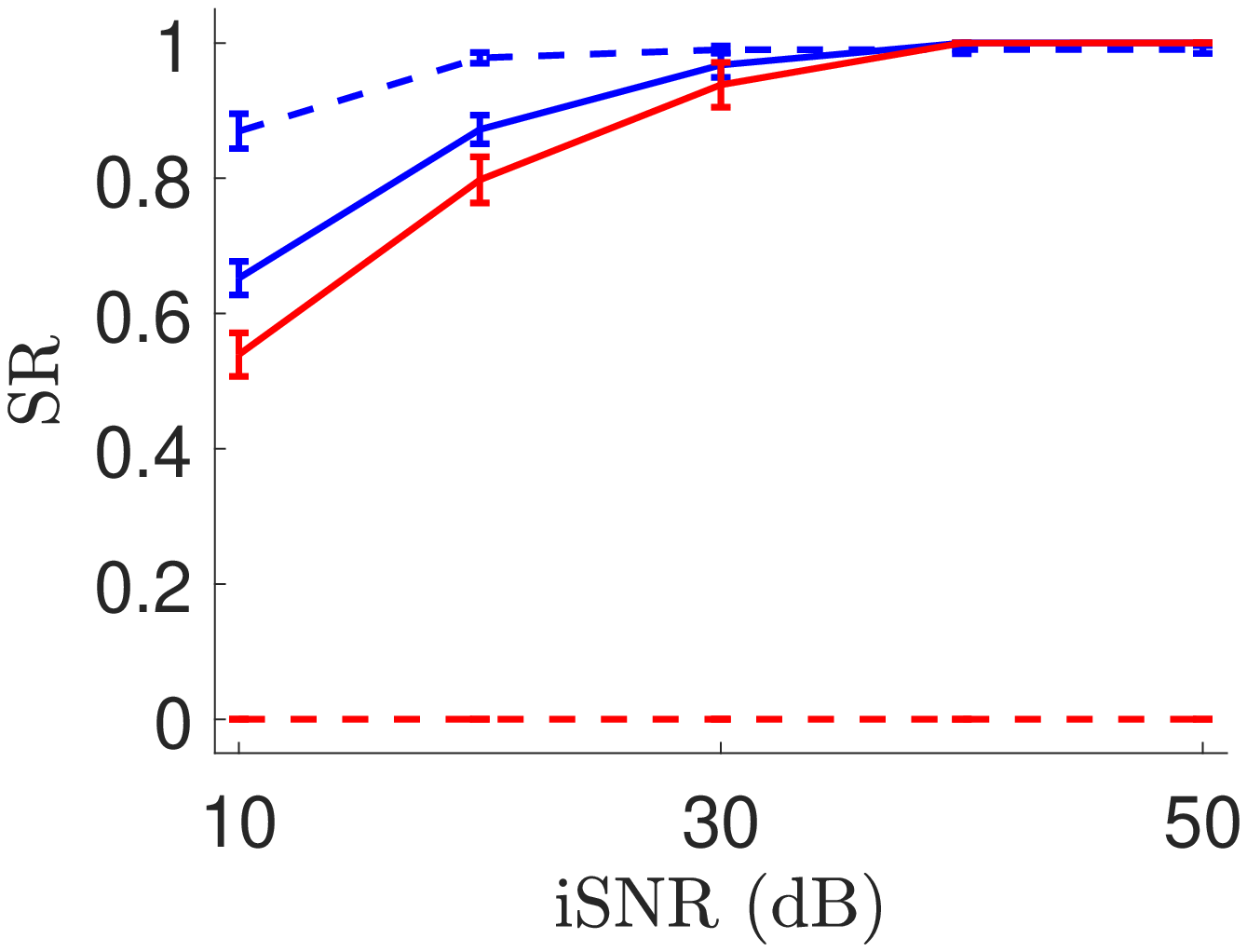}&\includegraphics[align=c,width=0.32\linewidth]{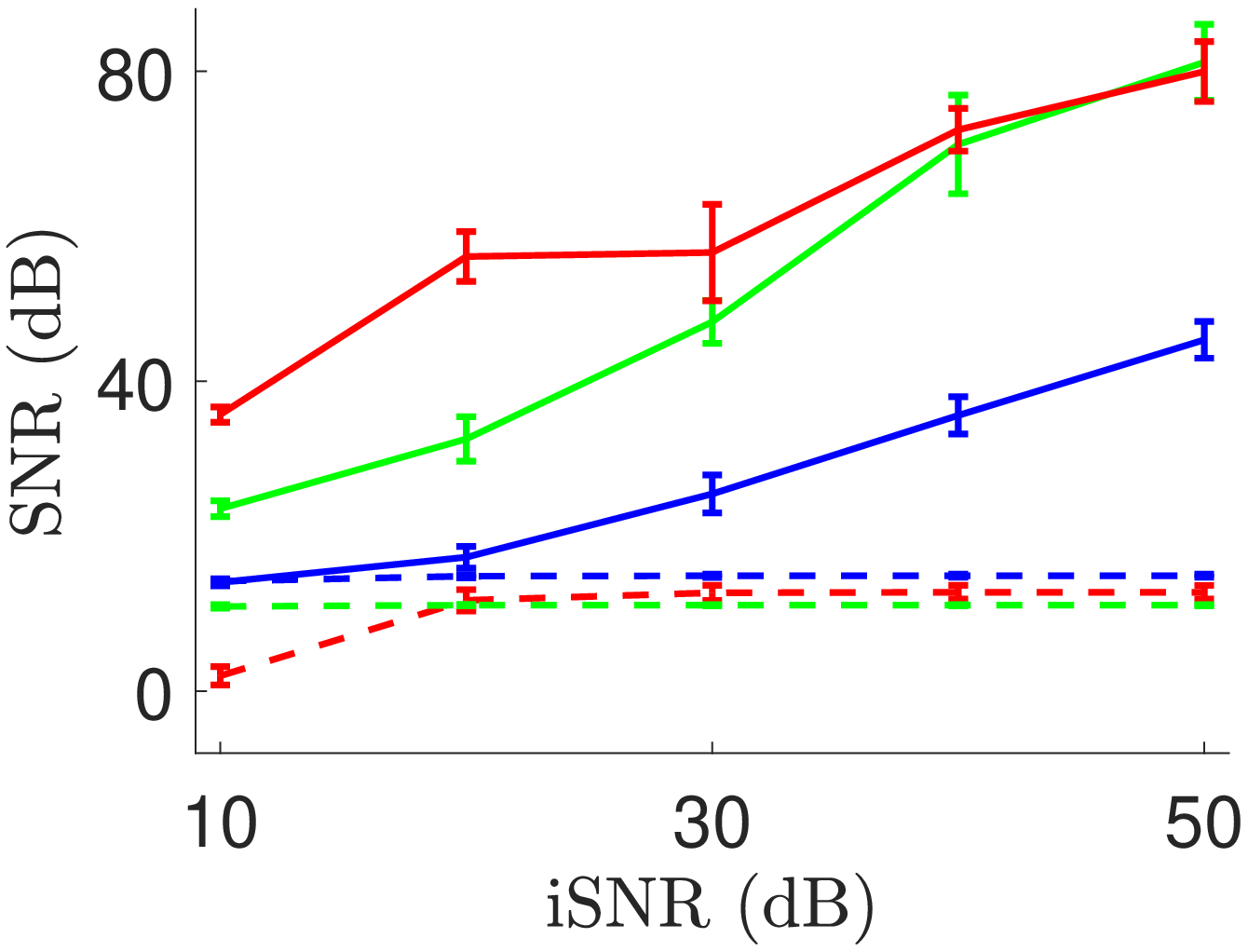}
	\end{tabular}
	 \endgroup
    \caption{\label{im_e2pdsnr}
	Experiment 1 - 
    Simulation results obtained with BLIP (dashed lines) and GAP-MRF (solid lines). 
    Left: Tissue proton density maps (Adipose, WM, Muscle, GM, CSF) and magnetisation sequence ($\mathbf{M}$) evaluation.
    Center: SR evaluation for the pure and PV voxels.
    Right: Dominant tissue parameter maps ($\rho$, $T_1$, $T_2$) evaluation.
    }
\end{figure}
The results of the proton density maps are shown in Fig.~\ref{im_e2pdsnr} (Left). GAP-MRF significantly outperform BLIP when the iSNR is greater than $30$dB. We can notice that GAP-MRF estimates correctly the number of true atoms when the iSNR is $30$dB or greater. The reconstruction of adipose tissue is more affected by the noise since there are significantly less pure voxels of this tissue. GAP-MRF magnetisation sequence reconstruction is significantly more accurate than BLIP reconstruction, because BLIP does not consider the PVE and also because of the dictionary inaccuracy. GAP-MRF magnetisation sequence SNR has a linear behaviour with respect to the iSNR. In Fig.~\ref{im_e2pdsnr} (Center), the SR with respect to the iSNR can be seen. We can observe that the SR is significantly affected by the iSNR.

The results for the dominant tissue parameter maps SNR can be seen in Fig.~\ref{im_e2pdsnr} (Right). GAP-MRF outperforms BLIP reconstructing the dominant tissue parameter maps. It is important to mention than GAP-MRF is more affected by noise because the linear combination of dictionary elements overfits the noise.

\begin{table}[!t]\footnotesize
\centering
\caption{Parameter values of example in Fig.~\ref{im_e2pdmaps} corresponding to Experiment~1 with an iSNR of $30$dB. The relaxation times are in ms.}
\label{table:SNRexperiment}
\begin{tabular}{l|r|r|r|r|r|r|}
\cline{2-7} & \multicolumn{2}{c|}{Ground Truth} & \multicolumn{2}{c|}{BLIP}      & \multicolumn{2}{c|}{GAP-MRF} \\ \cline{2-7}
             & \multicolumn{1}{|c|}{$T_1$}         &  \multicolumn{1}{|c|}{$T_2$}         & \multicolumn{1}{|c|}{$T_1$}          & \multicolumn{1}{|c|}{$T_2$}      & \multicolumn{1}{|c|}{$T_1$}       & \multicolumn{1}{|c|}{$T_2$}       \\\hline
\multicolumn{1}{|c|}{Adipose}& 530             & 77             & {[}460-590{]}   & {[}74-84{]} & 531.1        & 77.0    \\\hline
\multicolumn{1}{|c|}{\begin{tabular}{@{}c@{}}White\\Matter\end{tabular}}&811&77&{[}690-930{]}& {[}66-80{]} & 811.1& 77.0 \\\hline
\multicolumn{1}{|c|}{Muscle} & 1425  & 41    & {[}1220-1630{]} & {[}36-46{]} & 1424.0       & 41.0         \\\hline
\multicolumn{1}{|c|}{\begin{tabular}{@{}c@{}}Gray\\Matter\end{tabular}}& 1545& 83& {[}1320-1610{]}&{[}74-86{]}&1544.3& 83.1\\\hline
\multicolumn{1}{|c|}{CSF}& 5012   & 512            & {[}4400-5000{]} & 500         & 5013.1       & 512.1       \\\hline
\end{tabular}
\end{table}


We show an example of the proton density maps for each tissue in Fig.~\ref{im_e2pdmaps} when the iSNR is $30$dB. By visual inspection, we can observe that the GAP-MRF method outperforms the BLIP method for PV reconstructions for moderate noise scenarios. The values of BLIP in Table~\ref{table:SNRexperiment} are given in a range because multiple parameters were assigned to the corresponding ground truth tissue. On the contrary, GAP-MRF has a single value because only one value was assigned to the corresponding ground truth tissue. In this example, for BLIP and GAP-MRF respectively, the SNR values are as follows: 9.70dB and 11.94dB for Adipose, 9.14dB and 19.52dB for WM, 17.66dB and 39.29dB for Muscle, 8.31dB and 31.87dB for GM, 5.72dB and 52.60dB for CSF and for the magnetisation sequence 23.84dB and 48.18dB. The SR: 0.9944 and 0.9745 for pure voxels, and for PV voxels 0 and 0.9465. The GAP-MRF correctly estimates the manifold regions of interest. BLIP has a residual map formed by all the elements that are not sufficiently close to the true elements. Note that the residual map is quite similar to the distribution of the PV voxels shown in the last row in Fig.~\ref{im_e2pdmaps}, this shows that the parameter mismatch is due to the PVE. In the GAP-MRF reconstructions, the WM and adipose tissue are slightly mixed due to the noise since their parameters are close one to each other. By choosing a better $\boldsymbol{\Gamma}$ we can make the atoms of the dictionary more distant in the $\ell_2$-norm sense, this would provide noise robustness to the reconstructions. The dominant tissue parameter maps are shown in Fig.~\ref{fig:e2dominanttissues}. The $T_1$ and $T_2$ maps reconstructed by BLIP show a smooth transition from one tissue to another due to the partial volume. On the contrary, GAP-MRF reconstructions show abrupt transitions in the $T_1$ and $T_2$ maps delimiting the tissues. This is expected since each tissue is modelled with a unique set of parameters.
We can observe that the dominant tissue proton density reconstruction of GAP-MRF is significantly affected by the noise. Nevertheless, thanks to the constraint $\mathcal{S}_+$ handled by the proposed method, the $T_1$ and $T_2$ parameter maps are accurate. 

\subsection{Experiment 2 - Impact of $L$}
In this subsection, we compare the proposed GAP-MRF algorithm with the BLIP algorithm, for different number of acquisition instances $L$. The iSNR is set to $50$dB. The dictionary for BLIP is defined as in \cite{CSMRF} with $D=16170$. 
The results correspond to an average (with standard deviation) over 10 runs of each choice of L.

\begin{figure}[!t]
	\centering
	\begingroup
\setlength{\tabcolsep}{0pt} 
	\begin{tabular}{ccc}
	\includegraphics[align=c,width=0.3\linewidth]{fig/Llegend5.eps}& \includegraphics[align=c,width=0.28\linewidth]{fig/Llegend3.eps}& \includegraphics[align=c,width=0.2\linewidth]{fig/Llegend4.eps}\\
	     \includegraphics[align=c,width=0.32\linewidth]{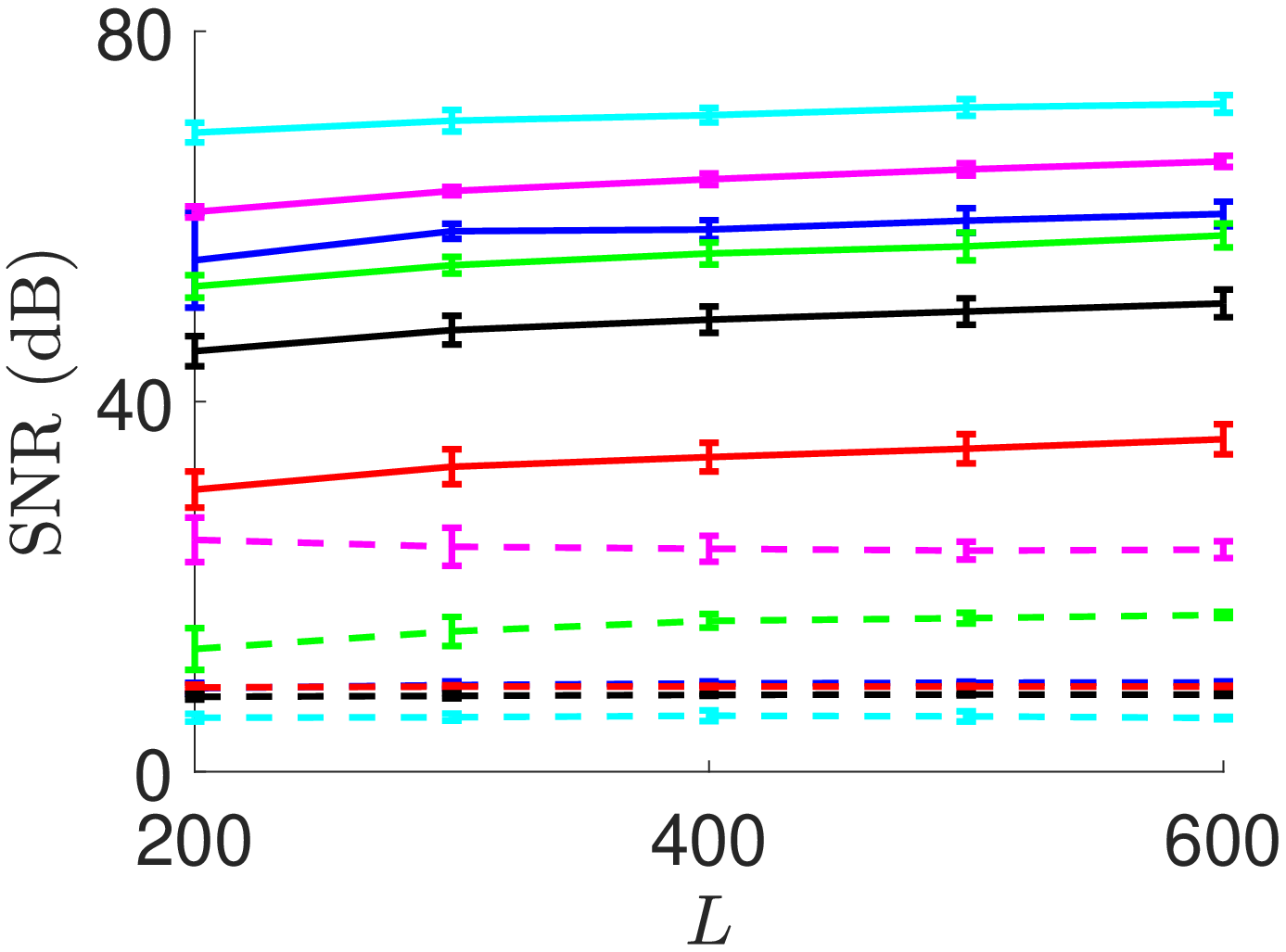}&\includegraphics[align=c,width=0.32\linewidth]{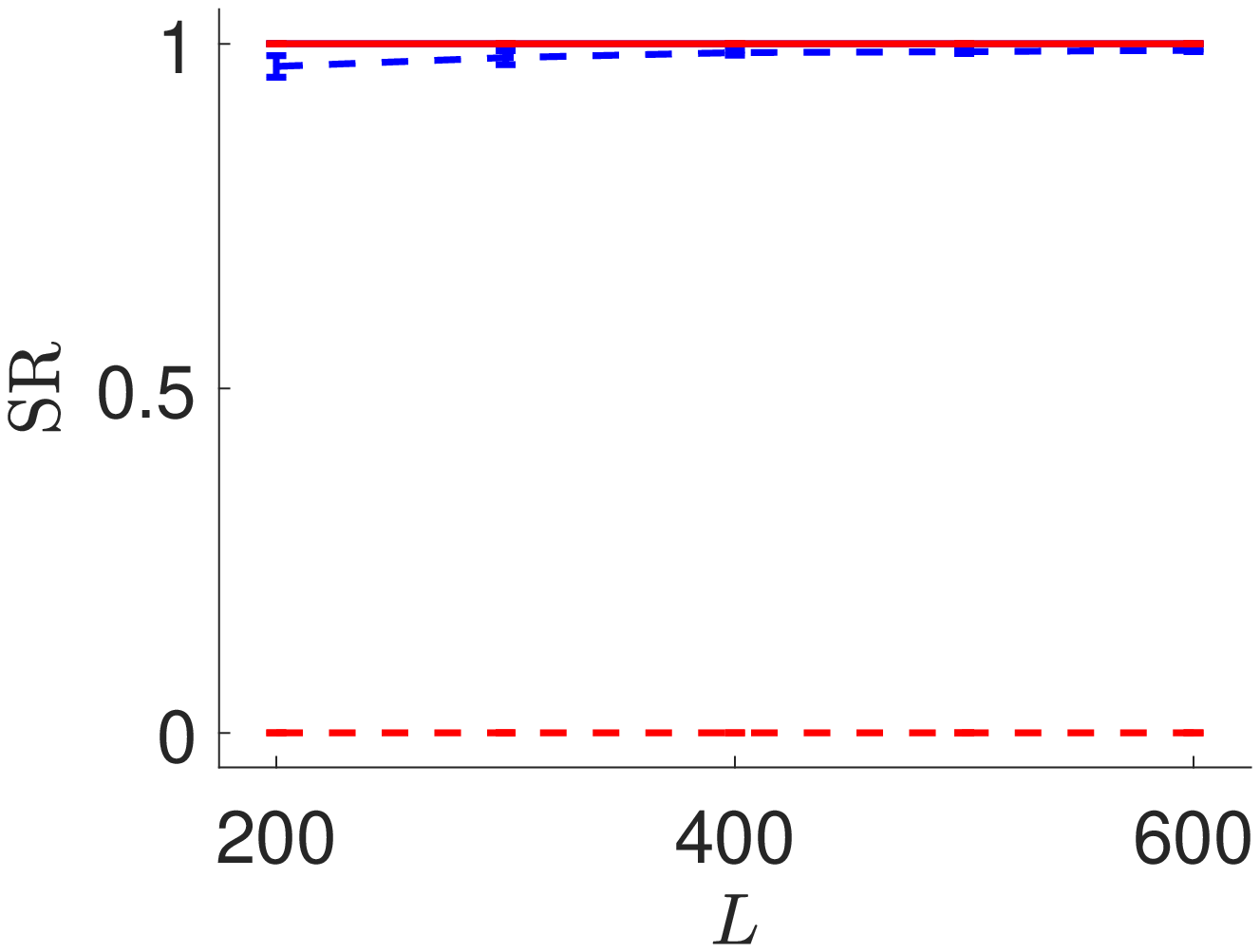}&\includegraphics[align=c,width=0.32\linewidth]{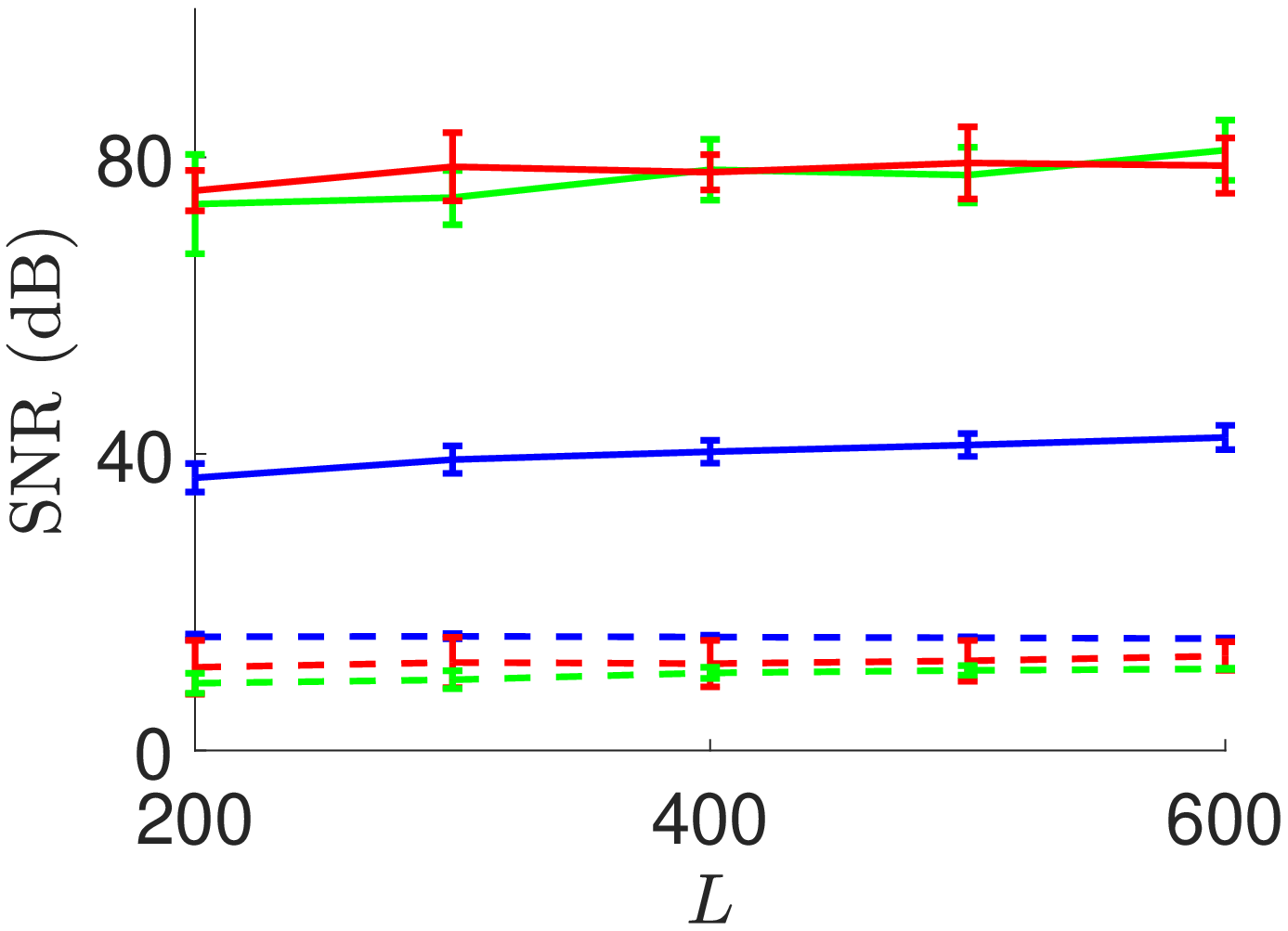}
	\end{tabular}
	 \endgroup
    \caption{
	\label{im_e1pdsnr}
	Experiment 2 - 
    Simulation results obtained with BLIP (dashed lines) and GAP-MRF (solid lines). 
    Left: Tissue proton density maps (Adipose, WM, Muscle, GM, CSF) and magnetisation sequence ($\mathbf{M}$) evaluation.
    Center: SR evaluation for the pure and PV voxels.
    Right: Dominant tissue parameter maps ($\rho$, $T_1$, $T_2$) evaluation.
    }
\end{figure}

Fig.~\ref{im_e1pdsnr} (Left) shows the evaluation of the proton density maps for each tissue (Adipose, WM, Muscle, GM, and CSF) and the magnetisation sequence. Note that GAP-MRF results are taken directly from the matrix $\widetilde{\mathbf{U}}$ without using any post-processing. We can observe that GAP-MRF outperforms BLIP in reconstructing $\mathbf{U}$. 
This can be explained by the fact that BLIP is restricted to the input dictionary, while our method estimates the dictionary. 
In addition, we can observe that the SNR values of magnetisation sequence reconstructed with BLIP slightly decreases while $L$ increases (while it is not the case for the proton density maps). This is expected since the linear combination of short fingerprints are less distinctive, hence it is easier to approximate it with fingerprints of other elements (allowing BLIP to fit better PV voxels with other elements). This behaviour is not observed with the proposed GAP-MRF method for which accurate proton density map estimates result in accurate magnetisation sequence reconstructions.
Fig.~\ref{im_e1pdsnr} (Center) gives the SR for both pure and PV voxels. Since BLIP can only reconstruct one element per voxel, its SR for PV is always equal to $0$. In a low noise scenario, GAP-MRF can identify the correct voxel elements even for short sequences. An important remark is that due to PV, the dictionary sampling and the number of acquisitions, the BLIP algorithm can mis-reconstruct pure voxels even in a low noise scenario.

    

For the dominant tissue parameter maps in the low noise scenario, GAP-MRF outperforms BLIP as shown in Fig.~\ref{im_e1pdsnr} (Right). The proton density map of BLIP is affected by the PV since it is not able to distinguish between the voxel tissues. The $T_1$ and $T_2$ maps are affected by the PV voxels and the dictionary inaccuracies.

BLIP reconstructions show a variation on $T_1$ and $T_2$ for the same tissue while GAP-MRF reconstructions are accurate. The GAP-MRF has the additional advantage that it simultaneously estimates the manifold regions of interest, resulting in better reconstructions.


\section{Real data results} \label{section_realdata}
In this section, we show the reconstructions on the EUROSPIN phantom and on two \textit{in vivo} datasets. The first and second datasets were acquired using spiral sampling scheme and the third dataset was acquired using EPI sampling scheme \cite{Arnold2018}. The parameters were chosen as discussed in Section~\ref{sec:Initialisation}. The obtained proton density maps were normalised as $\widetilde{\mathbf{U}}/ \max(\widetilde{\mathbf{U}})$ and only the proton densities greater than the $10\%$ of $\max(\widetilde{\mathbf{U}})$ are shown in the figures. The normalised proton density is in arbitrary units (a.u.) and the relaxation times are in ms. 
\subsection{EUROSPIN phantom dataset with spiral sampling}
In this subsection, we show the results obtained with both the proposed approach and BLIP, considering a dataset from 
a GE HDx MRI system with an 8 channel receive only head RF coil (GE Medical
Systems, Milwaukee, WI). The acquisition scheme uses a variable density spiral with 377 interleaves using FISP based $\alpha$ \cite{Jiang2015} and a constant $TR=10$ms. The excitation sequence length is $L=1000$. In this experiment, we have FOV~=~$22.5\times22.5$cm$^2$ with a $5$mm slice thickness. The EPG model is used for the reconstructions with an inversion time (TI) of $18$ms and an Echo Time $TE=1.902$ms. The scanned objects are the tubes $1$, $5$ and $9$ of the EUROSPIN phantom. We reconstruct the parameter maps with two spatial resolutions: the first one at $180\times 180$ with an undersampling ratio of $N/Q=44.8753$, and the second one at $40\times 40$ with an undersampling ratio of $N/Q=20.6869$ to introduce the PV. Reconstructing for higher spatial resolution would introduce high frequency artefacts as shown in \cite{AIR-MRF}. An acquisition without the tubes is performed to estimate $\sigma_{\mathbf{Y}}$ and compute a lower bound on the iSNR. More precisely, using the triangle inequality, since $\| \mathbf{Y}\|_2 \ge \| \boldsymbol{\eta}\|_2$, we have $\text{iSNR}\geq 20\log\left((\|\mathbf{Y}\|_2-\|\boldsymbol{\eta}\|_2)/(\sqrt{QLC}\sigma_{\mathbf{Y}})\right)=64.73\text{dB}$, where $\mathbf{Y}$ corresponds to the measurements with the tubes, $\boldsymbol{\eta}$ corresponds to the measurements without the tubes, and the value $\sigma_{\mathbf{Y}}$ is the standard deviation of $\boldsymbol{\eta}$.


\begin{figure}[!t]
	\centering
	\begingroup
    \setlength{\tabcolsep}{0.2mm}
\renewcommand{\arraystretch}{0.2}
\footnotesize
\begin{tabular}{ccccc}
    \multicolumn{2}{c}{BLIP} & \multicolumn{2}{c}{GAP-MRF}&\\
     $180\times 180$ & $40\times 40$ & $180\times 180$ & $40\times 40$\\
     \includegraphics[width=0.22\linewidth]{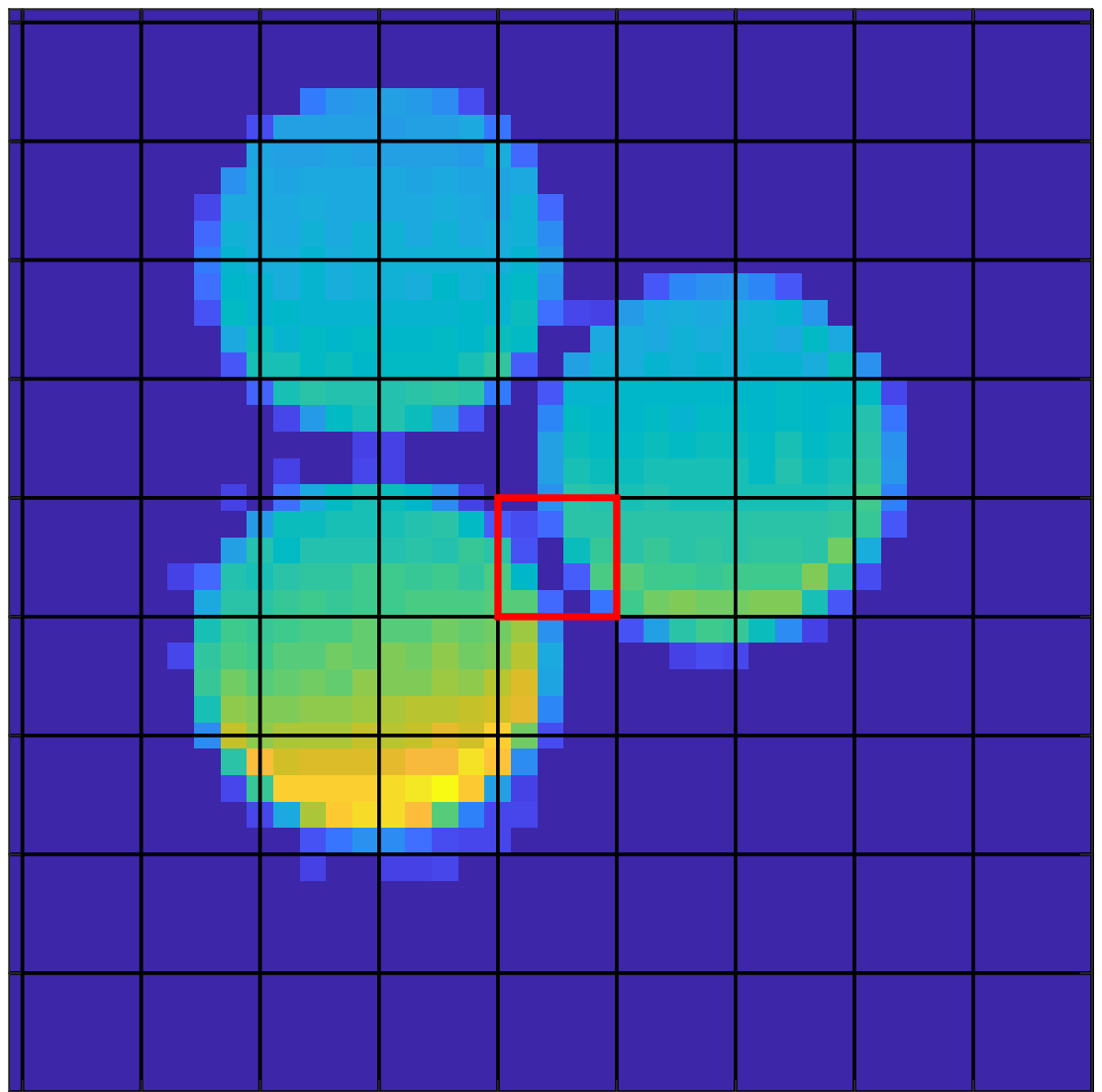}& \includegraphics[width=0.22\linewidth]{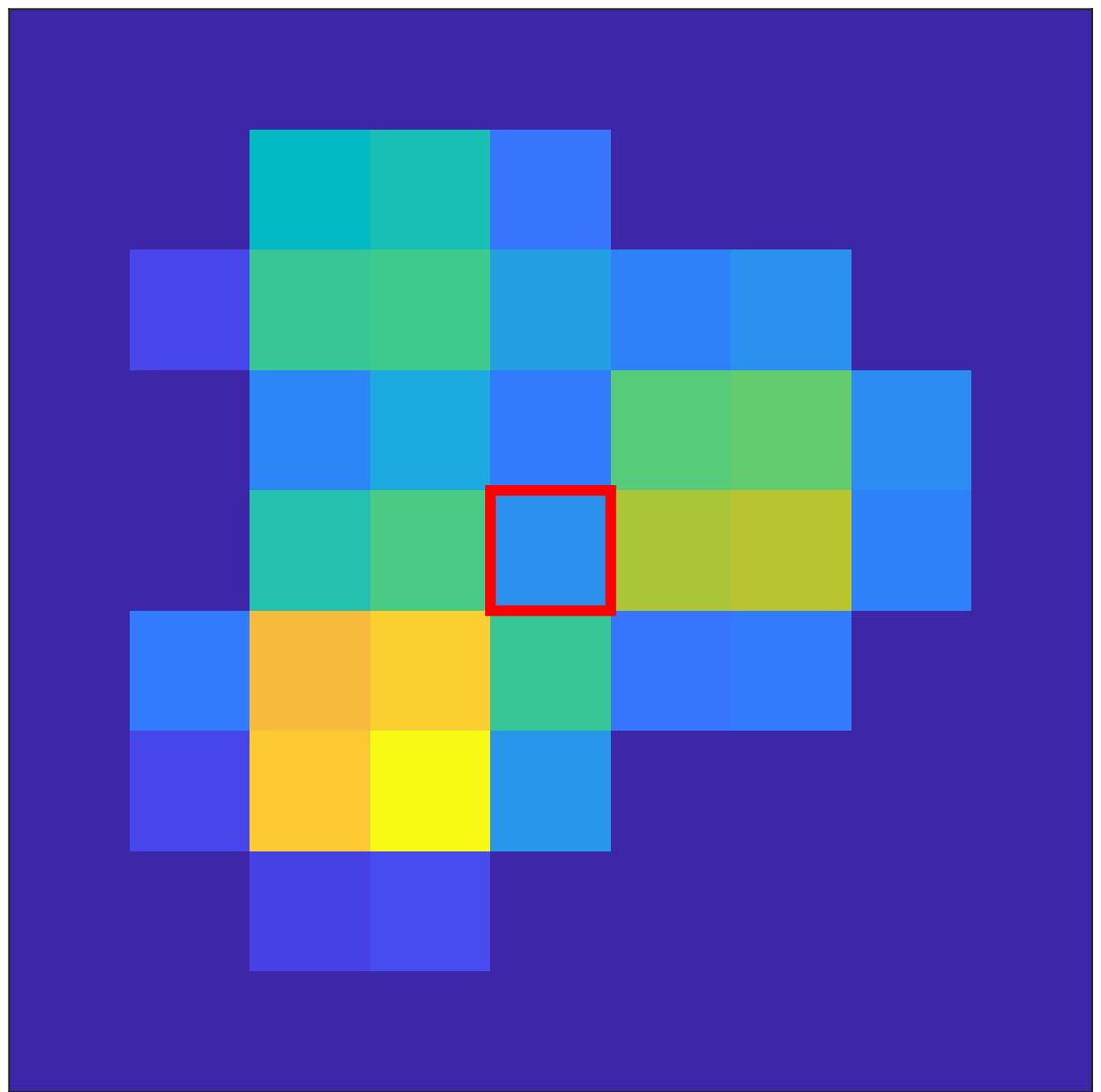}& \includegraphics[width=0.22\linewidth]{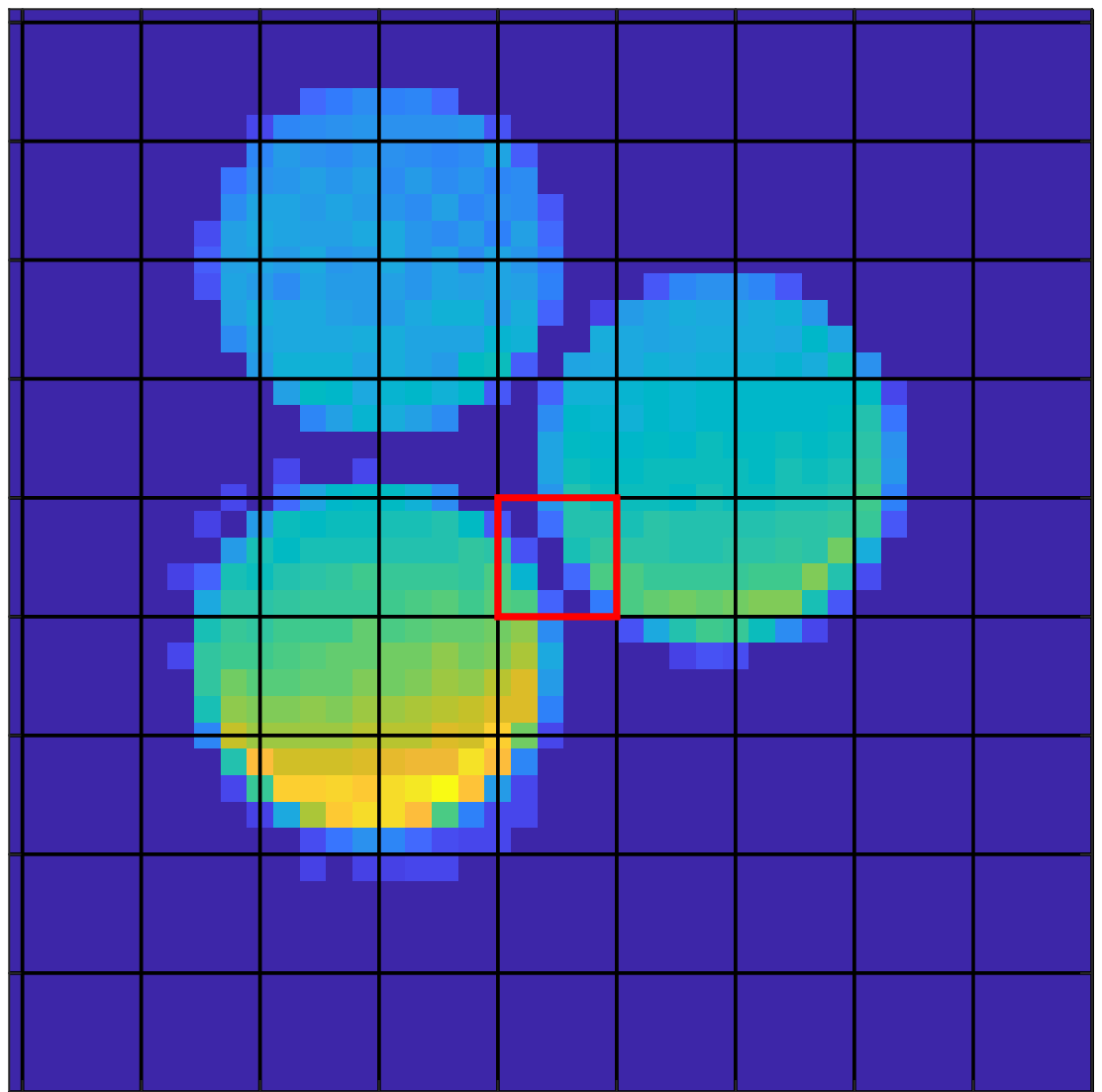}& \includegraphics[width=0.22\linewidth]{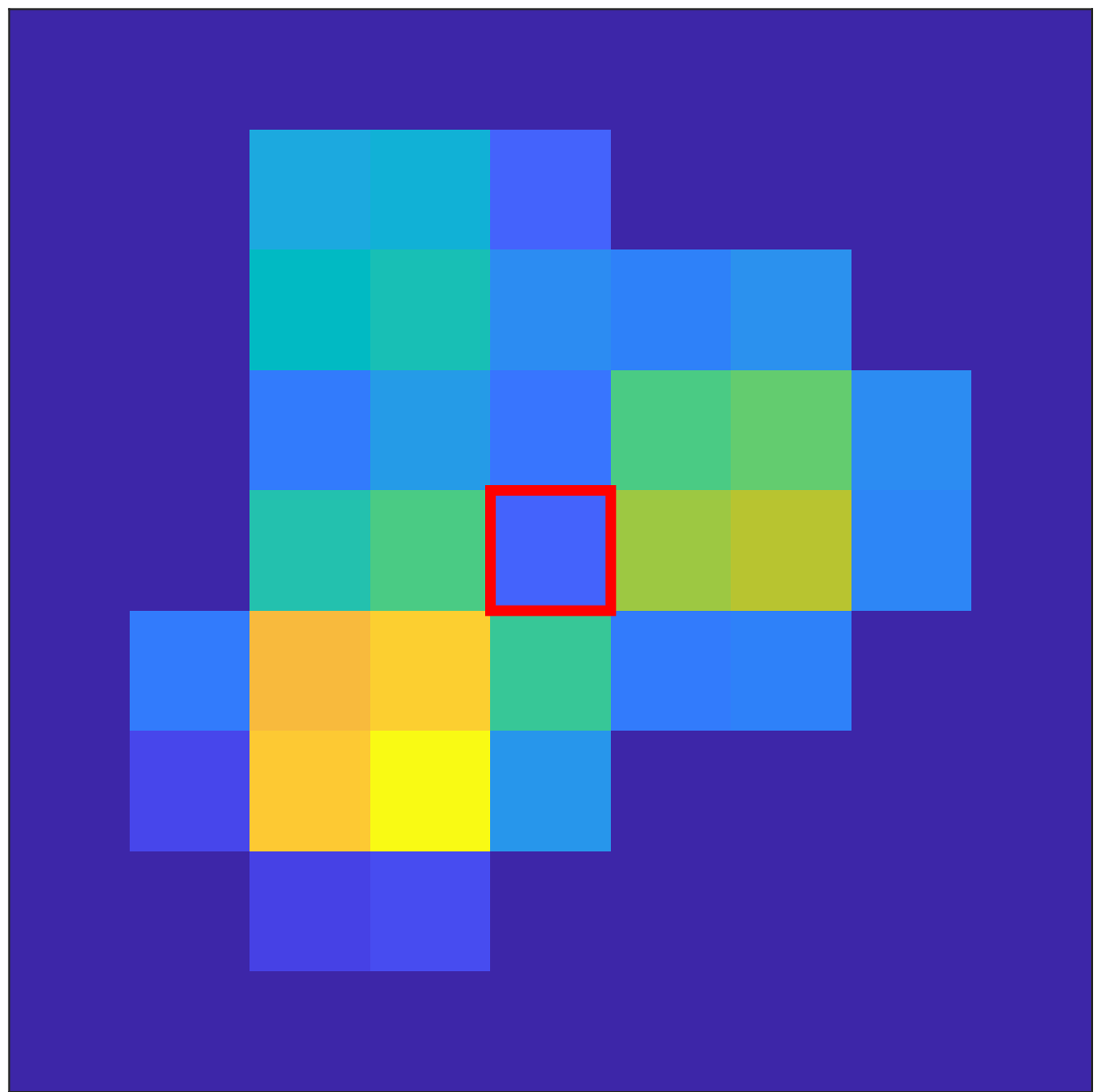}& \includegraphics[width=0.0315\linewidth]{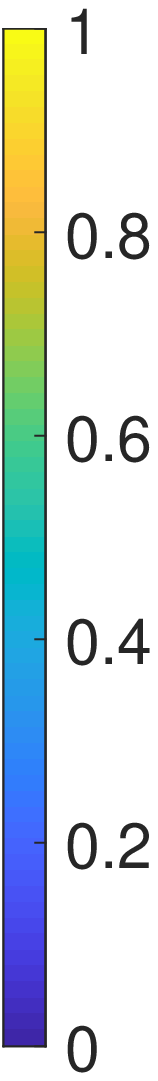}\\
	\includegraphics[width=0.22\linewidth]{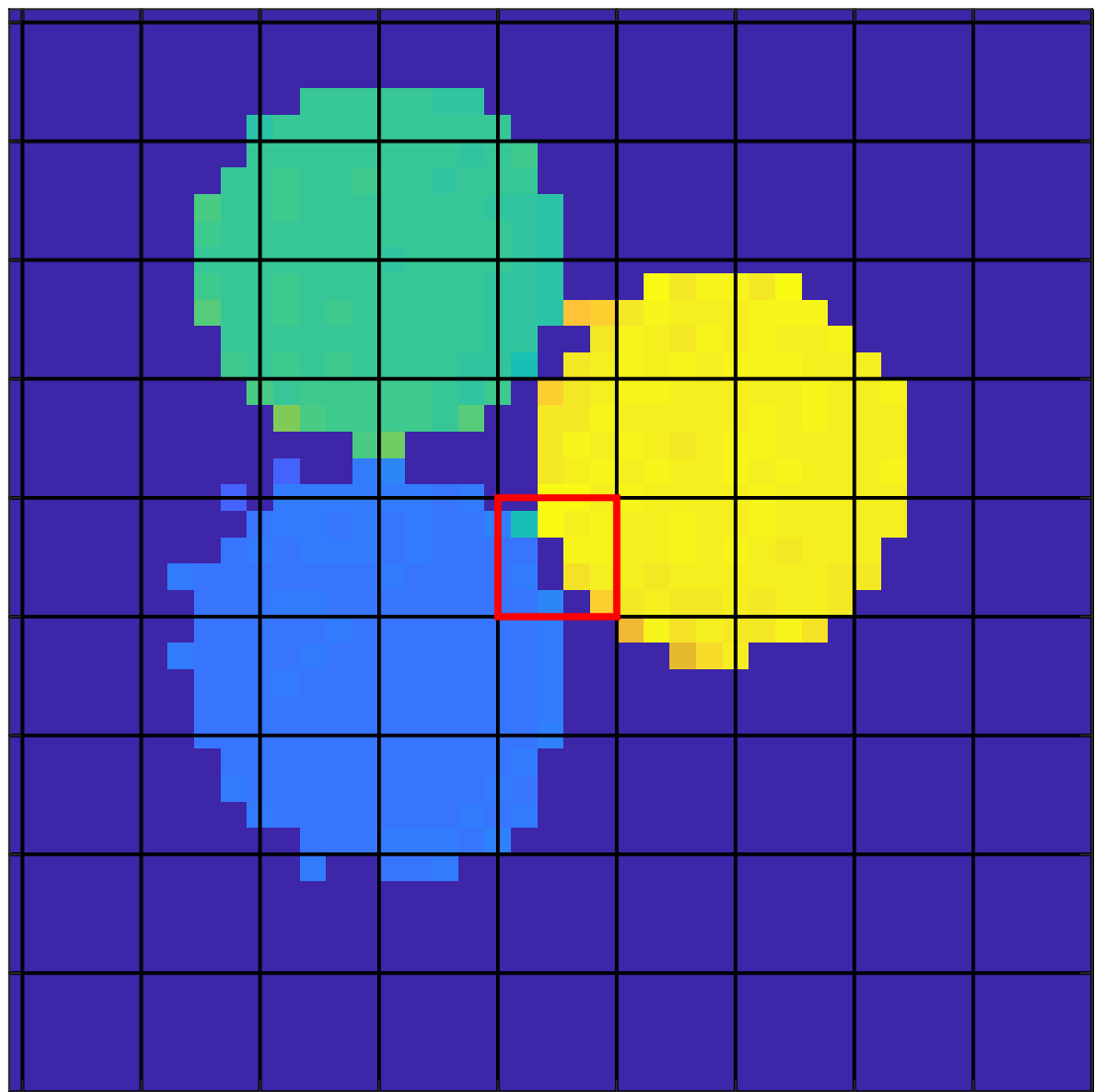}& \includegraphics[width=0.22\linewidth]{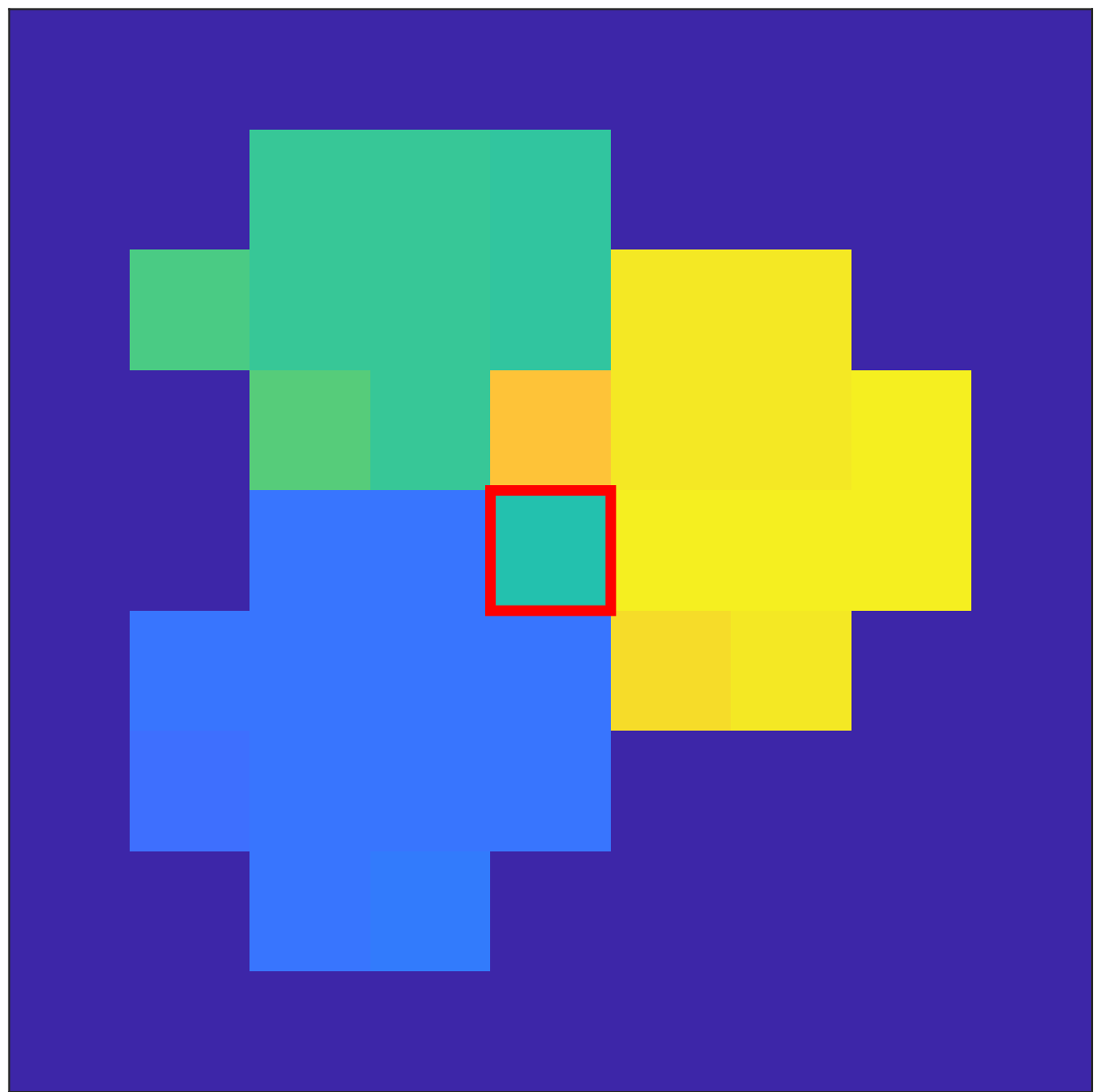}& \includegraphics[width=0.22\linewidth]{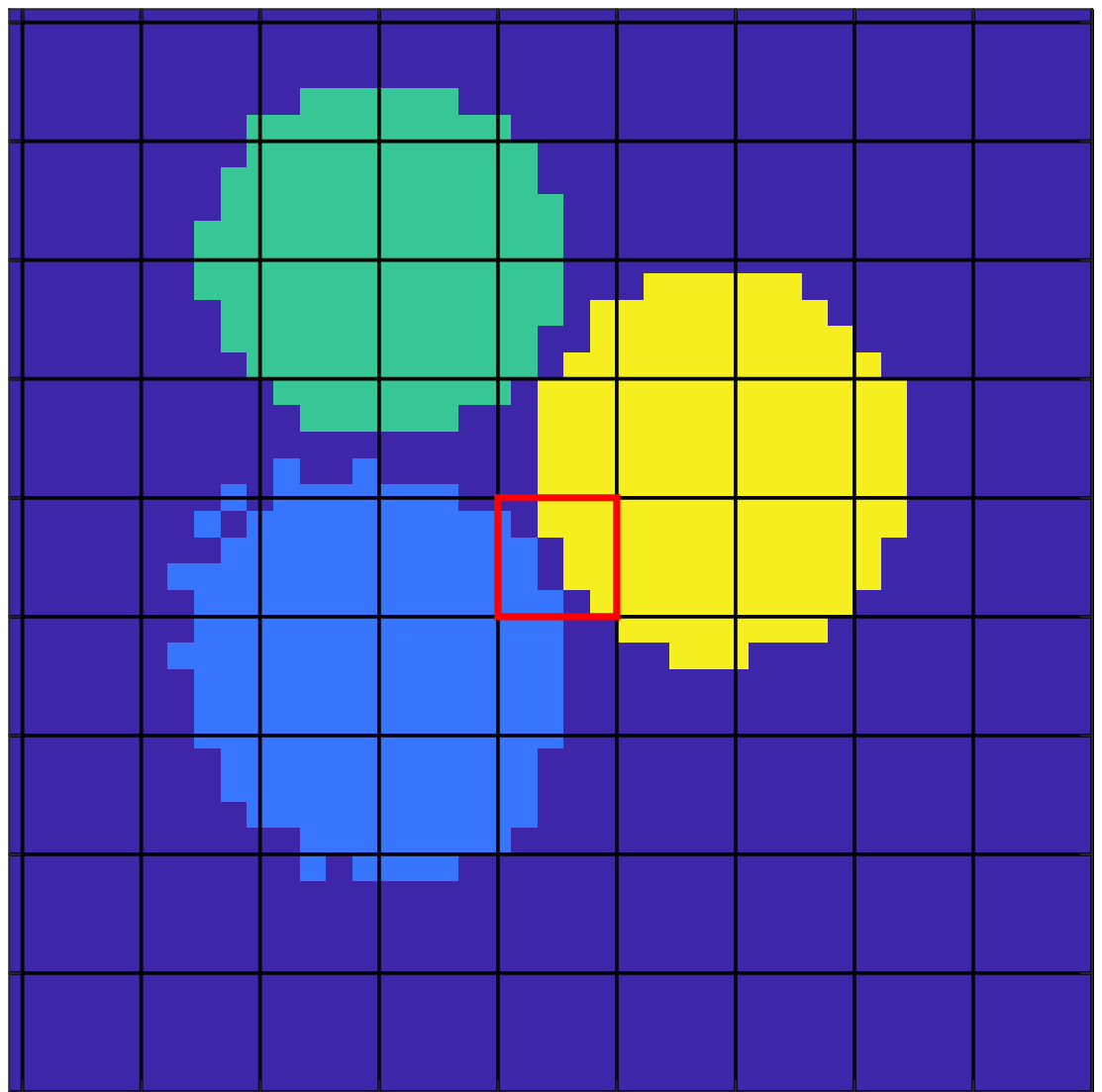}& \includegraphics[width=0.22\linewidth]{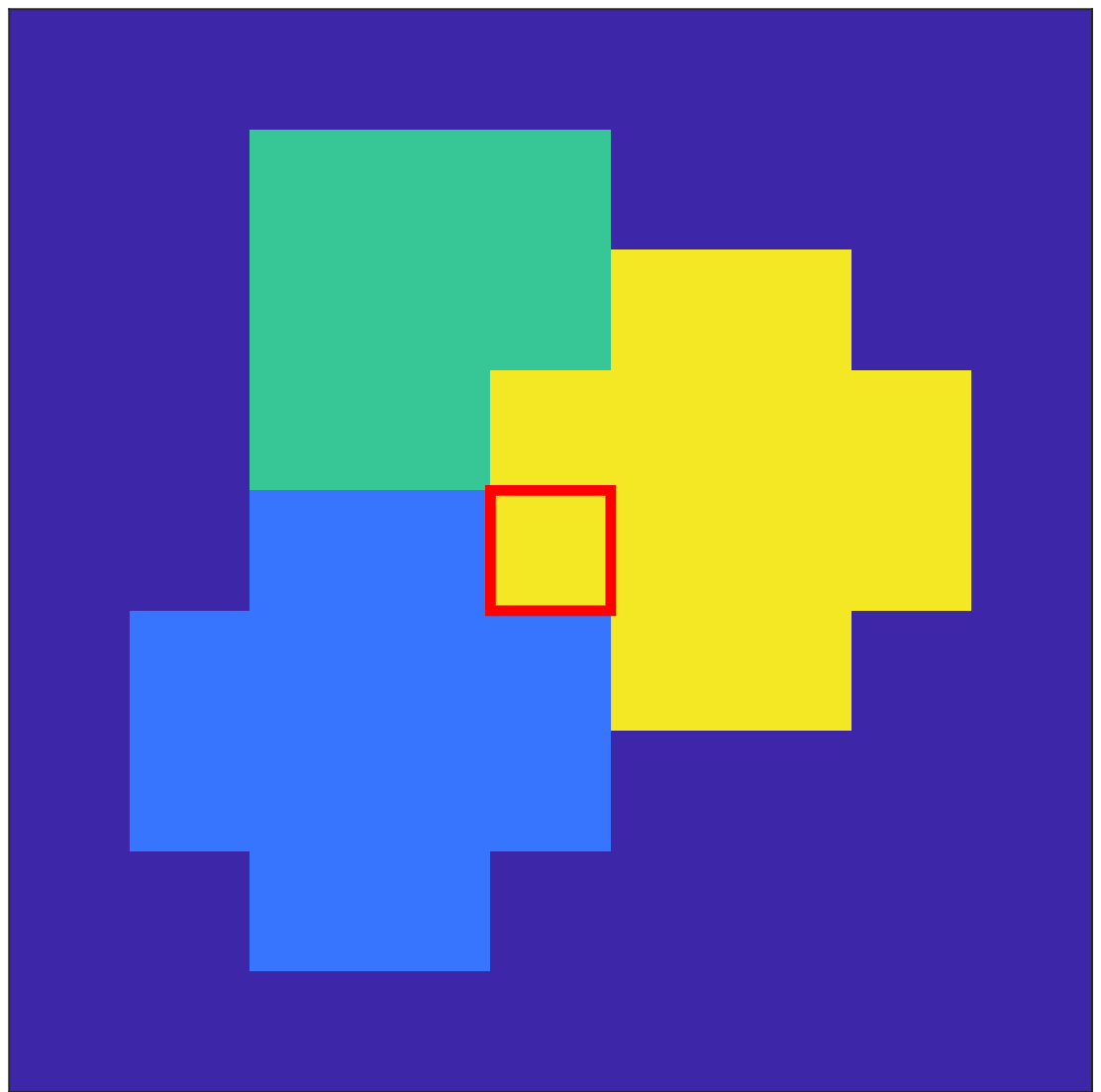}& \includegraphics[width=0.035\linewidth]{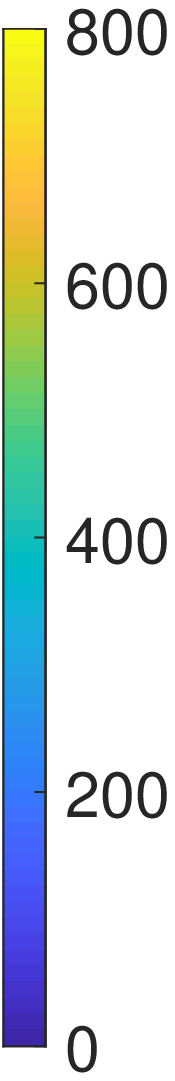}\\
	\includegraphics[width=0.22\linewidth]{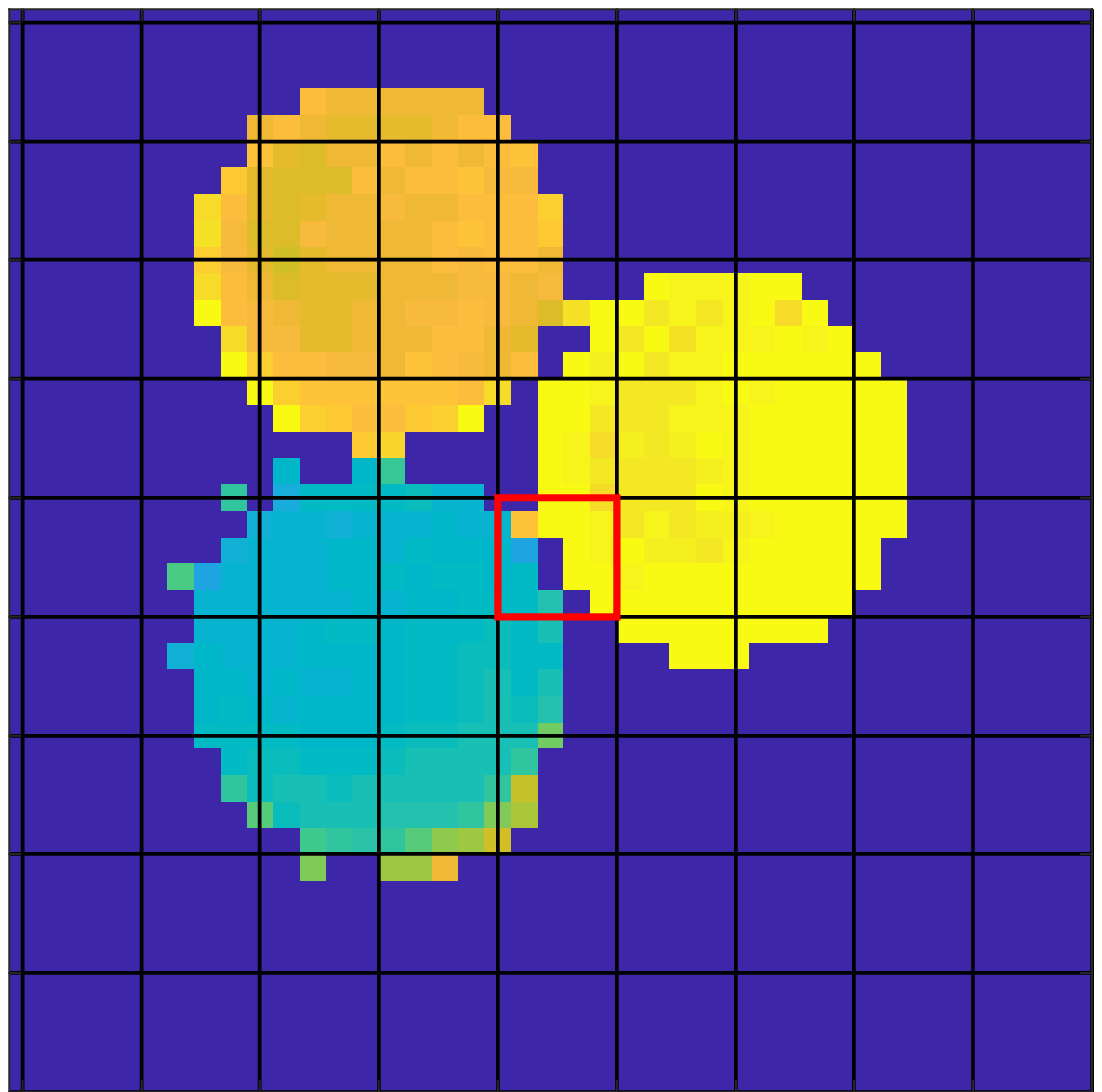}& \includegraphics[width=0.22\linewidth]{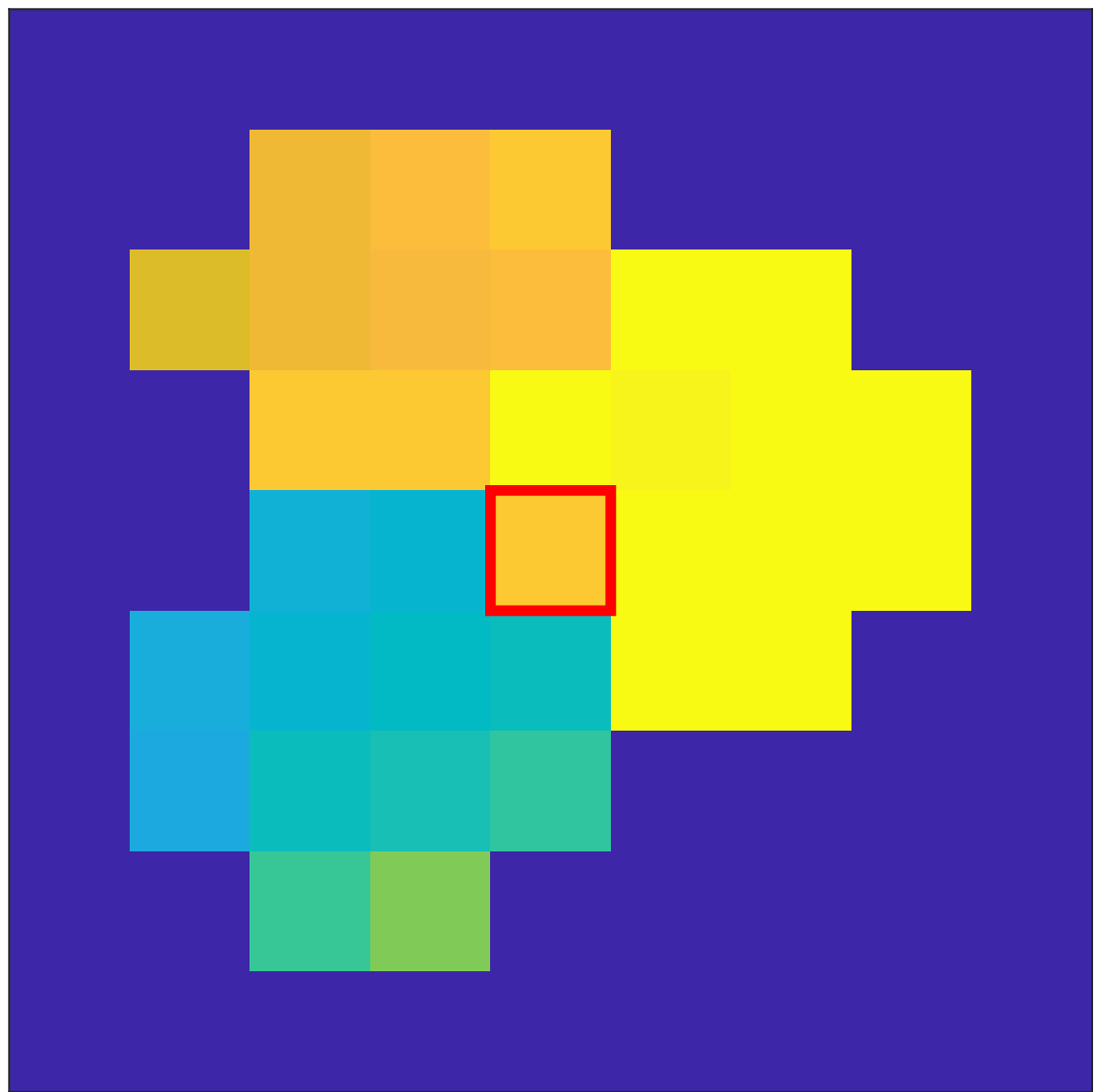}& \includegraphics[width=0.22\linewidth]{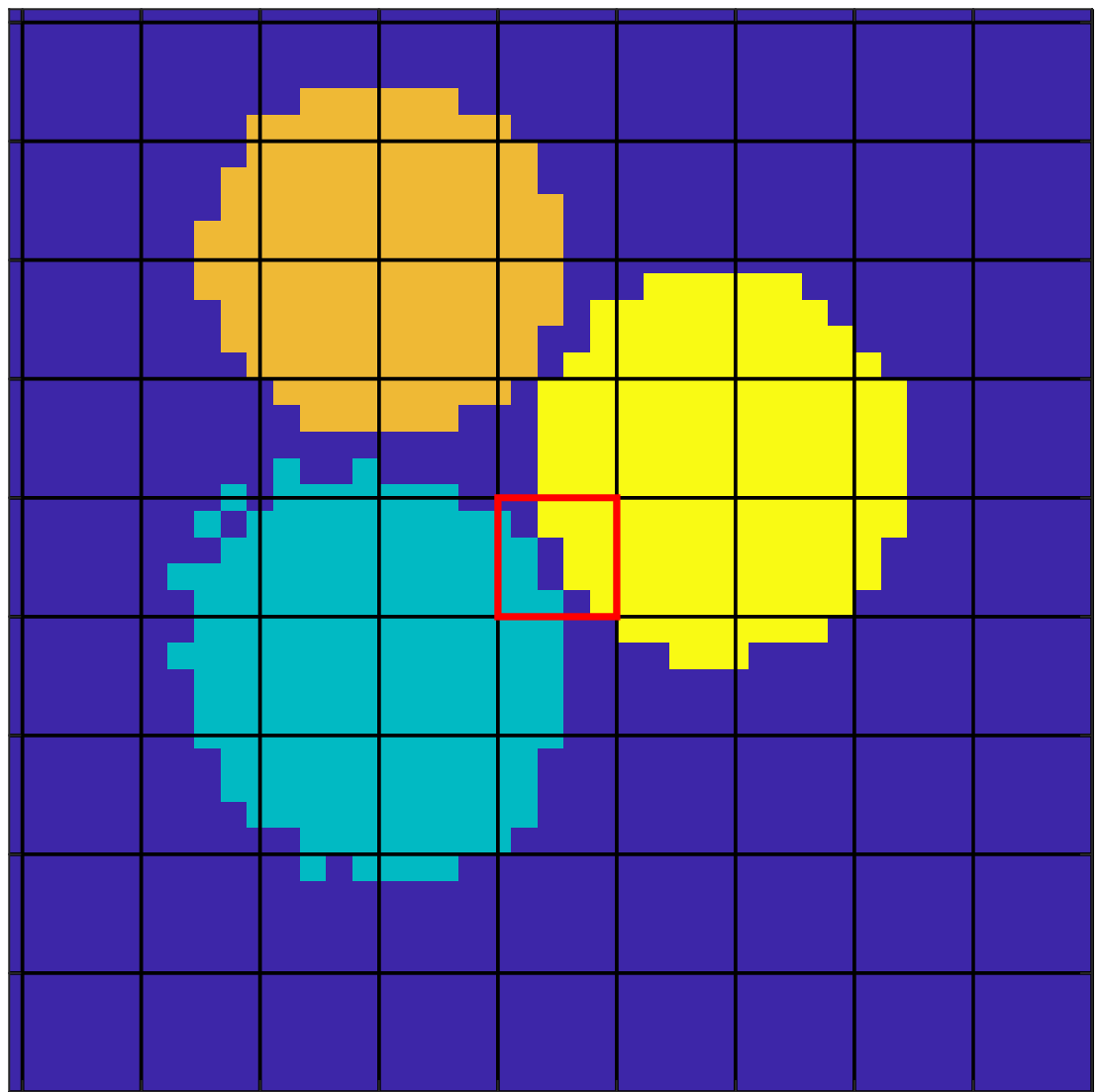}& \includegraphics[width=0.22\linewidth]{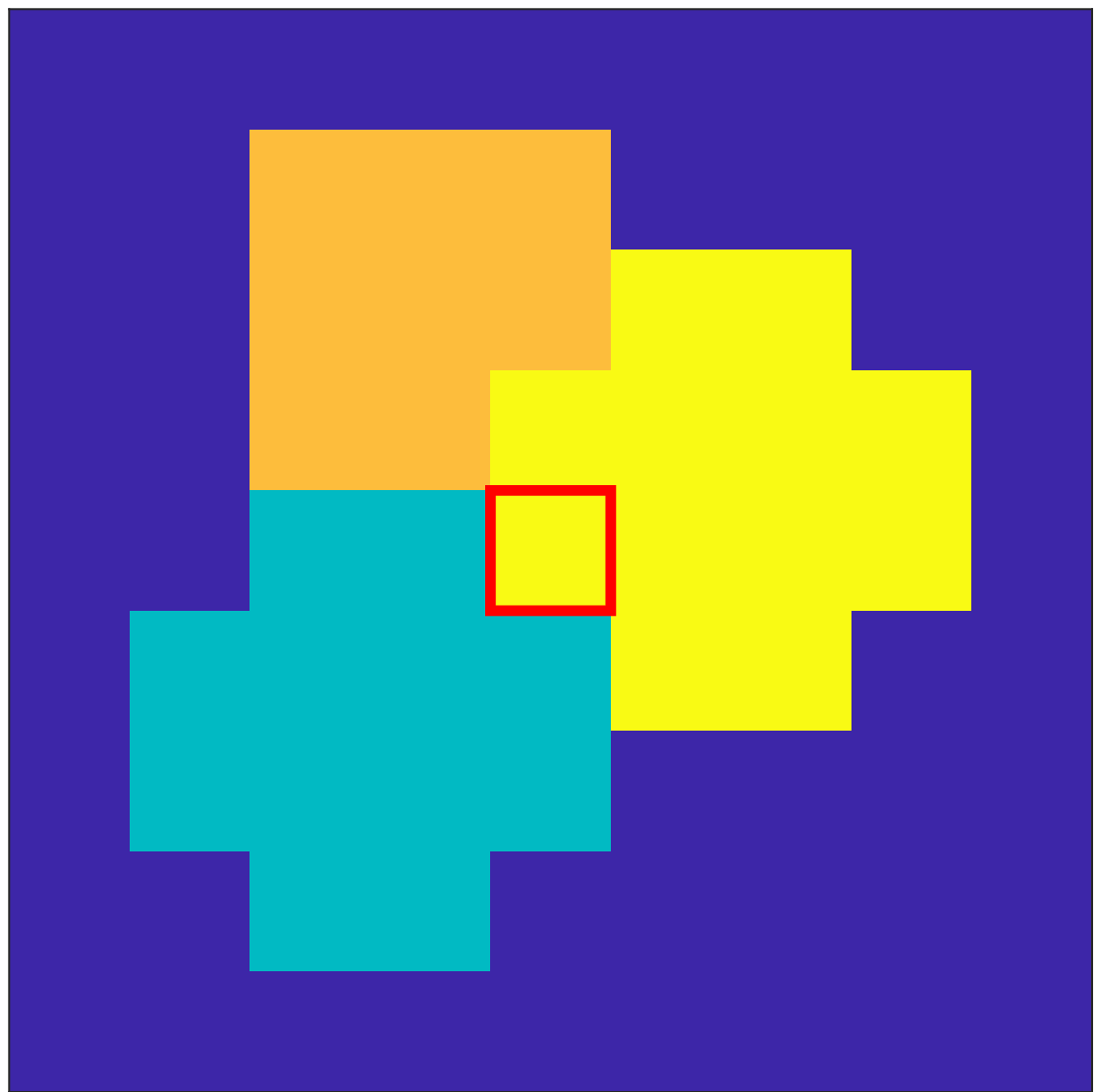}& \includegraphics[width=0.035\linewidth]{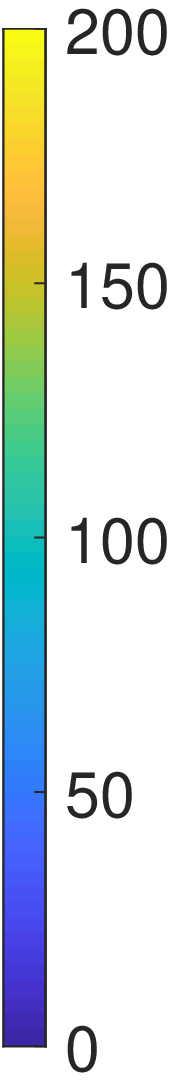}\\
\end{tabular}
	\endgroup
	 \caption{Dominant tissue parameter maps corresponding to the EUROSPIN phantom dataset. 
	 From first to last column: $180\times 180$ BLIP reconstruction trimmed to $41\times 41$ voxels, $40 \times 40$ BLIP reconstruction trimmed to $9\times 9$ voxels, $180\times 180$ GAP-MRF reconstruction trimmed to $41\times 41$ voxels, $40\times 40$ GAP-MRF reconstruction trimmed to $9\times 9$ voxels.
	 From first to last row: normalised proton density, $T_1$, and $T_2$. }
    \label{im_phantomdominant}
\end{figure}

\begin{figure}[!t]
\centering
\begingroup
    \setlength{\tabcolsep}{0.2mm}
\renewcommand{\arraystretch}{0.2}
\begin{tabular}{cccc}
	\includegraphics[width=0.3\linewidth]{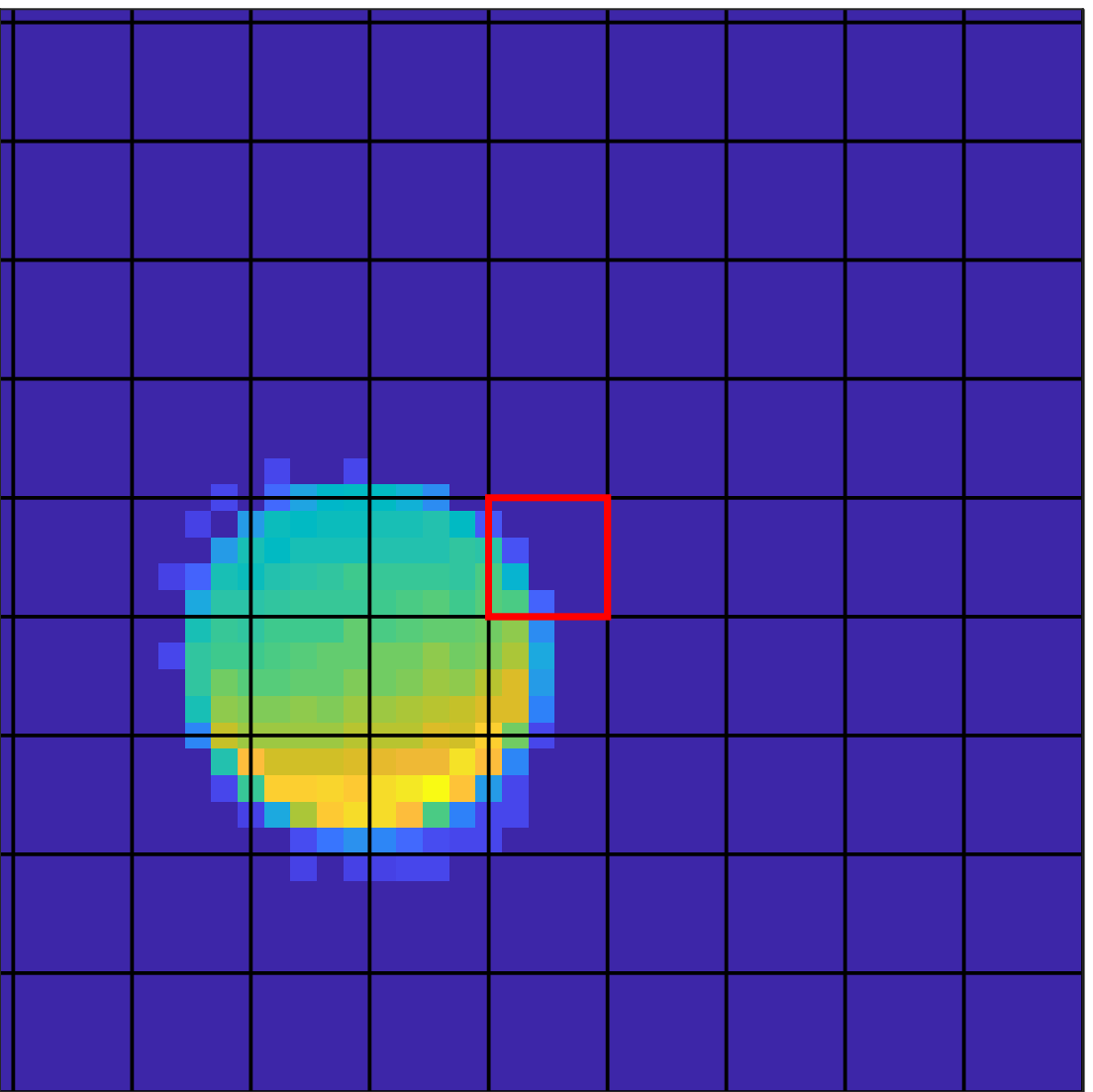}
 &	\includegraphics[width=0.3\linewidth]{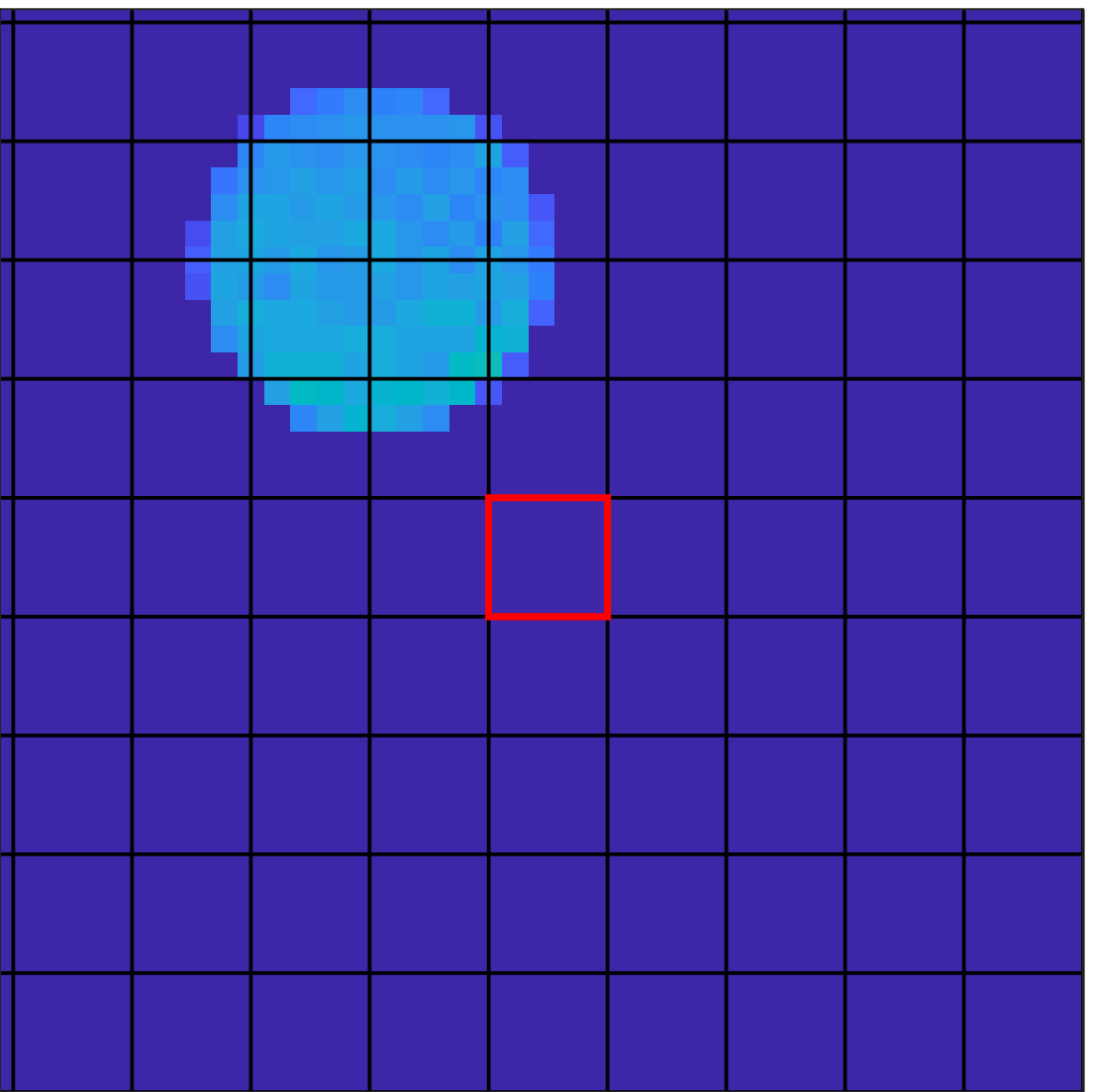}
 &	\includegraphics[width=0.3\linewidth]{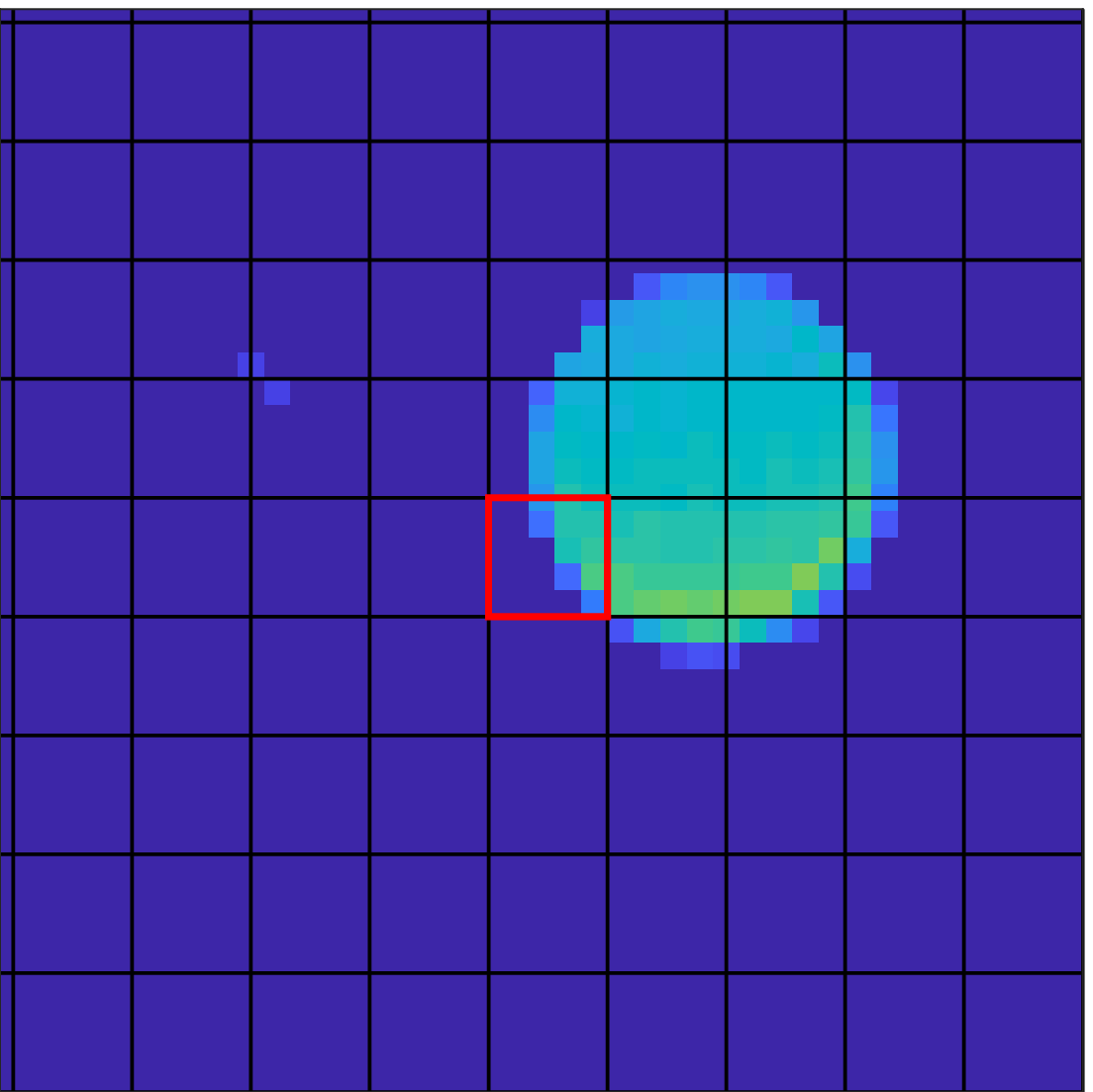}
 &	\includegraphics[width=0.38cm]{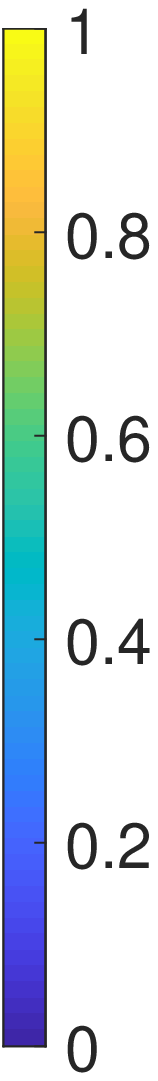}\\
\includegraphics[width=0.3\linewidth]{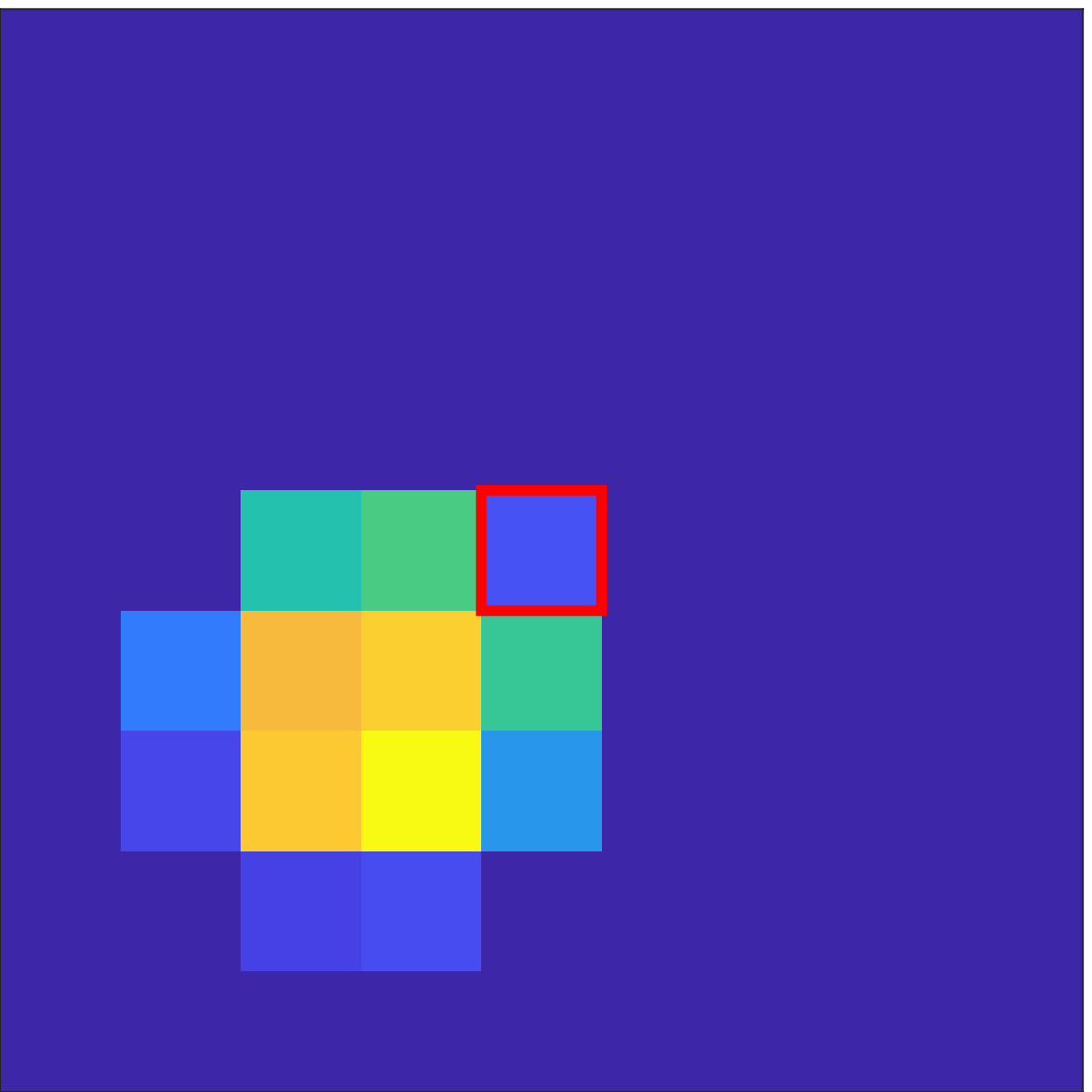}
&	\includegraphics[width=0.3\linewidth]{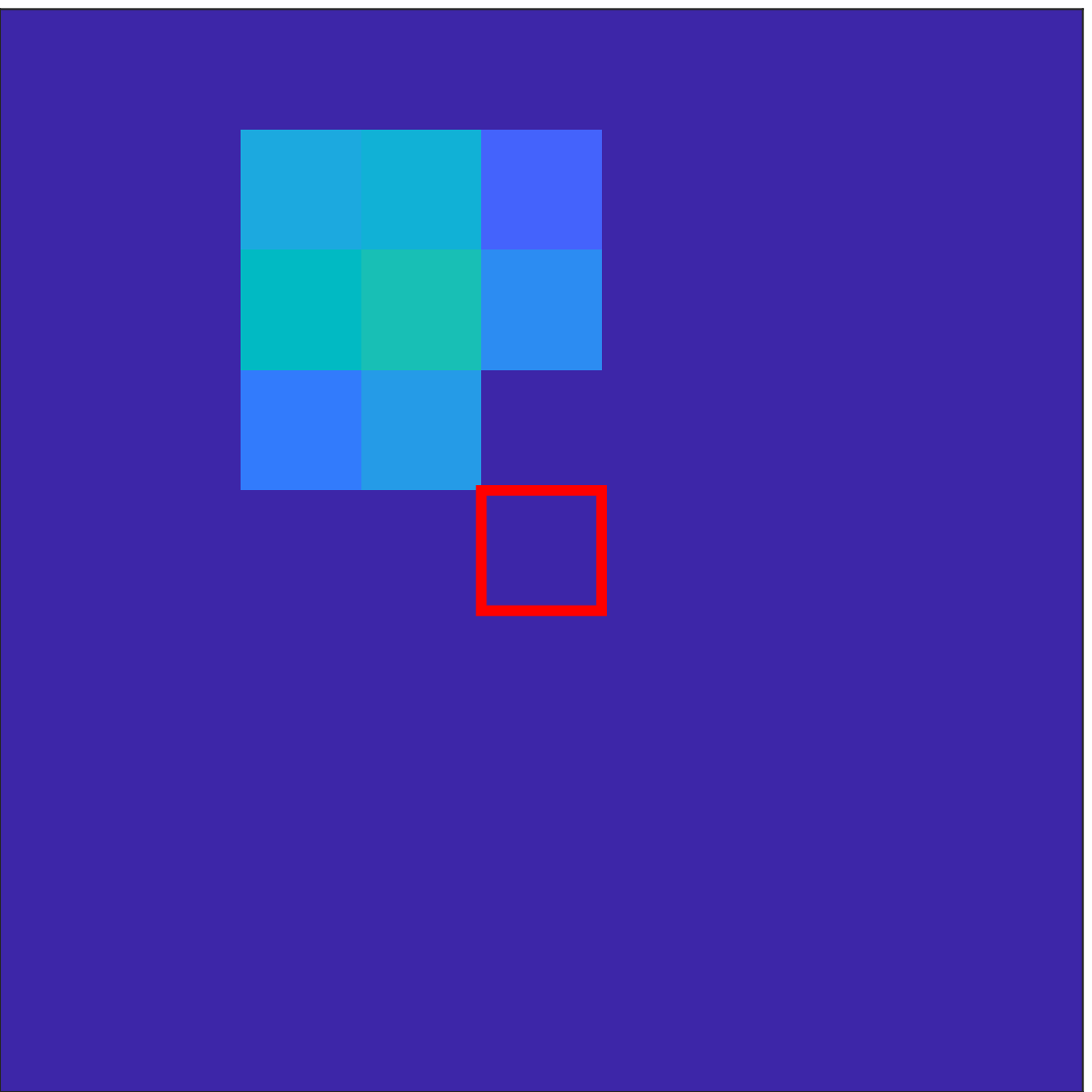}
&	\includegraphics[width=0.3\linewidth]{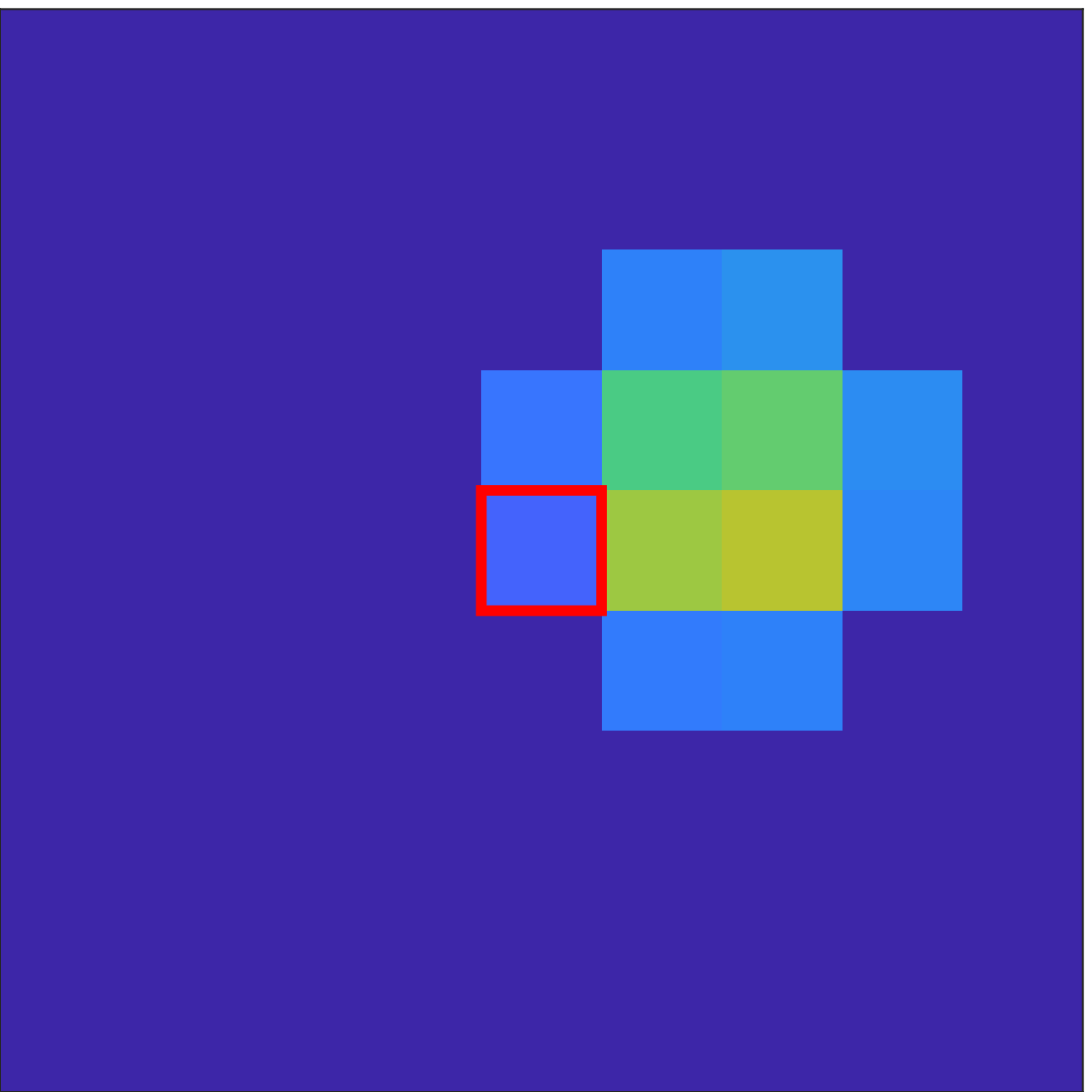}
&	\includegraphics[width=0.38cm]{fig/phantomcolorbar.eps}
\end{tabular}
\endgroup
	\caption{Normalised proton density maps corresponding to the EUROSPIN phantom dataset. The first and second row correspond to the $180\times 180$ reconstruction trimmed to $41\times 41$, and the $40\times 40$ reconstruction trimmed to $9\times 9$, respectively. The corresponding $T_1$ and $T_2$ can be seen in Table~\ref{tab:phantom}. From left to right the columns correspond to Tube~1, Tube 5 and Tube 9 of the EUROSPIN phantom.}
    \label{im_phantomresults}
\end{figure}

\begin{table}[!t]\footnotesize
\centering
\caption{Comparison between the parameters obtained with GAP-MRF and the EUROSPIN phantom values.}
\label{tab:phantom}
\begin{tabular}{l|r|r|r|r|r|r|}
\cline{2-7}
                                   & \multicolumn{2}{c|}{Phantom Values} & \multicolumn{2}{c|}{$180\times 180$} & \multicolumn{2}{c|}{$40\times 40$} \\ \cline{2-7} 
                                   & \multicolumn{1}{|c|}{$T_1$}           & \multicolumn{1}{|c|}{$T_2$}        & \multicolumn{1}{|c|}{$T_1$}        & \multicolumn{1}{|c|}{$T_2$} & \multicolumn{1}{|c|}{$T_1$}        & \multicolumn{1}{|c|}{$T_2$}       \\ \hline
\multicolumn{1}{|l|}{Tube 1} & $200\pm 6$          & $52\pm 1.6$          & 197.0       & 93.9      &195.4 & 96.0 \\ \hline
\multicolumn{1}{|l|}{Tube 5}  & $450\pm 13.5$         & $94\pm 2.8$        & 455.9       & 159.8    &459.5 &  168.1 \\ \hline
\multicolumn{1}{|l|}{Tube 9}   &   $754\pm 22.6$   &   $116\pm 3.5$             & 766.3       & 199.2   &757.5 & 199.9   \\ \hline
\end{tabular}
\end{table}

The box in red shows a PV voxel artificially created by reconstructing a lower resolution image. As predicted by the corresponding high resolution maps, this voxel is formed by a linear combination of the Tubes 1 and~9.

In Fig.~\ref{im_phantomdominant}, we show a comparison of the reconstructions with two different spatial resolutions. The $T_1$ and $T_2$ lower resolution maps of BLIP show a variation introduced by the PVE. A clear example of the PVE is the voxel in the red box where two tissues appear, the BLIP reconstruction shows a parameter mismatch. Note that the parameters predicted by BLIP suggest that the voxel contains the same substance as Tube 5, contrary to the true composition (Tubes 1 and 9). GAP-MRF reconstructions do not show this behaviour since we take the PV into account in the model. Note that GAP-MRF is more sensitive to noise as shown in the simulations, this may explain small artefacts in the proton density maps. 

The $T_1$ values in Table~\ref{tab:phantom} are in agreement with the values of the EUROSPIN phantom. The $T_2$ values are higher than expected. As seen in Fig.~\ref{im_phantomdominant}, BLIP results show the same increased $T_2$, suggesting that the errors may be related to the datasets, unless both methods introduce the same bias.

In Fig.~ \ref{im_phantomresults}, the normalised proton density maps reconstructed by GAP-MRF with two different resolutions can be seen. As highlighted by the red box, the PV voxel in the low resolution reconstruction has values different than 0 in the maps corresponding to Tube 1 and 9, this is in agreement with the high resolution maps.

\subsection{\textit{In vivo} brain dataset with spiral sampling}
The scanning for this dataset was performed on a GE HDx MRI system with an 8 channel receive only head RF coil (GE Medical
Systems, Milwaukee, WI). The acquisition scheme uses a variable density spiral with 89 interleaves using FISP based $\alpha$ and TR as in \cite{Jiang2015}. The excitation sequence length is $L=1000$. In this experiment, we have FOV~=~$22.5\times22.5$cm$^2$ and the spatial resolution is $180\times 180$ voxels, with a $5$mm slice thickness. The undersampling ratio is $N/Q=89.53$. The EPG model is used for the reconstructions with a TI of $18$ms and a TE of $2$ms. The reconstruction for BLIP and GAP-MRF was accelerated with the SVD compression in the time domain described in \cite{mcgivney2014} using $30$ eigenvectors.
\begin{figure}[!t]
\centering
\begingroup
\setlength{\tabcolsep}{0.1mm}
\renewcommand{\arraystretch}{0.5}
\begin{tabular}{cccc}
\includegraphics[width=0.3\linewidth]{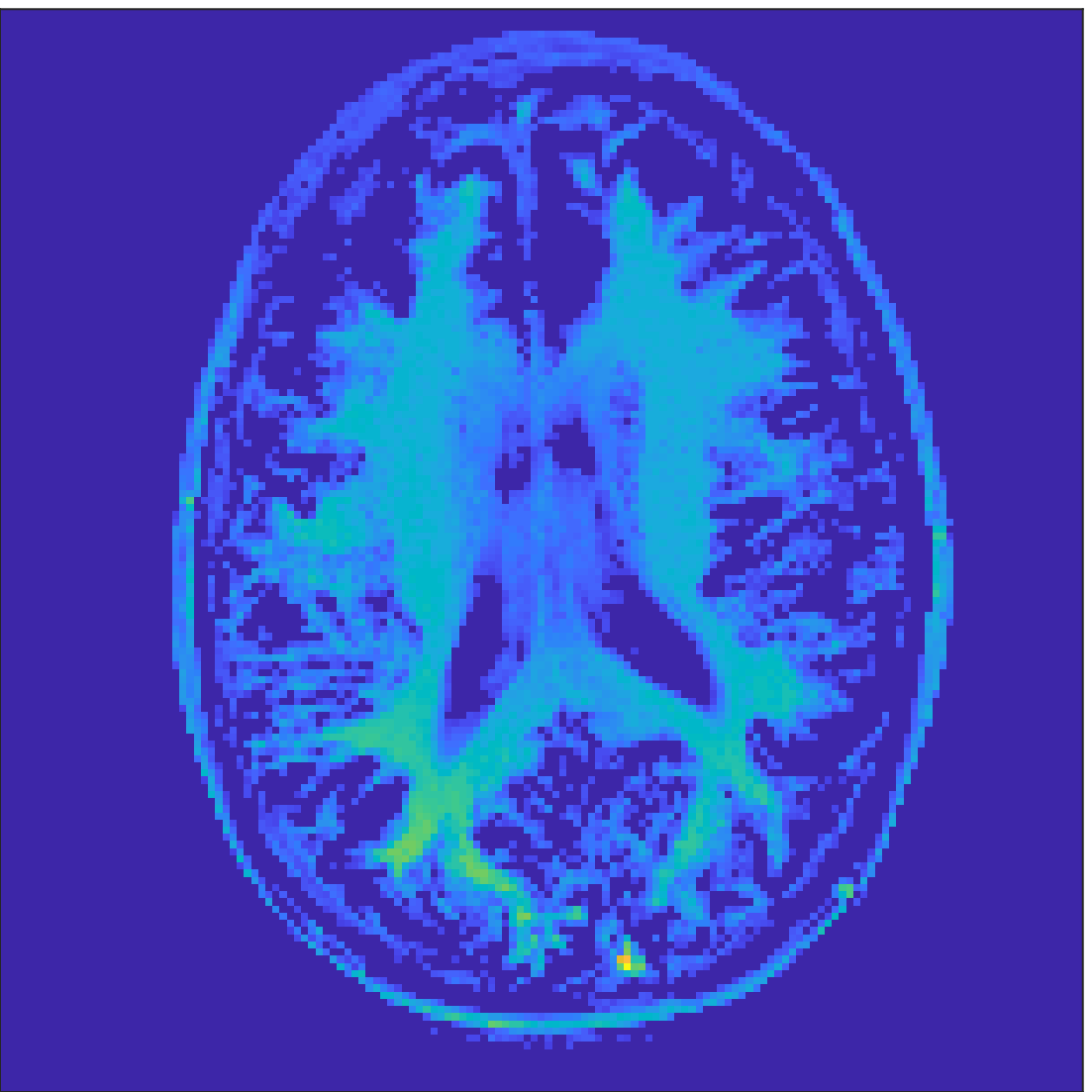} &	\includegraphics[width=0.3\linewidth]{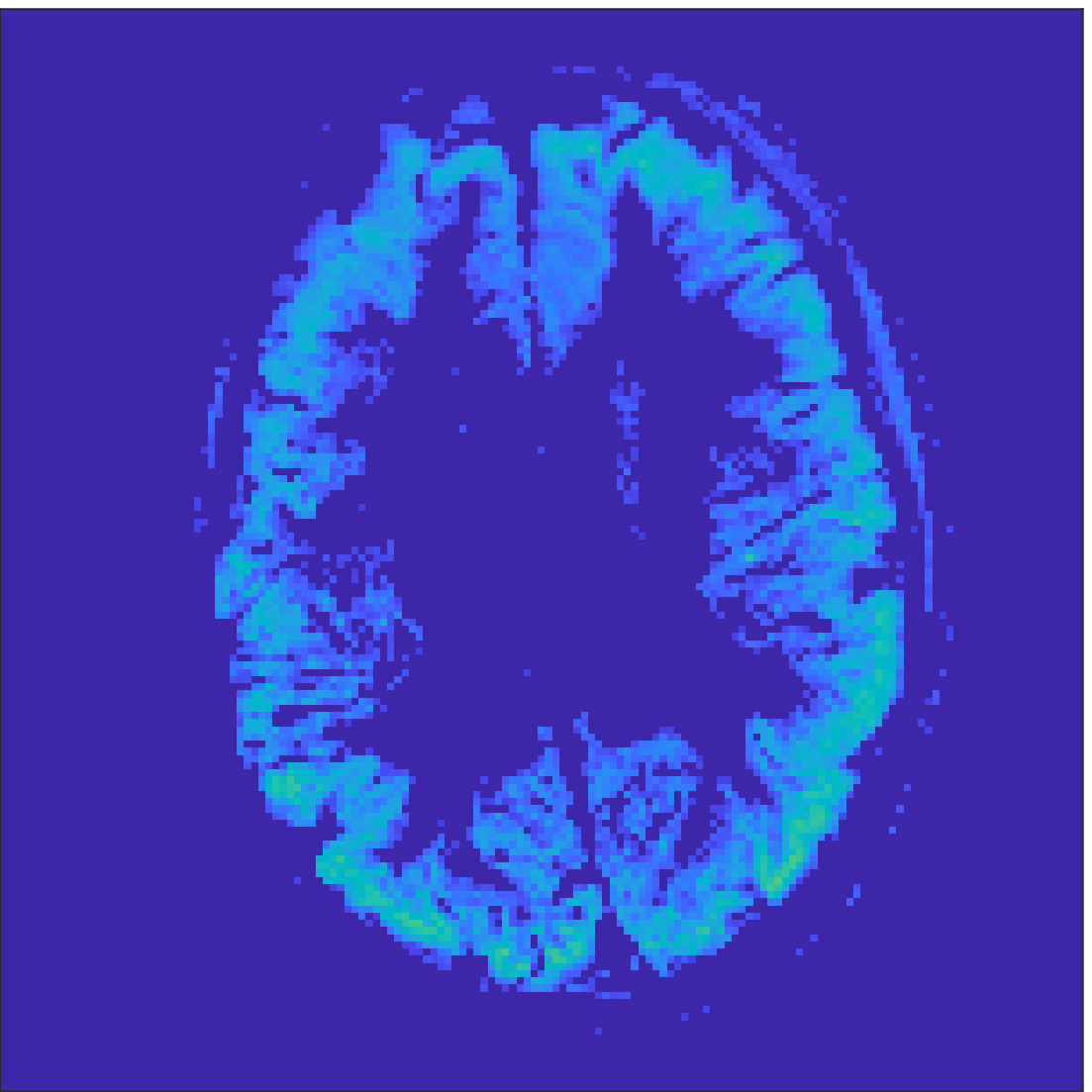}
&	\includegraphics[width=0.3\linewidth]{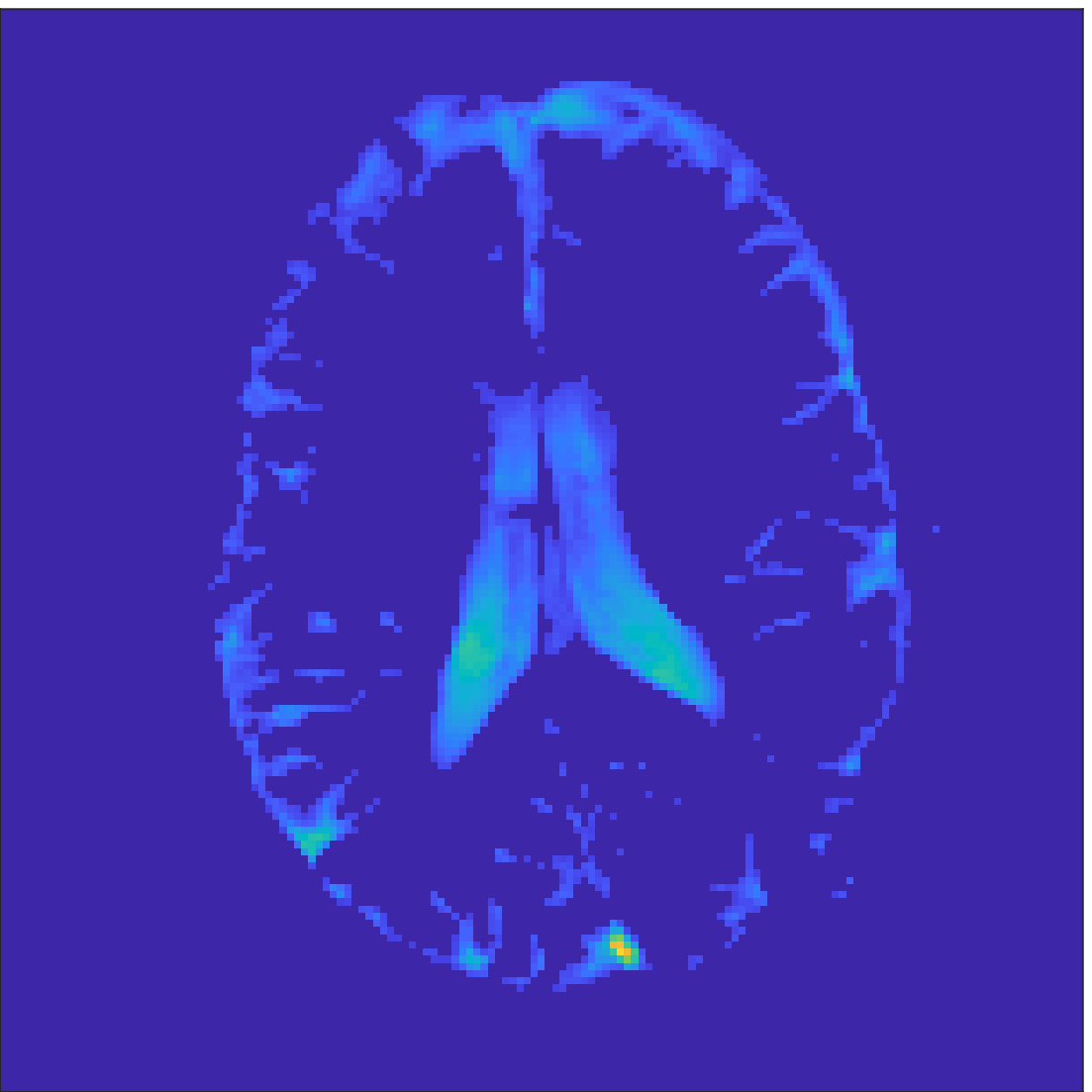}
&	\includegraphics[width=0.38cm]{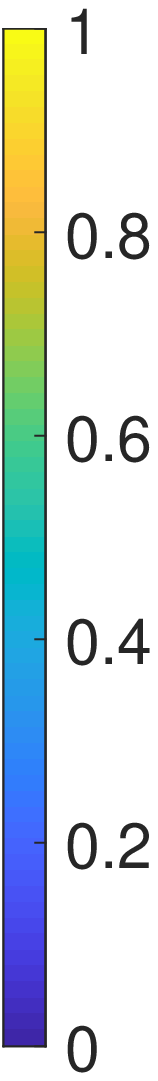}\\
\end{tabular}
\begin{tabular}{ccc}
 \includegraphics[width=0.3\linewidth]{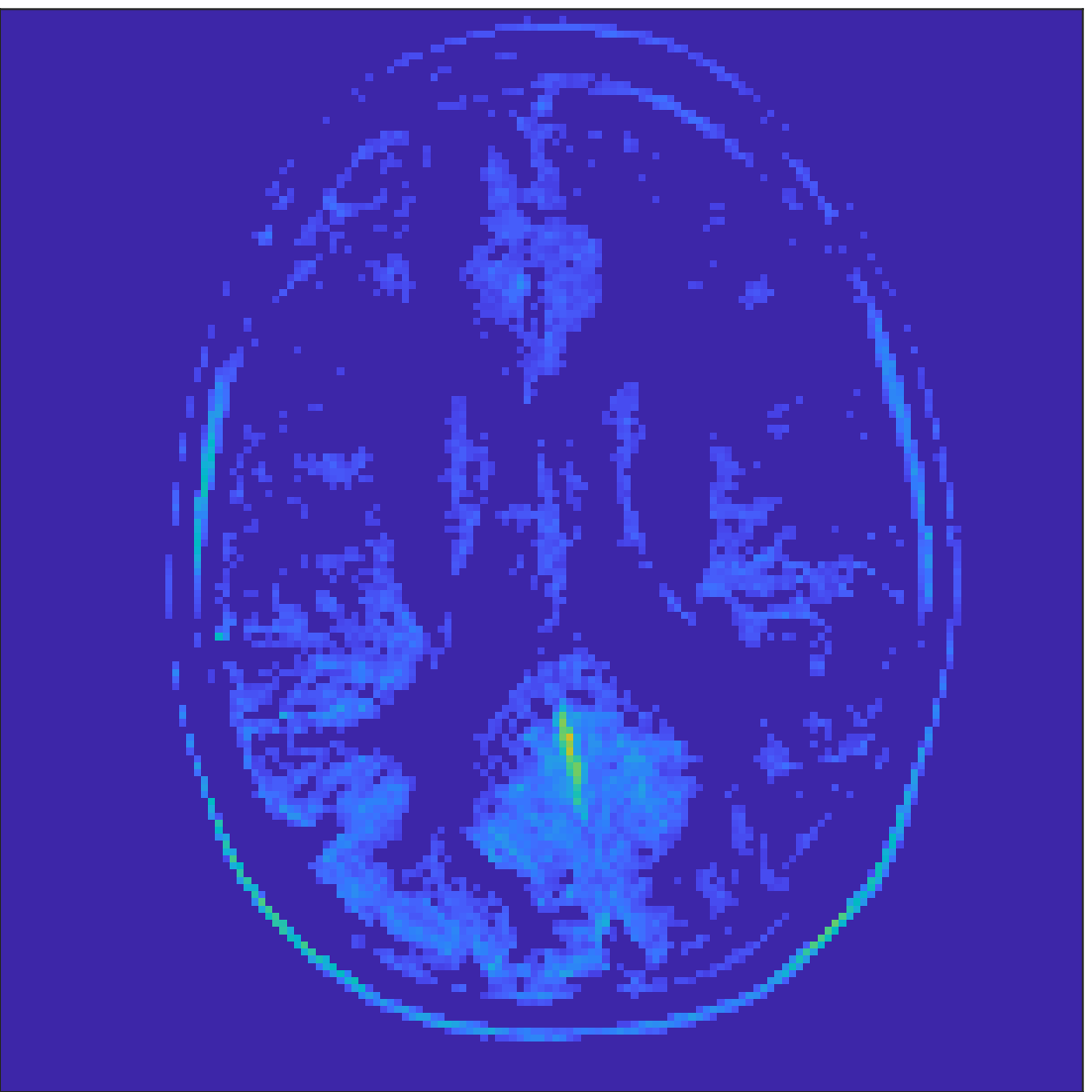}
&	\includegraphics[width=0.3\linewidth]{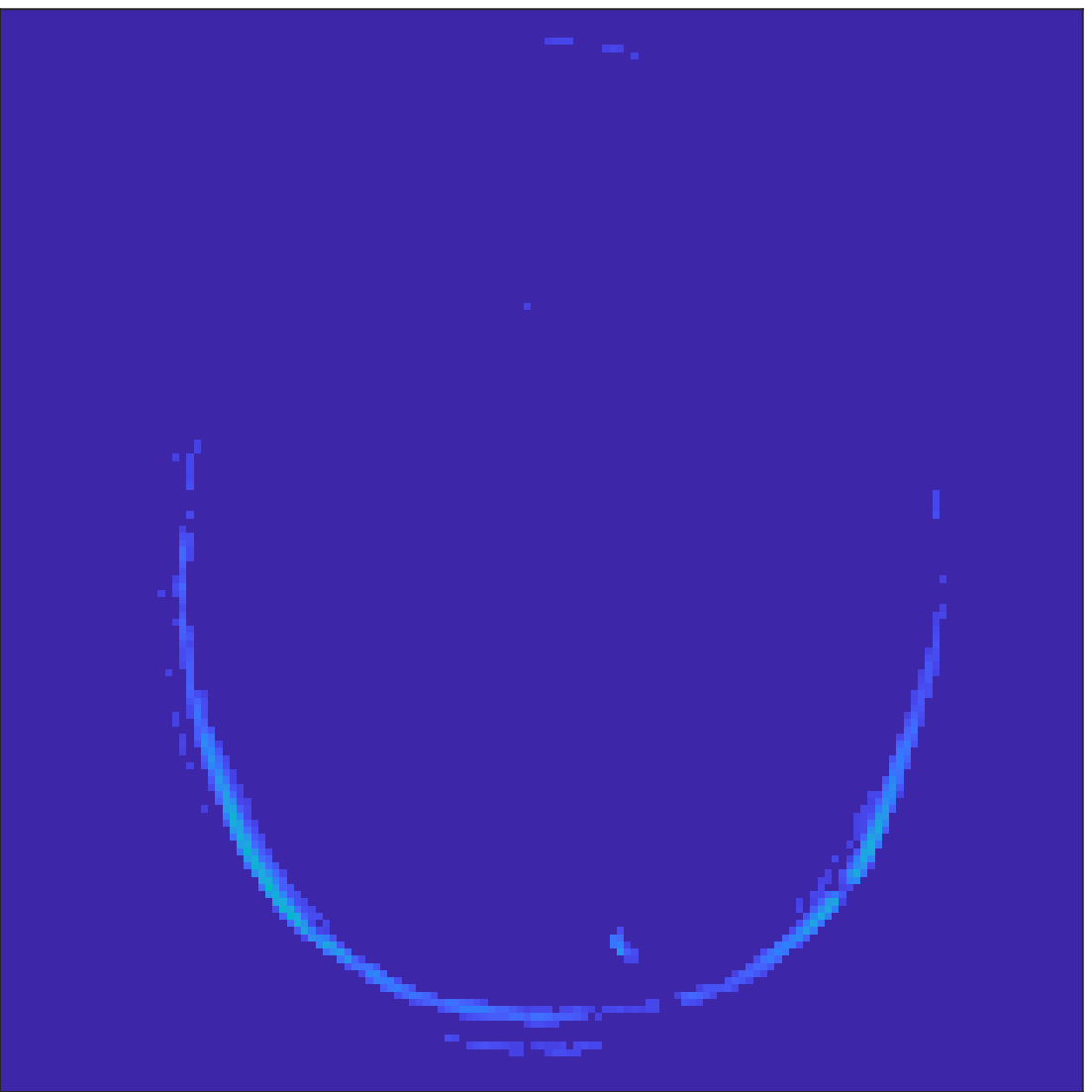}	
&	\includegraphics[width=0.38cm]{fig/spiralcolorbar5.eps}\\
\end{tabular}
\endgroup
	\caption{Normalised proton density maps corresponding to the brain dataset with spiral sampling. The corresponding $T_1$ and $T_2$ can be seen in Table~\ref{tab:spiral}. The reconstructions were trimmed from $180\times 180$ to $151\times 151$ voxels. From left to right and top to bottom, the figures correspond to WM, GM, CSF, muscle and fat.}
    \label{im_spiralresults}
\end{figure}
\begin{table}[!t]\footnotesize
\centering
\caption{Comparison between the parameters obtain with GAP-MRF and the reported values in \cite{BOJORQUEZ201769} (MRF FISP sequence) for the brain dataset with spiral sampling.}
\label{tab:spiral}
\begin{tabular}{l|r|r|r|r|}
\cline{2-5}
                                   & \multicolumn{2}{c|}{Values reported in \cite{BOJORQUEZ201769}} & \multicolumn{2}{c|}{GAP-MRF} \\ \cline{2-5} 
                                  & \multicolumn{1}{|c|}{$T_1$}           & \multicolumn{1}{|c|}{$T_2$}        & \multicolumn{1}{|c|}{$T_1$}        & \multicolumn{1}{|c|}{$T_2$}   \\ \hline
\multicolumn{1}{|l|}{WM} & 781$\pm$61      & 65$\pm$6      & 758.7        & 42.1         \\ \hline
\multicolumn{1}{|l|}{GM}  & 1193$\pm$65     & 109$\pm$11    & 872.4        & 67.3         \\ \hline
\multicolumn{1}{|l|}{CSF}          &     &               & 1658.5        & 799.8       \\ \hline
\multicolumn{1}{|l|}{Muscle}       & 1100$\pm$59     & 44$\pm$9      & 1218.0        & 23.2         \\ \hline
\multicolumn{1}{|l|}{Fat}          & 253$\pm$42      & 68$\pm$4      & 325.5       & 68.1        \\ \hline
\end{tabular}
\end{table}

In Fig.~\ref{im_spiralresults}, we can observe the resulting proton density maps provided by the GAP-MRF algorithm and the Table~\ref{tab:spiral} shows a comparison between the parameters reported in \cite{BOJORQUEZ201769} and the parameters obtained by GAP-MRF. The WM, GM and Fat parameters obtained by GAP-MRF differ slightly to the ones reported in \cite{BOJORQUEZ201769}. The muscle parameters are far from the expected values. This could be due to the small number of pure voxels that are not sufficient to accurately estimate the parameters. We believe that choosing better acquisition parameters $\boldsymbol{\Gamma}$ to make the elements of the dictionary more distant in the $\ell_2$-norm sense can significantly improve the accuracy of the parameters. Also, inaccuracies in the model such as calibration or motion in the acquisition can produce artefacts in the reconstruction.
\begin{figure}[!t]
	\centering
	\begingroup
    \setlength{\tabcolsep}{0.2mm}
\renewcommand{\arraystretch}{0.2}
\begin{tabular}{cccc}
     \includegraphics[width=0.3\linewidth]{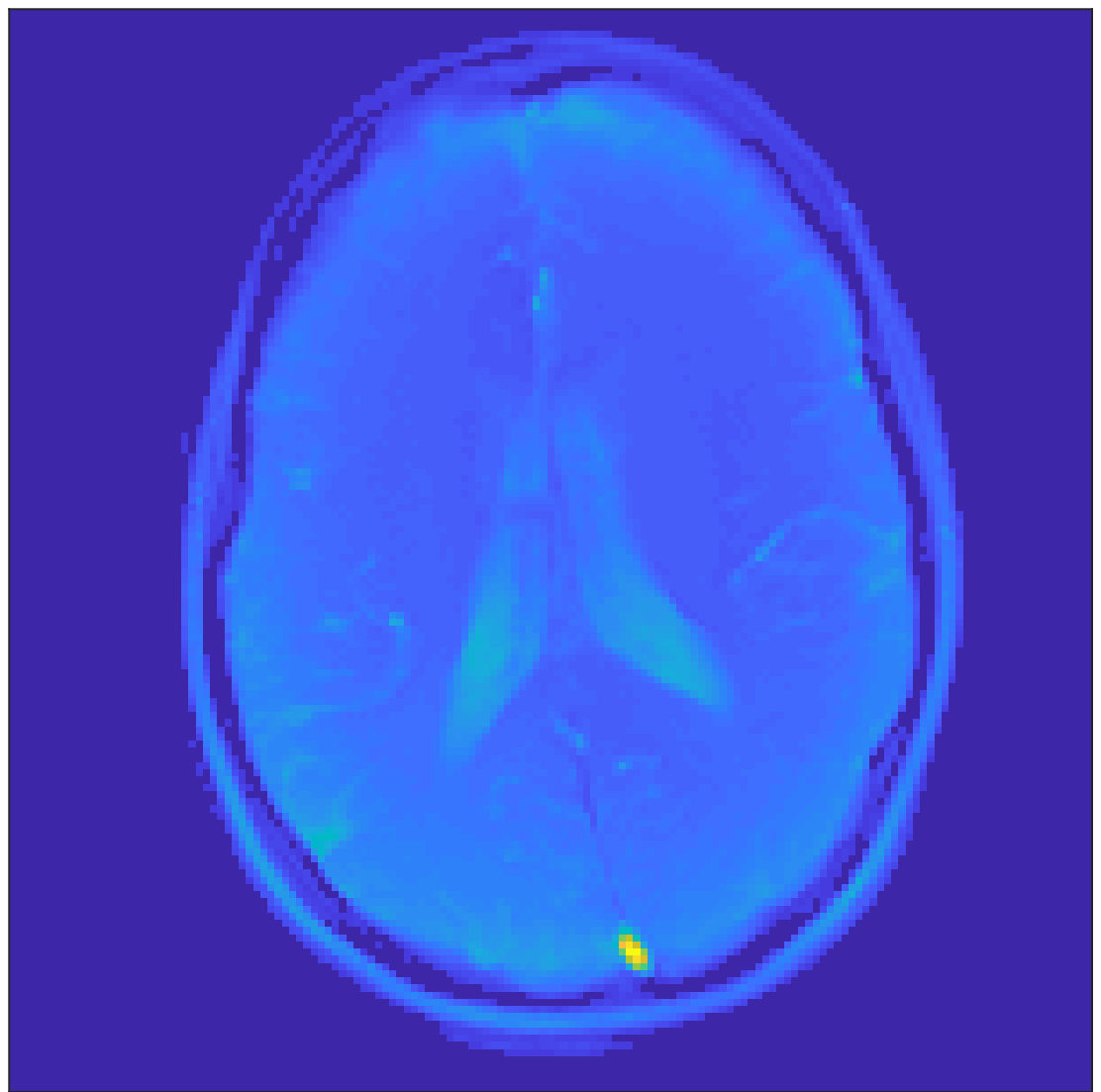}& \includegraphics[width=0.3\linewidth]{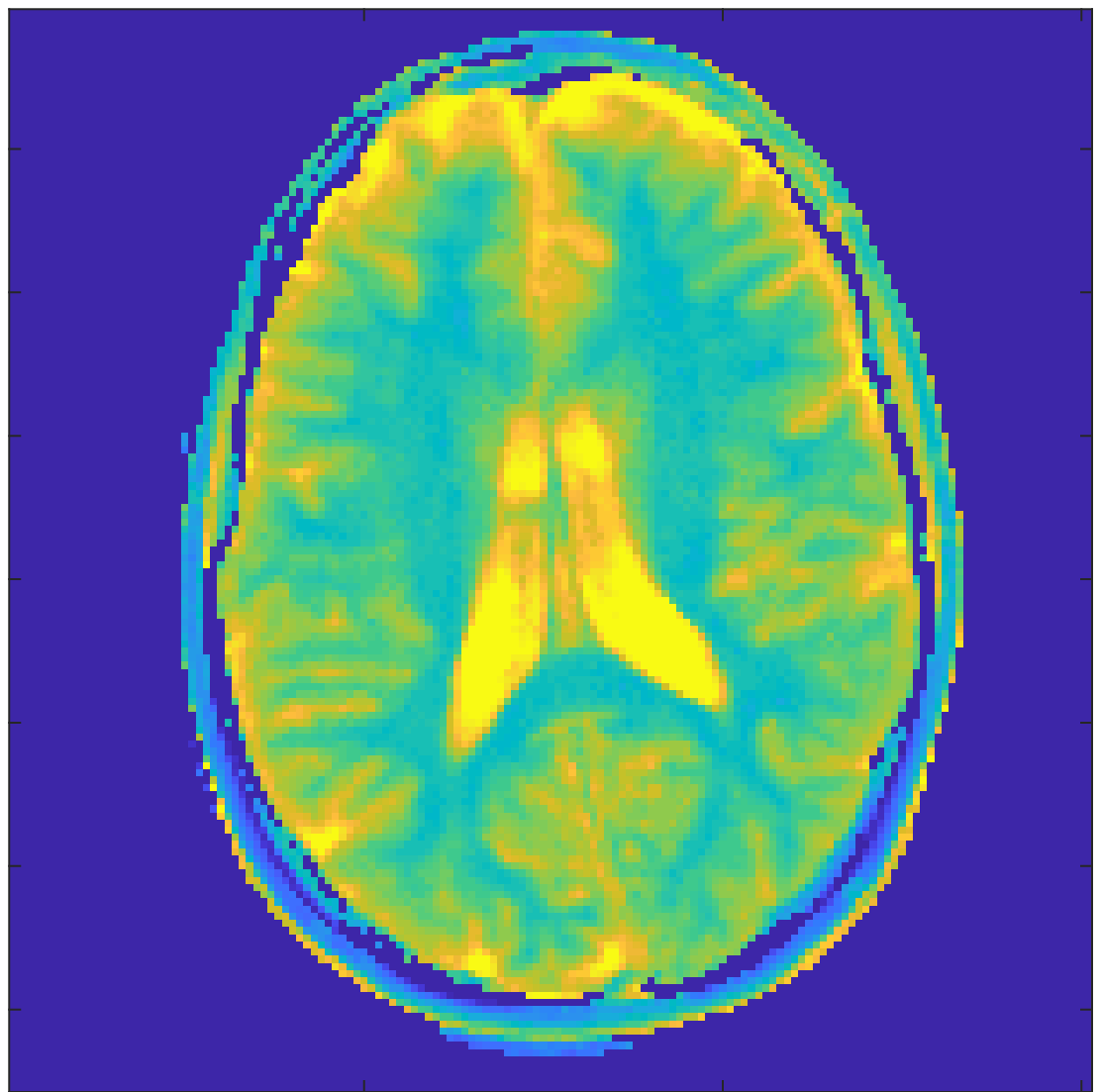}& \includegraphics[width=0.3\linewidth]{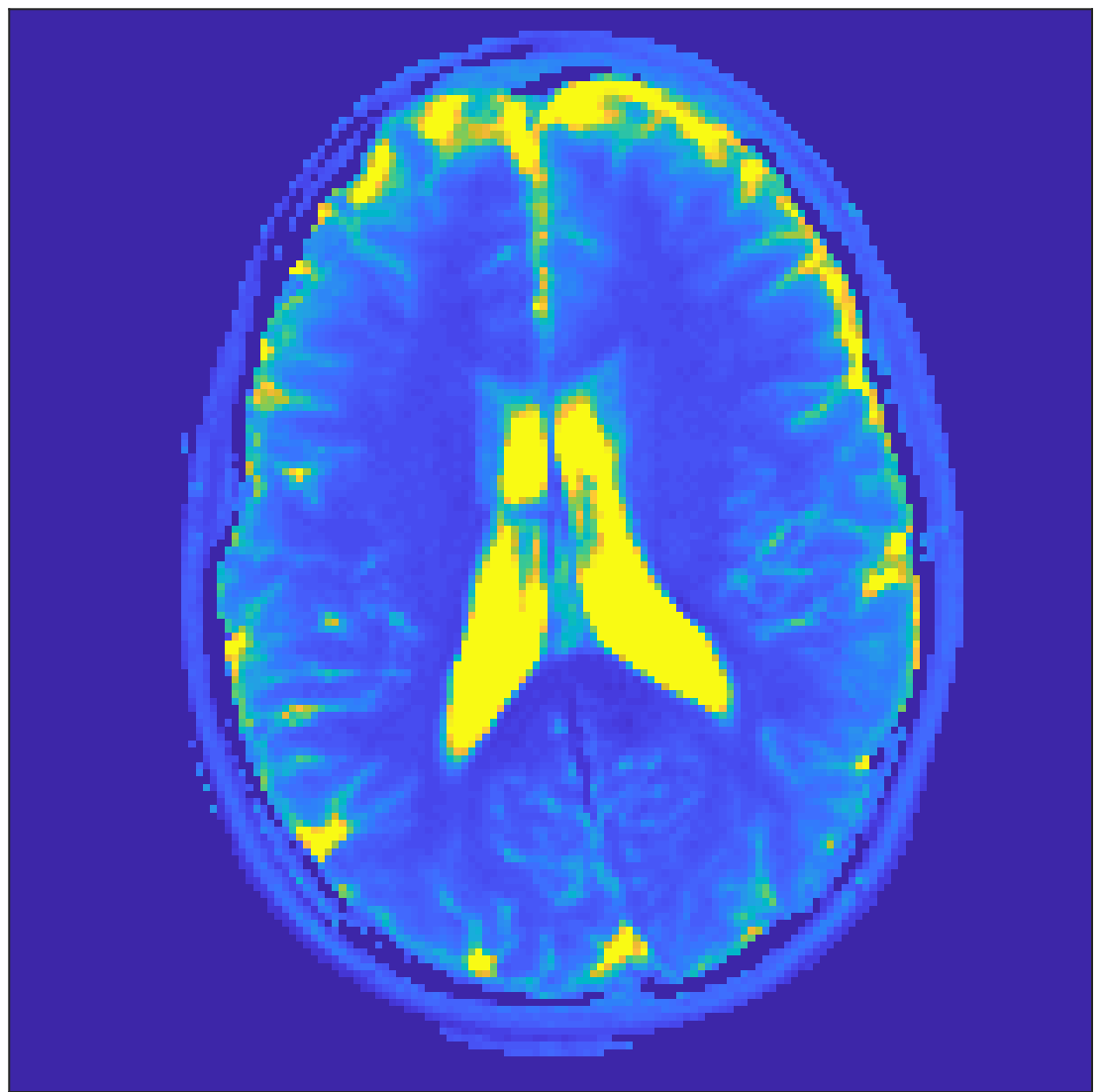}\\ 
	\includegraphics[width=0.3\linewidth]{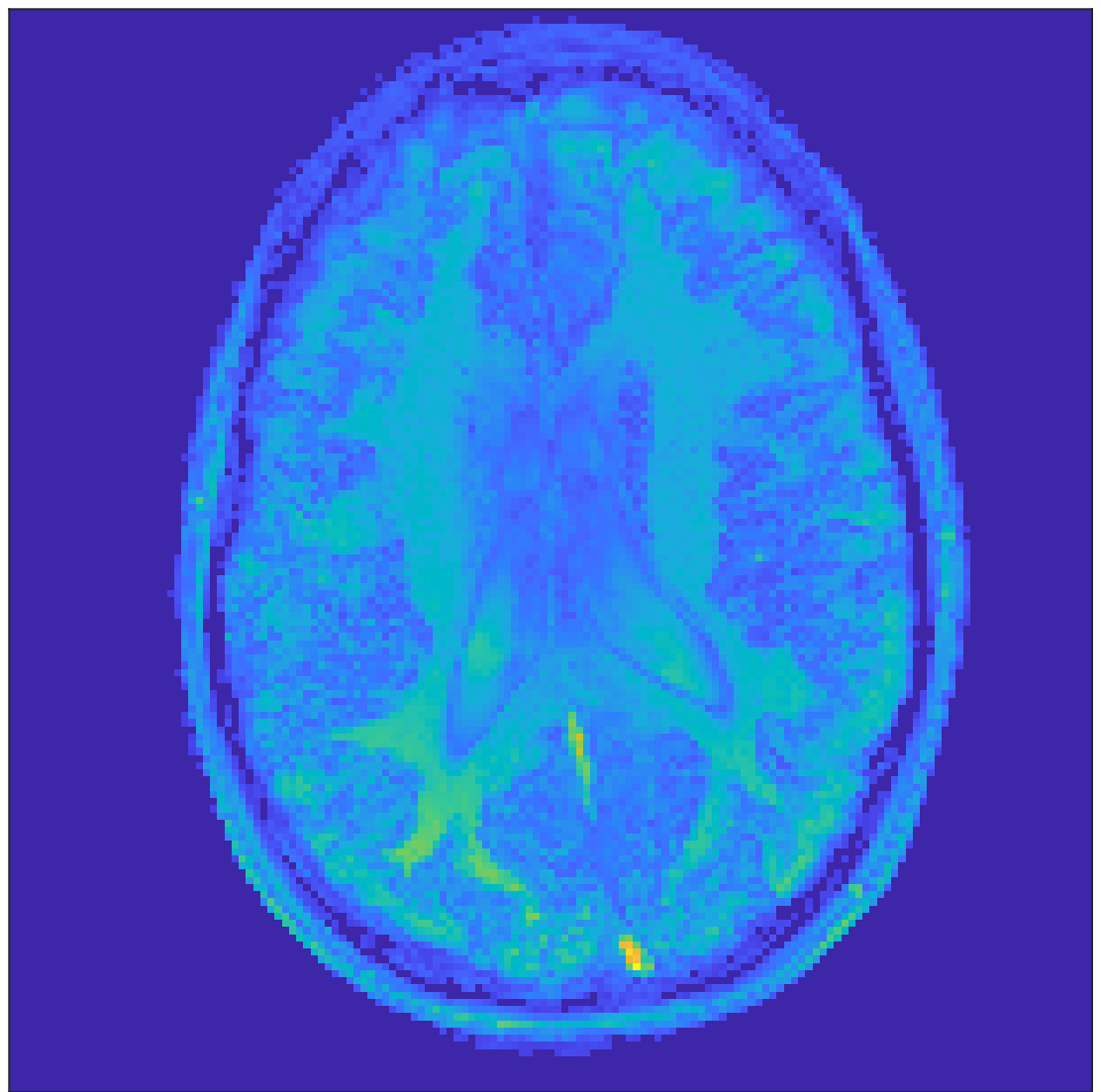}& \includegraphics[width=0.3\linewidth]{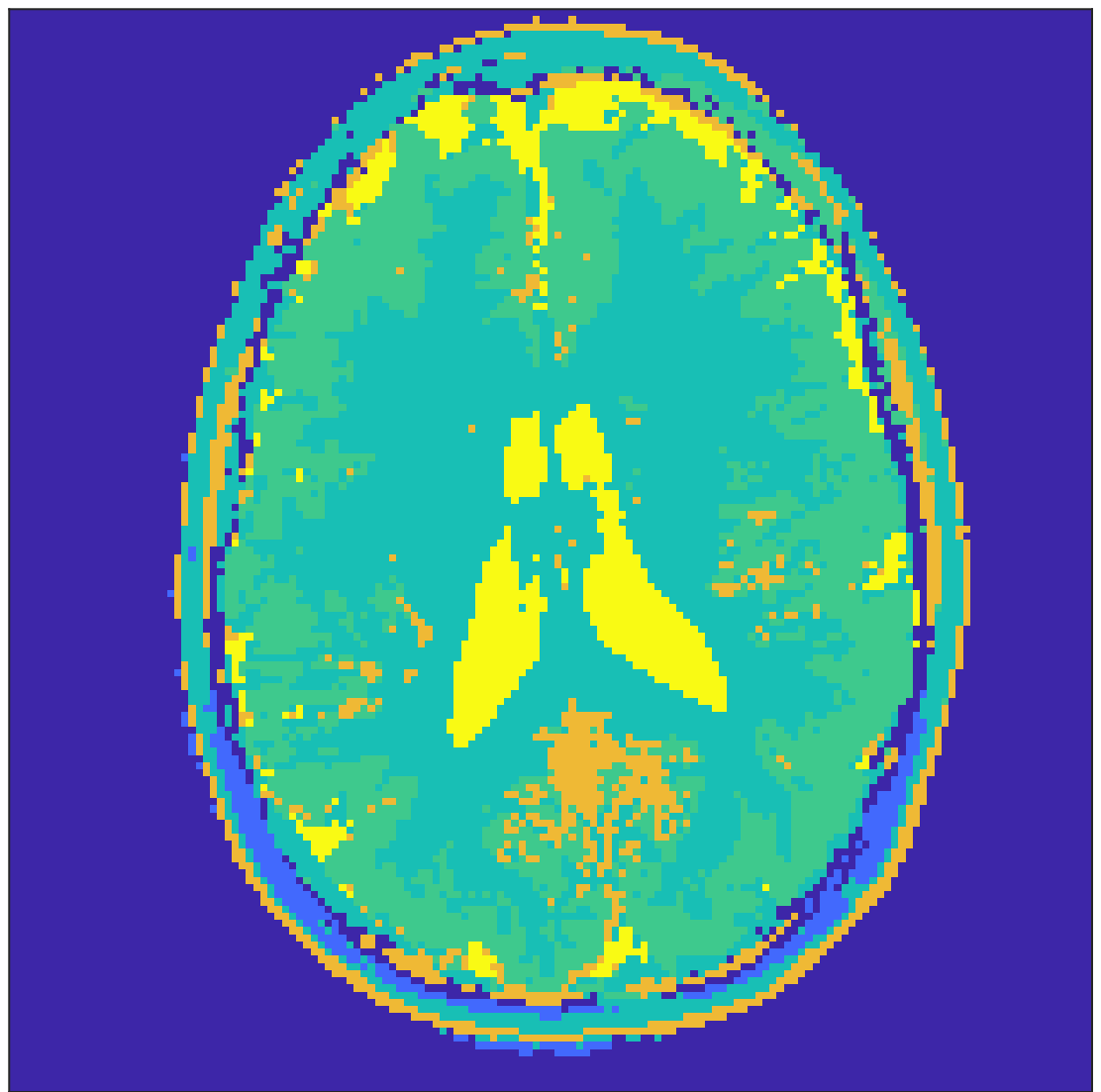}& \includegraphics[width=0.3\linewidth]{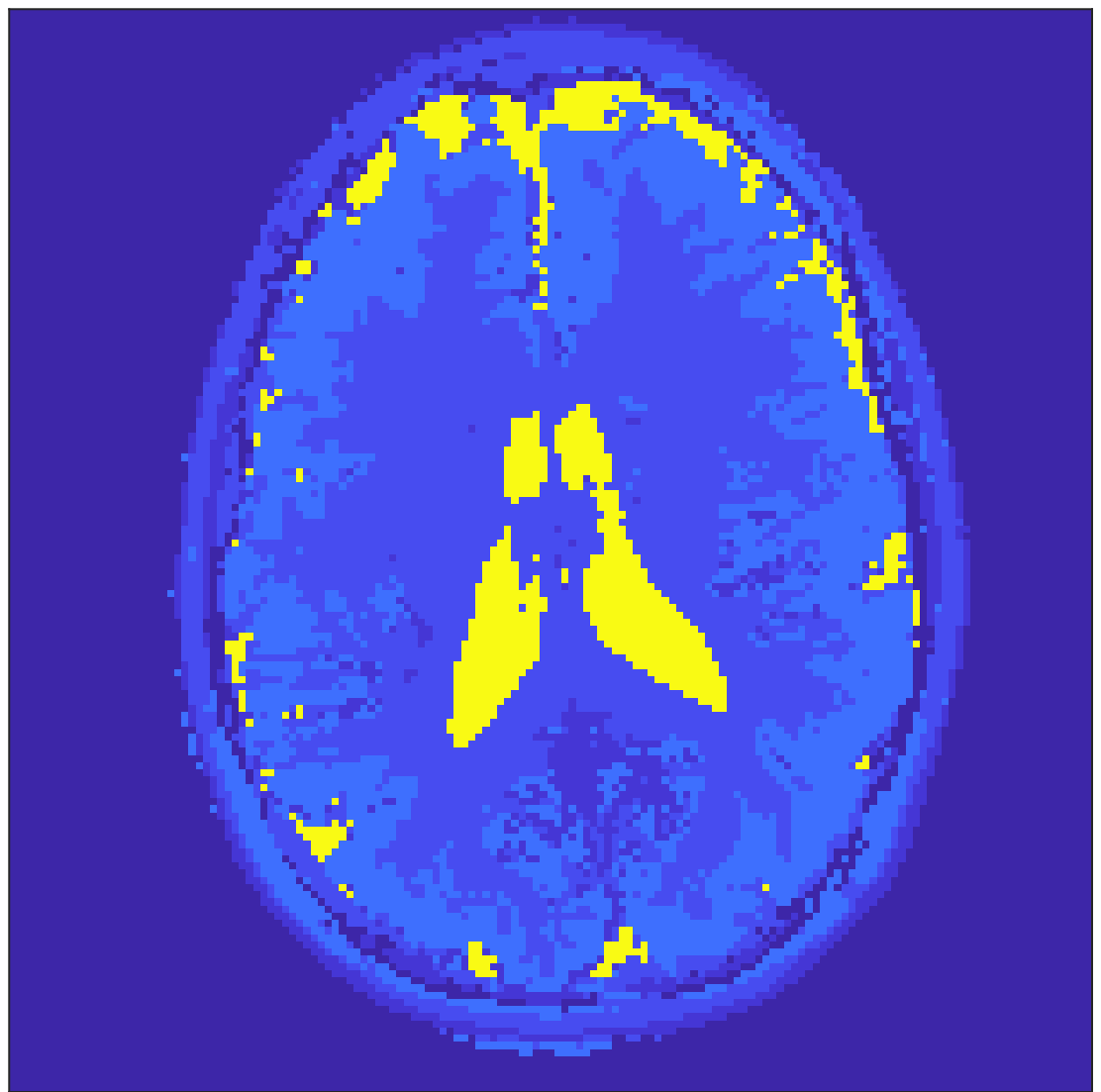}\\
	\includegraphics[width=0.3\linewidth]{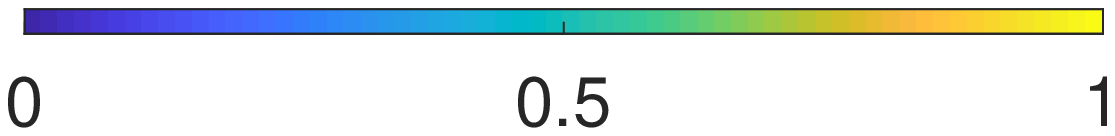}&
	 \includegraphics[width=0.3\linewidth]{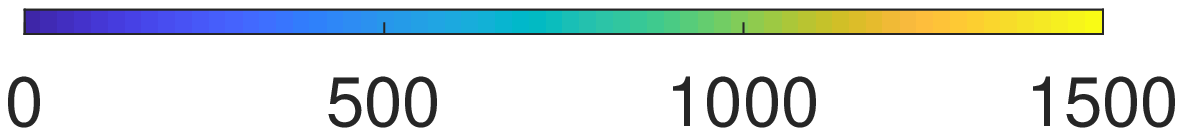}
	& \includegraphics[width=0.3\linewidth]{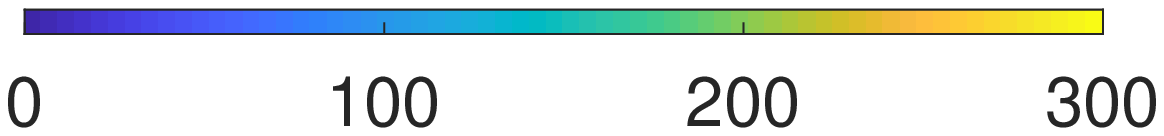}\\
\end{tabular}
	\endgroup
	 \caption{Dominant tissue parameter maps corresponding to the brain dataset with spiral sampling. The First row corresponds to the BLIP reconstructions and the second row to the GAP-MRF reconstructions. The reconstructions were trimmed from $180\times 180$ to $151\times 151$ voxels.. From left to right, the columns correspond to the normalised proton density, $T_1$ and $T_2$. The values of $T_1$ and $T_2$ are capped to $1500$ms and $300$ms respectively.}
    \label{im_spiraldominant}
\end{figure}

In Fig.~\ref{im_spiraldominant}, the $T_1$ and $T_2$  maps reconstructed by BLIP show a smooth transition from one tissue to another (similar to the simulated phantom). Moreover, the proton density map reconstructed by BLIP does not provide any information on the tissue distribution. On the contrary, GAP-MRF reconstructions show abrupt transitions in the $T_1$ and $T_2$ maps of the dominant tissues. In addition, the proton density map shows more structure than BLIP, but not all the tissue structures are appreciated compared to the normalised proton density maps.

\subsection{\textit{In vivo} brain dataset with EPI sampling}
The scanning for this dataset has been performed on a 3T GE MR750w scanner with a 12 channel receive only head RF coil (GE Medical
Systems, Milwaukee, WI). The study was approved by the local ethics committee. The used acquisition scheme was 16-shot EPI-MRF on a healthy volunteer using a variable flip angle $\alpha$ ramp, ranging from $1^{\circ}$ to $70^{\circ}$. The excitation sequence length is $L=500$. The repetition time TR was set to $16$ms. In \cite{gomez2017}, it was shown to be as effective at estimating the MRF parameters but had better sensitivity than the FISP sequence in \cite{Jiang2015}. The acquisition bandwidth (BW) = $5$kHz and the Field of View (FOV)~=~$22.5\times22.5$cm$^2$. The spatial resolution is $128\times 128$ voxels, with a $5$mm slice thickness. The undersampling ratio is $N/Q=16$. The EPG model is used for the reconstructions with an Inversion Time (TI) of $18$ms and an Echo Time (TE) of $3.5$ms. The acquisition time for the slice was $9$s. A reference scan with null Gy gradient was performed for phase correction of EPI raw data.

\begin{figure}[!t]
\centering
\begingroup
\setlength{\tabcolsep}{0.2mm}
\renewcommand{\arraystretch}{0.5}
\begin{tabular}{cccc}
	\includegraphics[width=0.3\linewidth]{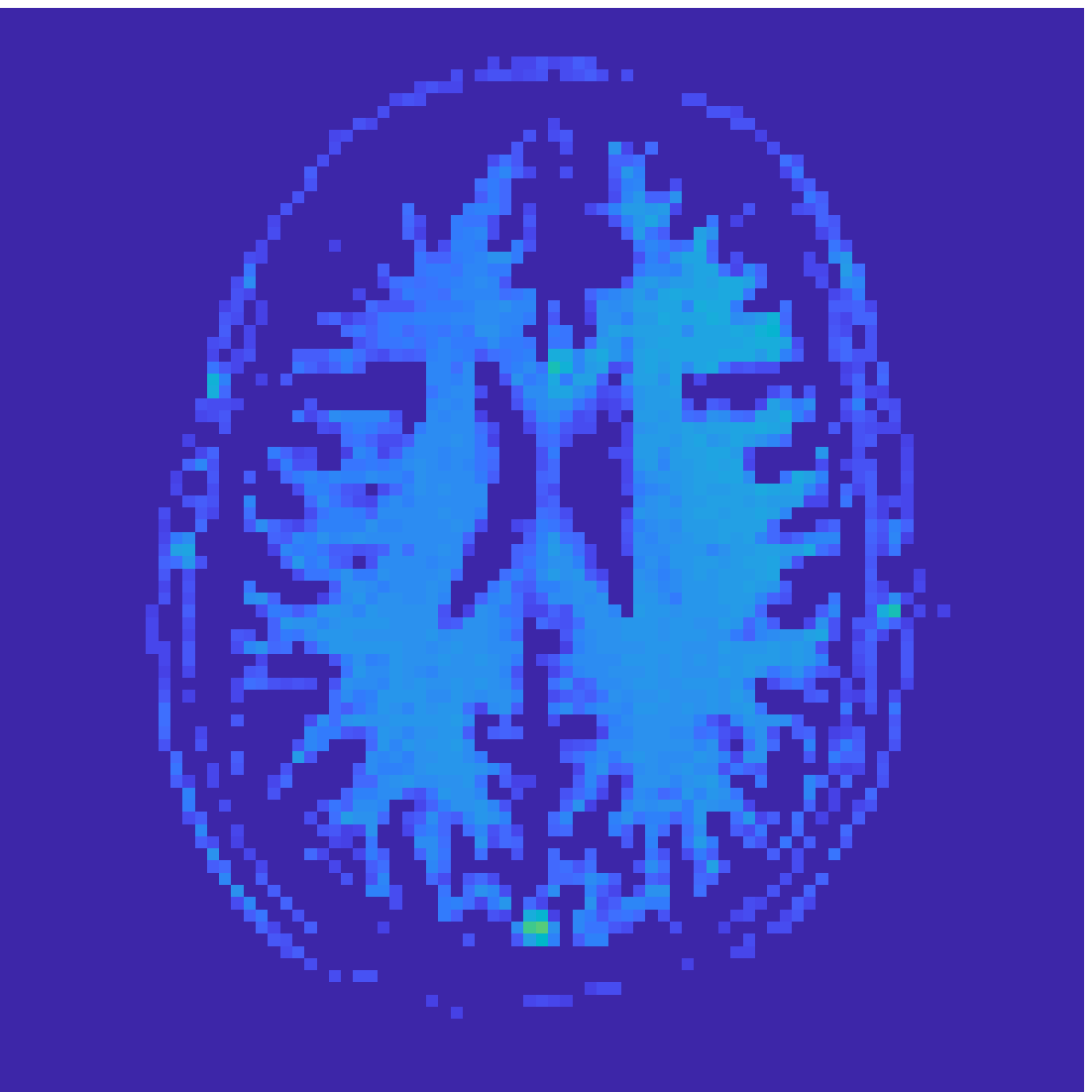}
&	\includegraphics[width=0.3\linewidth]{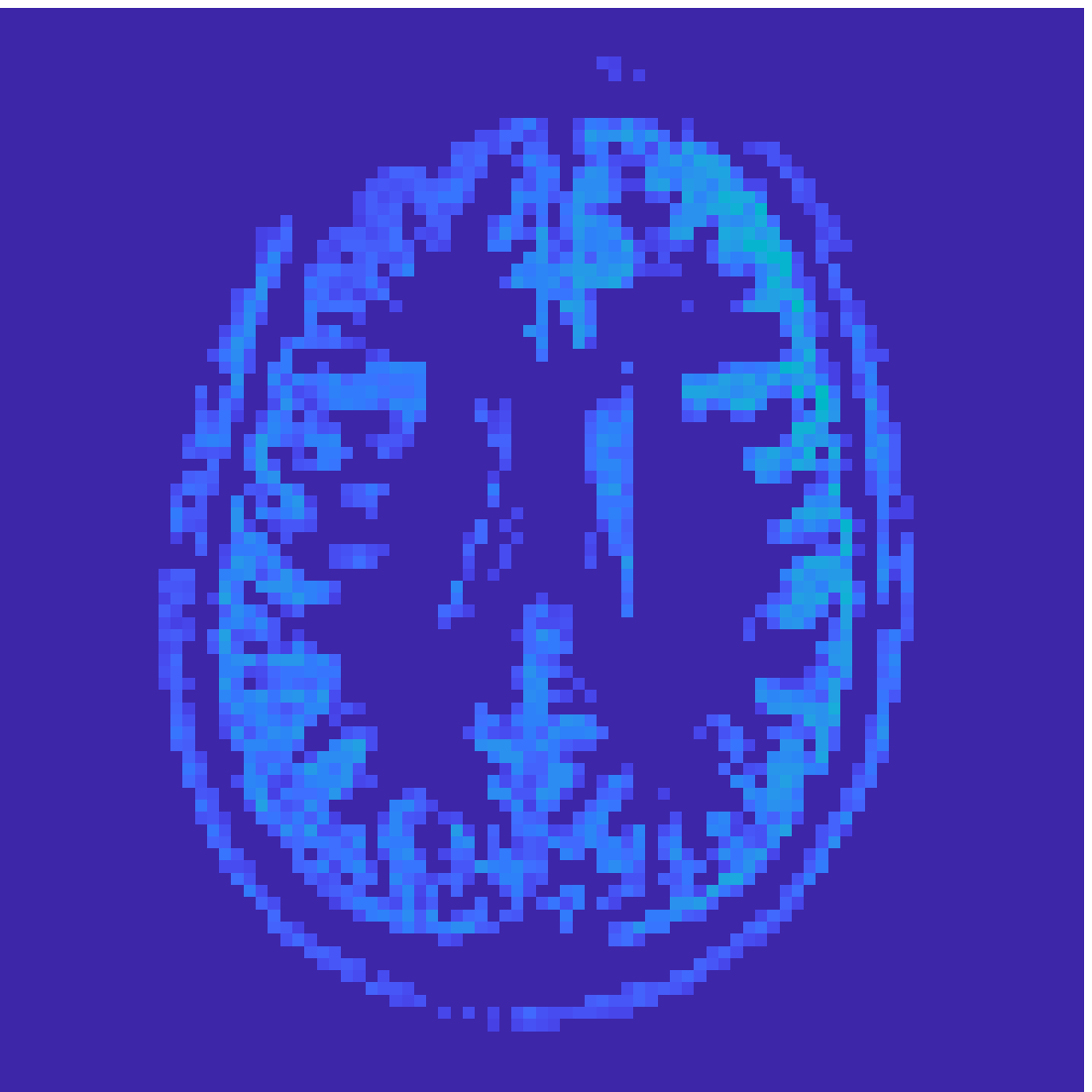}
&	\includegraphics[width=0.3\linewidth]{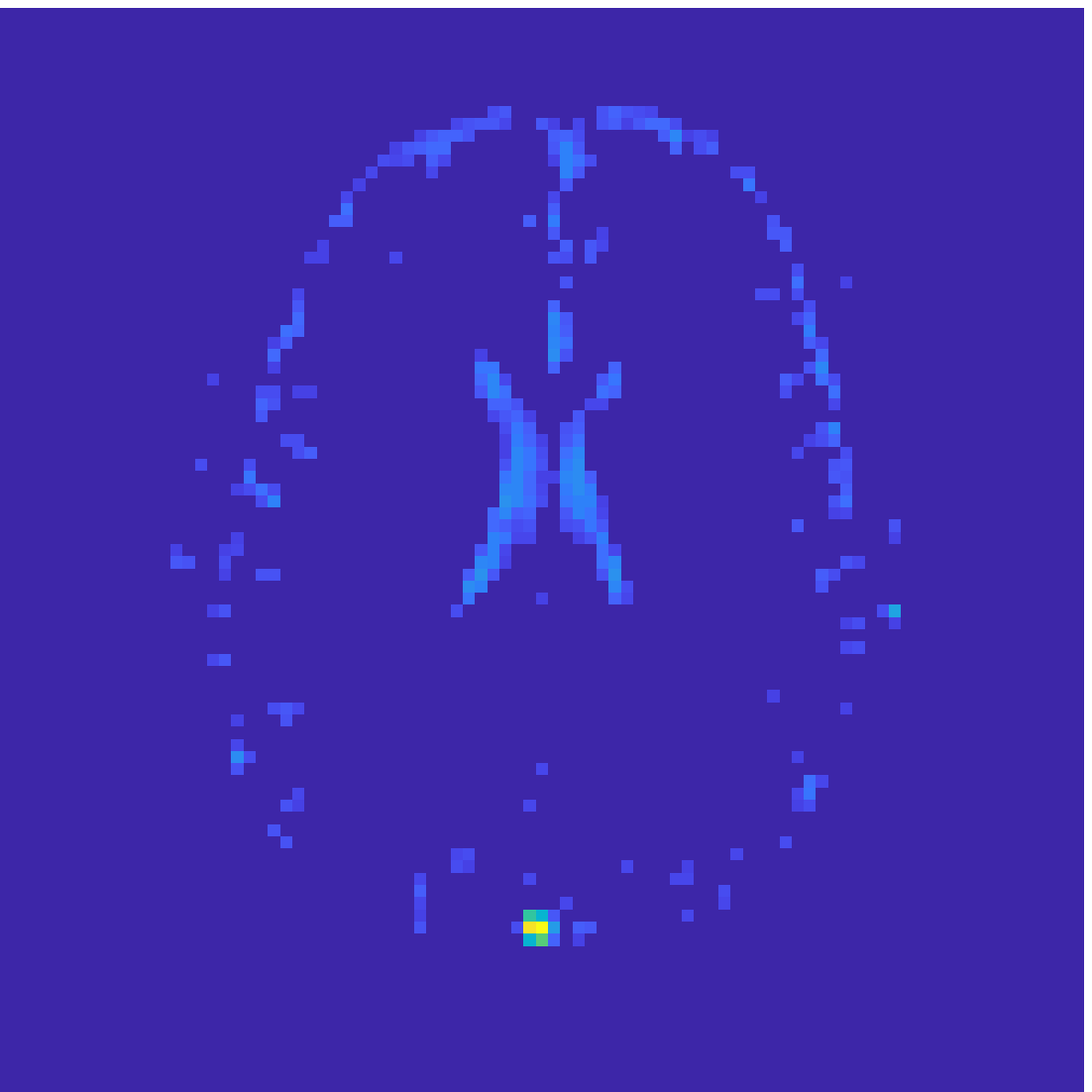}
&	\includegraphics[width=0.38cm]{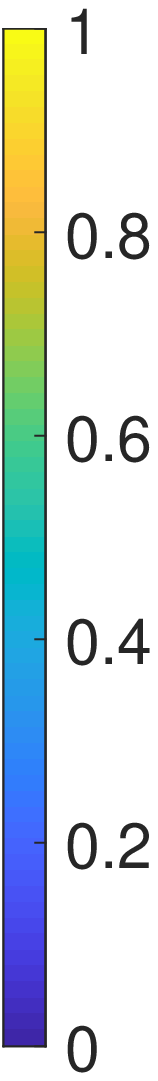}
\end{tabular}
\endgroup
	\caption{Normalised proton density maps corresponding to the brain dataset with EPI sampling.  The reconstructions were trimmed from $128\times 128$ to $89\times 89$ voxels. The corresponding $T_1$ and $T_2$ can be seen in Table~\ref{tab:epi}. From left to right, the figures correspond to WM, GM and CSF.}
    \label{im_epiresults}
\end{figure}
\begin{table}[!t]\footnotesize
\centering
\caption{Comparison between the parameters obtain with GAP-MRF and the reported values in \cite{BOJORQUEZ201769} (MRF FISP sequence) for the brain dataset with EPI sampling.}
\label{tab:epi}
\begin{tabular}{l|r|r|r|r|}
\cline{2-5}
                                   & \multicolumn{2}{c|}{Values reported in \cite{BOJORQUEZ201769}} & \multicolumn{2}{c|}{GAP-MRF} \\ \cline{2-5} 
                                   & \multicolumn{1}{|c|}{$T_1$}           & \multicolumn{1}{|c|}{$T_2$}        & \multicolumn{1}{|c|}{$T_1$}        & \multicolumn{1}{|c|}{$T_2$}       \\ \hline
\multicolumn{1}{|l|}{WM} & 781$\pm$61          & 65$\pm$6          & 762.6        & 67.2        \\ \hline
\multicolumn{1}{|l|}{GM}  & 1193$\pm$65         & 109$\pm$11        & 1116.6       & 107.1       \\ \hline
\multicolumn{1}{|l|}{CSF}          &         &               & 2391.1       & 856.2      \\ \hline
\end{tabular}
\end{table}

In Fig.~\ref{im_epiresults}, we can observe the resulting proton density maps provided by the GAP-MRF algorithm and the Table~\ref{tab:epi} shows a comparison between the parameters reported in \cite{BOJORQUEZ201769} for MRF FISP sequences and the parameters obtained by GAP-MRF. CSF values are not reported for the MRF FISP sequence. The WM parameters are similar to the ones reported in \cite{BOJORQUEZ201769} and the GM $T_1$ is slightly lower than the reported one. We believe that the lack of pure voxels (due the spatial resolution) made the approach unable to find the other tissues.

\begin{figure}[!t]
	\centering
	\begingroup
    \setlength{\tabcolsep}{0.2mm}
\renewcommand{\arraystretch}{0.2}
\begin{tabular}{cccc}
     \includegraphics[width=0.3\linewidth]{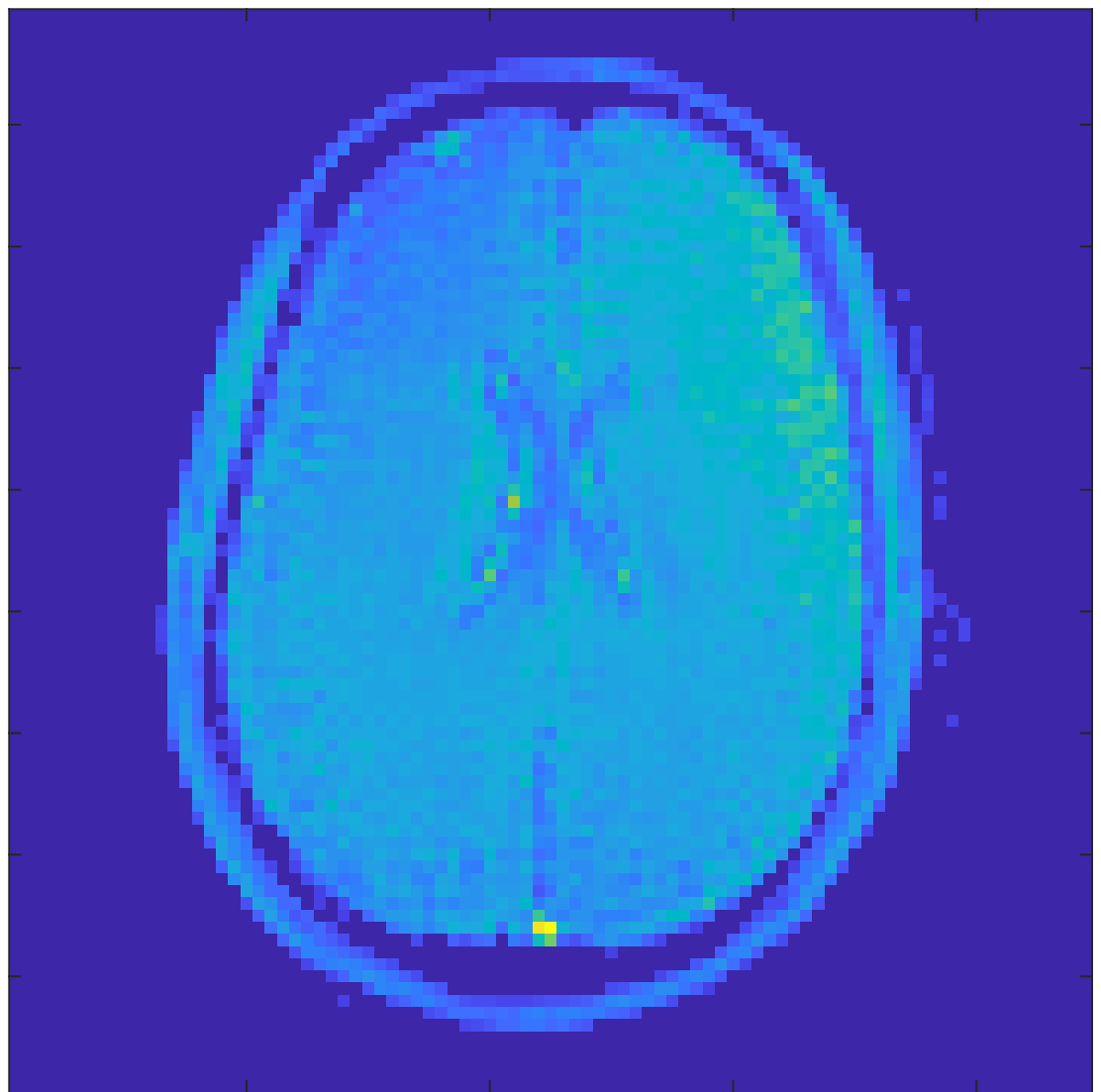}& \includegraphics[width=0.3\linewidth]{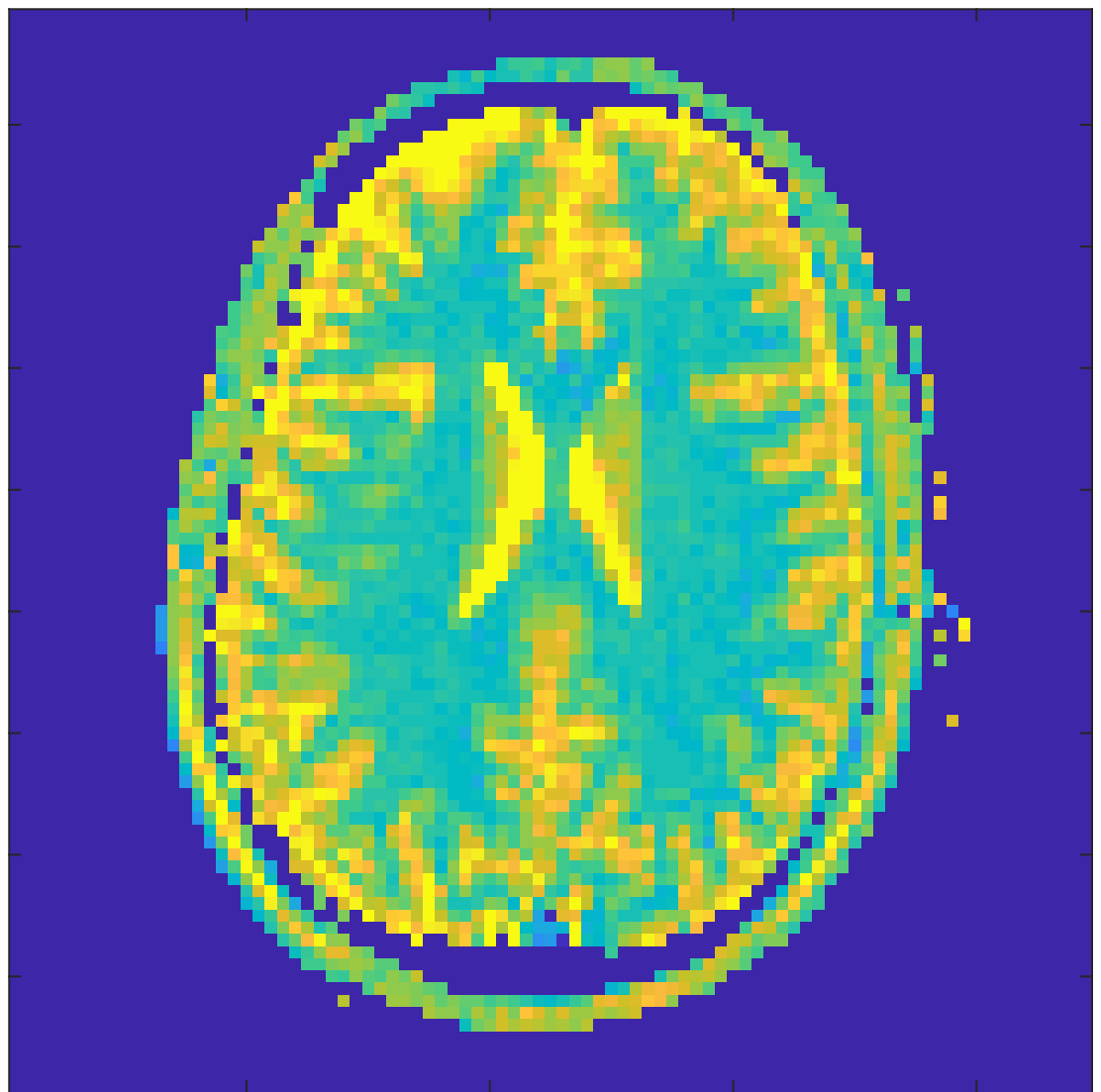}& \includegraphics[width=0.3\linewidth]{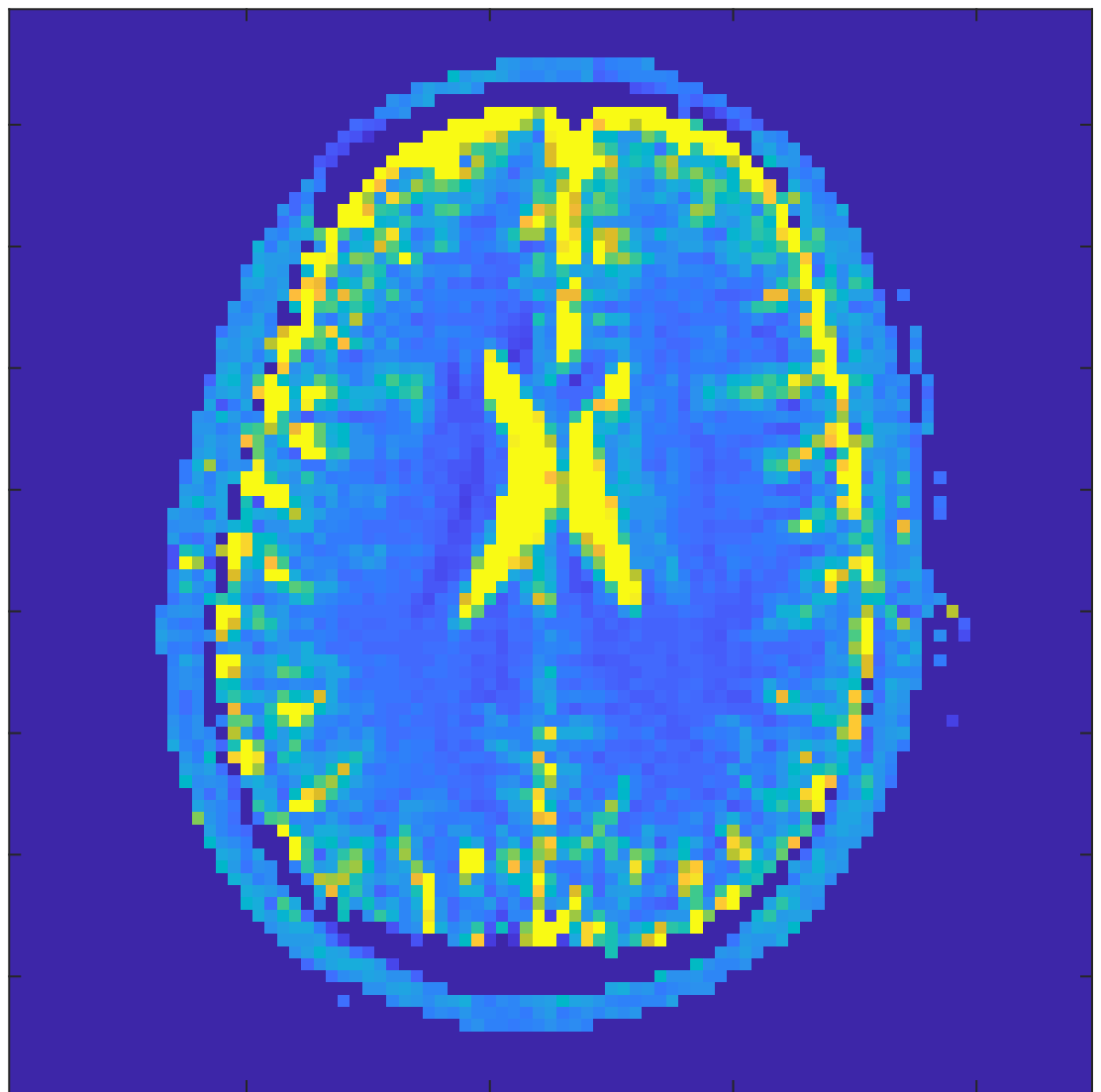}\\ 
	\includegraphics[width=0.3\linewidth]{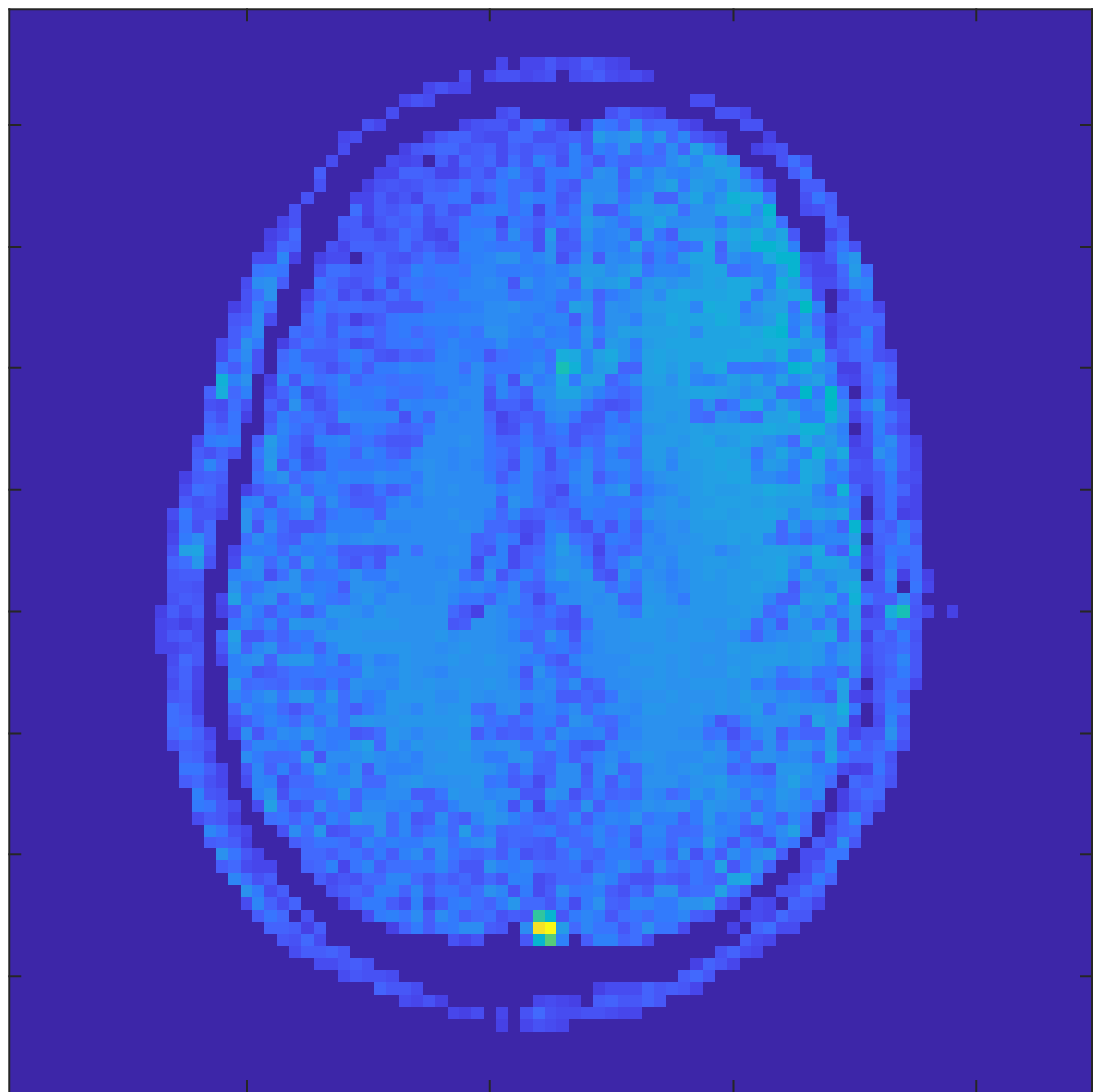}& \includegraphics[width=0.3\linewidth]{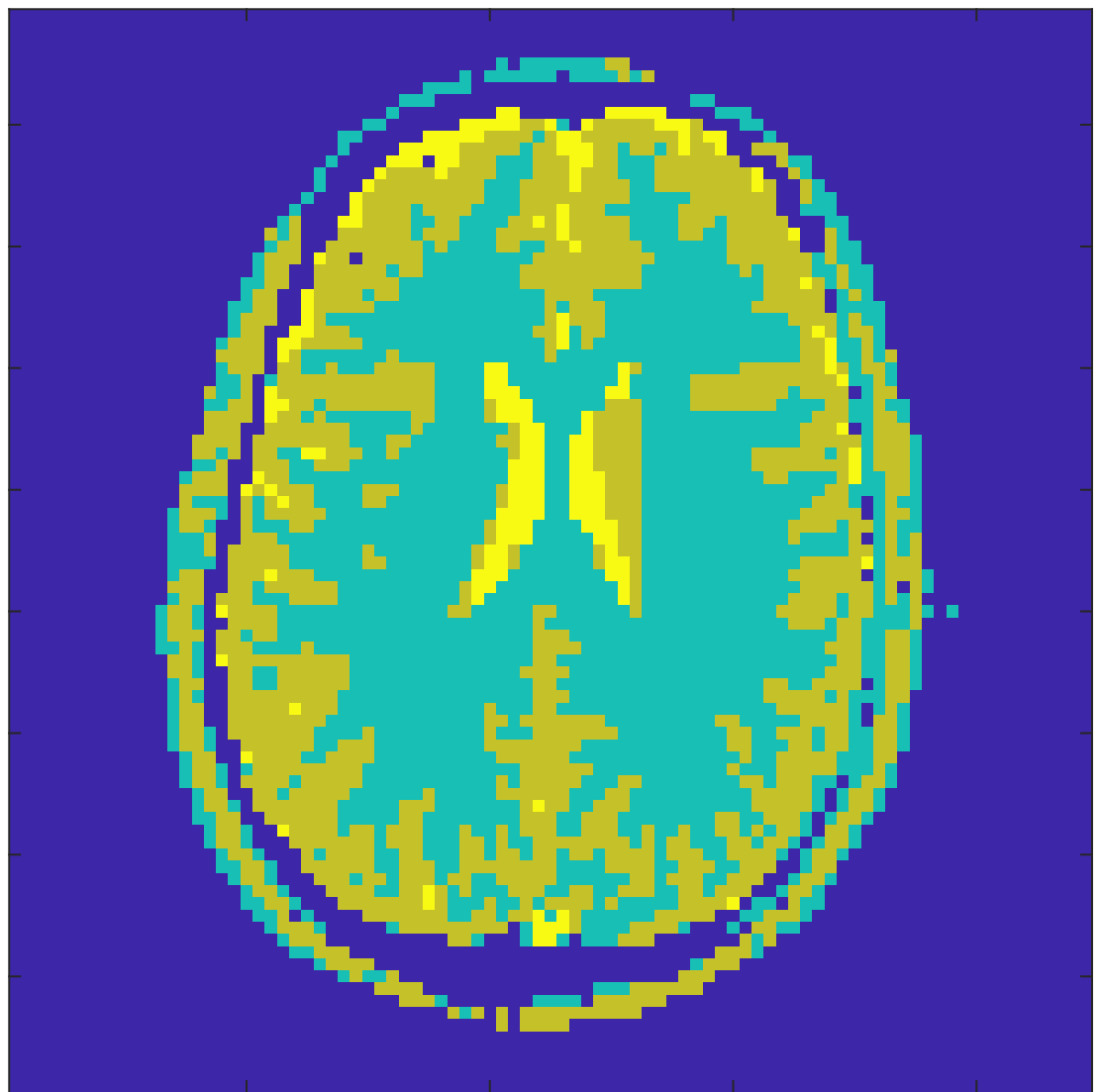}& \includegraphics[width=0.3\linewidth]{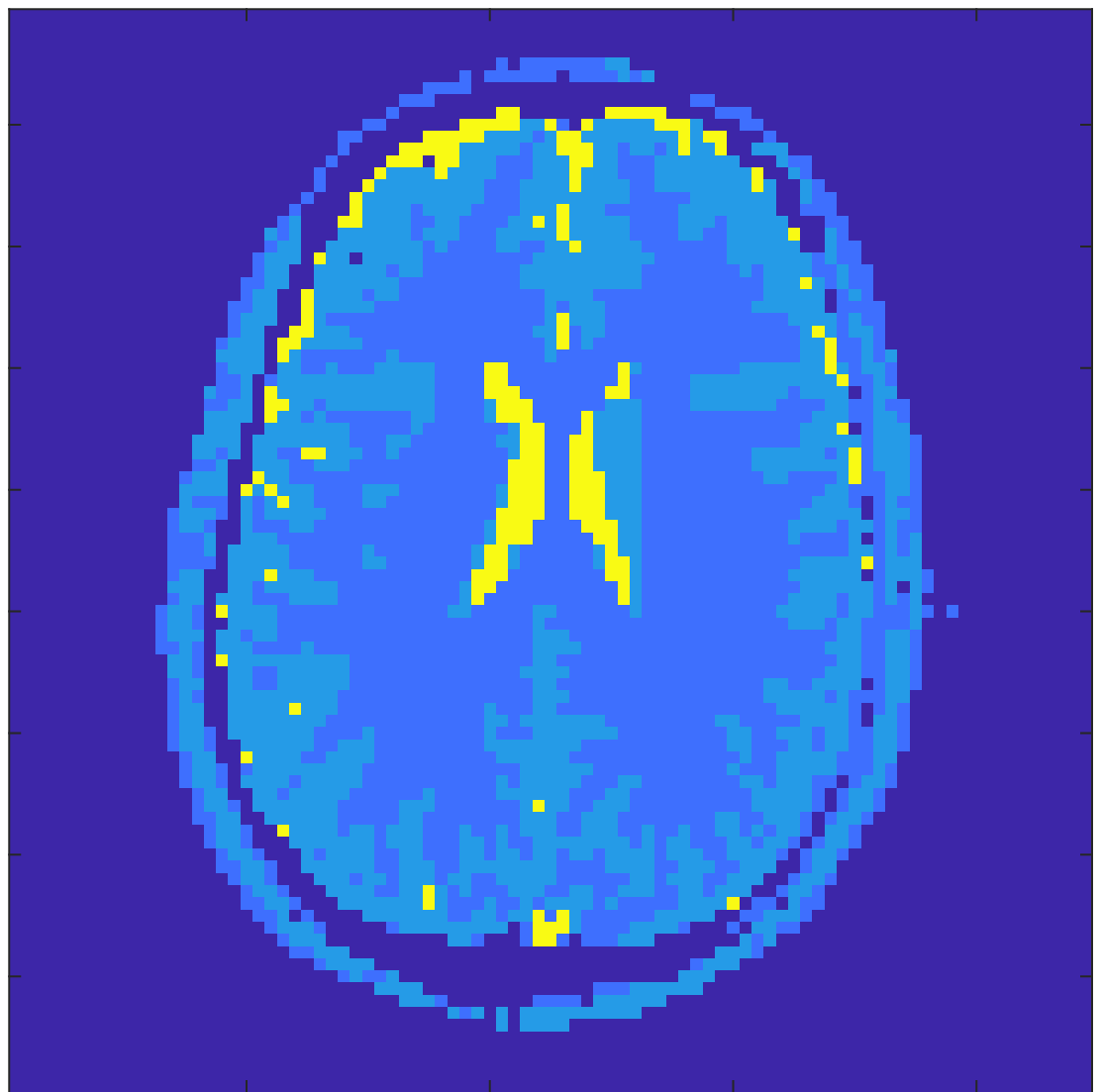}\\
	\includegraphics[width=0.3\linewidth]{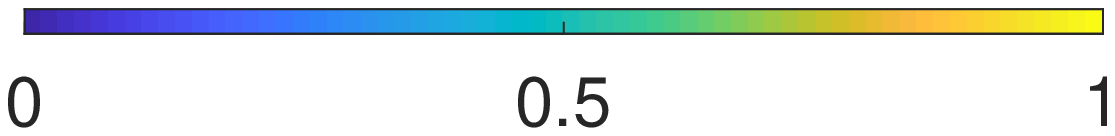}&
	 \includegraphics[width=0.3\linewidth]{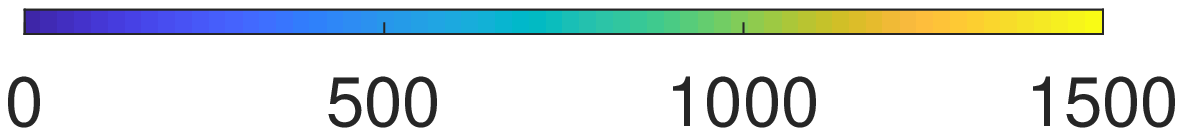}
	& \includegraphics[width=0.3\linewidth]{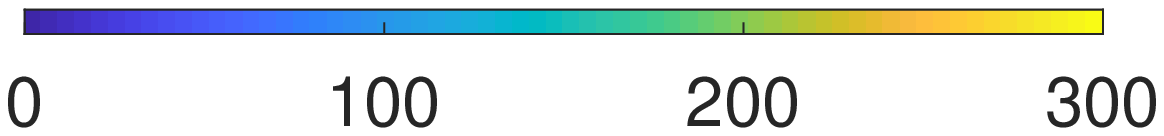}\\
\end{tabular}
	\endgroup
	 \caption{Dominant tissue parameter maps corresponding to the brain dataset with EPI sampling. The First row corresponds to the BLIP reconstructions and the second row to the GAP-MRF reconstructions. The reconstructions were trimmed from $128\times 128$ to $89\times 89$ voxels. From left to right, the columns correspond to the normalised proton density, $T_1$ and $T_2$. The values of $T_1$ and $T_2$ are capped to $1500$ms and $300$ms respectively.}
    \label{im_epidominant}
\end{figure}

In Fig.~\ref{im_epidominant}, the $T_1$ and $T_2$ maps reconstructed by BLIP shows a smooth transition from one tissue to another. On the contrary, GAP-MRF reconstructions show abrupt transitions in the $T_1$ and $T_2$ maps of the dominant tissues.


\section{Conclusions and future work}\label{Sec:Conclusions}
We have presented an extension of the model in \cite{CSMRF} to PV reconstructions in the context of MRF. Our algorithm provides a way to explore the manifold of magnetic resonance fingerprints without densely sampling $\mathcal{M}$. For this reason, the algorithm is memory efficient and the algorithmic structure allows parallel implementations.
The proposed model assumes that the number of independent tissues in the imaged volume is upper bounded, and that each tissue has a minimum number of pure voxels. Also, the parameters of each tissue should be sufficiently different to be distinguished. Finally, we assume that the combination of the sampling patterns should cover most of the $k$-space to avoid high frequency artefacts.

The simulation results presented in Section~\ref{Sec:simulations} show that the proposed GAP-MRF method can achieve accurate reconstructions with very short pulse sequences in the low input noise scenario. It also performs well when the iSNR is greater than $30$dB. We also present in Section~\ref{section_realdata} the results obtained with \textit{in vivo} datasets. Some parameters differ slightly to the reported in the literature, but the structure seen in the proton densities maps suggests that this approach can provide additional information that can be useful for diagnosis. 

The next step is to evaluate the PV reconstructions with a real PV phantom in the scanner and a full brain reconstruction to provide enough pure voxels to accurately estimate the true parameters. In particular, an interesting point would be to evaluate the behaviour of GAP-MRF in presence of a pathology. A pathology can be seen as a distinct additional tissue. Therefore, since the number of tissues is estimated along the iterations, if the pathology is represented by enough pure voxels, it should be detected by the algorithm exactly in the same way as for the other tissues. 
In addition, we plan to incorporate spatial regularisation in the objective function to improve the robustness of the method. A joint calibration and imaging problem will also be developed in order to provide both phase estimation and compensation.

\section*{Acknowledgements}
R. D. would like to thank CONACYT and Heriot-Watt University for the PhD funding. This work was supported by the UK Engineering and Physical
Sciences Research Council (EPSRC, grants  EP/M019306/1 and EP/M019802/1). We would like to thank GE for the access to the \textit{in vivo} datasets and Benjamin Arnold for acquiring the EPI dataset. We also would like to thank Zhouye Chen and Mohammad Golbabaee for their help in the code and the BASP Group in Heriot-Watt University for the insightful discussions.

{\small
\bibliographystyle{IEEEtran}
\bibliography{refs}}

\end{document}